\documentclass[preprint,showpacs,preprintnumbers,amsmath,amssymb,nofootinbib]{revtex4}

\usepackage{fancyhdr}
\usepackage{graphics,amssymb,amsmath,amsthm,amscd,array}
\usepackage{graphicx,subfigure}
\usepackage{dcolumn}
\usepackage{bm}
\usepackage{amsmath,dsfont}
\usepackage{color}
\usepackage{hyperref}

  [2007/06/14 v6.76i
  Hypertext links for LaTeX]
   [2008/07/16 v60 Package minitoc (JPFD)] 
\RequirePackage{mtcmess}[2006/03/14]
\mtcPackageInfo[I0001]{minitoc}%
   {*** minitoc package, version 60 ***\@gobble}

\newcommand{\const}{\mathop{\rm const}\nolimits}

\def\IR{{\mathds{R}}}
\def\IZ{{\mathds{Z}}}
\def\p{{\partial}}
\def\vnabla{{\vec{\nabla}}}

\def\vp{{\vec{p}}}
\def\vx{{\vec{r}}}
\def\hx{{\hat{r}}}
\def\vn{{\hat{n}}}

\def\vJ{{\vec{J}}}
\def\vL{{\vec{L}}}
\def\vK{{\vec{K}}}

\def\vA{{\vec{A}}}
\def\vB{{\vec{B}}}
\def\vD{{\vec{D}}}
\def\vC{{\vec{C}}}
\def\vX{{\vec{X}}}

\def\cD{{\cal{D}}}
\def\vW{{\vec{\Psi}}}

\def\vpi{{\vec{\pi}}}
\def\vN{{\vec{N}}}
\def\su{{\rm su}}

\def\so{{\rm so}}
\def\SO{{\rm SO}}
\def\smallover#1/#2{\hbox{$\textstyle\frac{#1}{#2}$}}
\def\IR{{\mathds{R}}}
\def\vr{{\vec{r}}}
\def\vA{{\vec{A}}}
\def\vK{{\vec{K}}}
\def\vJ{{\vec{J}}}
\def\vPi{{\vec{\Pi}}}
\def\a{{\alpha}}
\def\b{{\beta}}

\def\vv{{\vec{v}}}
\def\vp{{\vec{p}}}
\def\vx{{\vec{x}}}
\def\vn{{\vec{n}}}
\def\va{{\vec{a}}}
\def\vB{{\vec{B}}}
\def\vnabla{{\vec{\nabla}}}
\def\hx{{\frac{\vec{x}}{r}}}

\def\IR{{\mathds{R}}}
\def\IN{{\mathds{N}}}

\def\E{{\mathcal{E}}}
\def\I{{\mathcal{I}}}
\def\H{{\mathcal{H}}}

\def\Q{{\mathcal{Q}}}
\def\F{{\mathcal{F}}}
\def\U{{\mathcal{U}}}

\def\N{{\mathcal{N}}}
\def\C{{\mathcal{C}}}
\def\IZ{{\mathds{Z}}}
\def\p{{\partial}}

\def\vp{{\vec{p}}}
\def\vj{{\vec{j}}}
\def\vP{{\vec{P}}}
\def\hx{{\hat{r}}}
\def\hp{{\hat{p}}}
\def\vR{{\vec{R}}}
\def\vn{{\hat{n}}}

\def\vJ{{\vec{J}}}
\def\vL{{\vec{L}}}
\def\vo{{\vec{\Omega}}}
\def\vK{{\vec{K}}}

\def\vA{{\vec{A}}}
\def\vB{{\vec{B}}}
\def\vD{{\vec{D}}}
\def\vG{{\vec{G}}}
\def\vC{{\vec{C}}}
\def\vX{{\vec{X}}}

\def\cD{{\cal{D}}}
\def\vW{{\vec{\Psi}}}
\def\vpsi{{\vec{\psi}}}
\def\vxi{{\vec{\xi}}}

\def\vsigma{{\vec{\sigma}}}

\def\vpi{{\vec{\pi}}}
\def\vN{{\vec{N}}}
\def\su{{\rm su}}

\def\so{{\rm so}}
\def\SO{{\rm SO}}
\def\smallover#1/#2{\hbox{$\textstyle\frac{#1}{#2}$}}

\def\M{{\mathcal{M}}}

\def\S{{\mathcal{S}}}
\def\G{{\mathcal{G}}}
\def\V{{\mathcal{V}}}
\def\R{{\mathcal{R}}}
\def\vr{{\vec{r}}}
\def\b{{\beta}}
\def\Id{{1\!\!1}}
\def\a{{\alpha}}
\def\k{{\kappa}}
\def\vA{{\vec{A}}}
\def\vK{{\vec{K}}}
\def\vJ{{\vec{J}}}
\def\vPi{{\vec{\Pi}}}
\def\g{{\mathcal{g}}}

\def\L{{\mathcal{L}}}
\def\gyro{{g}}
\def\b{{\beta}}
\def\d{{\delta}}

\def\vv{{\vec{v}}}
\def\vp{{\vec{p}}}
\def\vx{{\vec{x}}}
\def\vc{{\vec{c}}}
\def\vn{{\vec{n}}}
\def\va{{\vec{a}}}
\def\vk{{\vec{k}}}
\def\vE{{\vec{E}}}
\def\vb{{\vec{b}}}
\def\vB{{\vec{B}}}
\def\vS{{\vec{S}}}
\def\vnabla{{\vec{\nabla}}}
\def\hx{{\frac{\vec{x}}{r}}}

\def\beq{\begin{equation}}
\def\eeq{\end{equation}}
\def\beqa{\begin{eqnarray}}
\def\eeqa{\end{eqnarray}}
\def\g{\gamma}

\def\nn{\nonumber}
\def\osp{{\mathfrak{osp}}}

\def\vTheta{{\vec{\Theta}}}

\def\ort{{\rm o}}
\newtheorem{prop}{Theorem}[section]
\newtheorem{Axiom}{Axiom}[section]

\def\gyro{{g}}
\def\beq{\begin{equation}}
\def\eeq{\end{equation}}
\def\beqa{\begin{eqnarray}}
\def\eeqa{\end{eqnarray}}
\def\nn{\nonumber}

\begin{document}


\title{(Super)symmetries of semiclassical models in theoretical and condensed matter physics}

\author{\large
 J.-P.~Ngome}
\affiliation{
Laboratoire de Math\'ematiques et de Physique Th\'eorique, 
Universit\'e Fran\c cois-Rabelais de Tours,
F\'ed\'eration Denis Poisson - CNRS
Parc de Grandmont, 37200 Tours, France.
}
\email{ 
ngome-at-lmpt.univ-tours.fr .}

\date{\today}

\begin{abstract}
Van Holten's covariant algorithm for deriving conserved quantities is
presented, with particular attention paid to Runge-Lenz-type vectors.
The classical dynamics of isospin-carrying particles is reviewed.
Physical applications including non-Abelian monopole-type systems in diatoms, introduced by Moody, Shapere and Wilczek, are considered.
Applied to curved space, the formalism of van Holten allows us to describe the dynamical symmetries of 
generalized Kaluza-Klein monopoles. The framework is extended to
supersymmetry and applied to the SUSY of the monopoles. Yet another application concerns
the three-dimensional non-commutative oscillator.\end{abstract}


\maketitle
\section*{Acknowledgements}

I foremost express my heartfelt thanks to my advisor Peter Horv\'athy whose support, guidance and patience made this thesis possible. His insight, passion for physics and quest for perfection inspired me all along our collaboration.

I am thankful to my thesis reviewers L\'asl\'o Feh\'er and Richard Kerner for their careful work in reading this thesis. Their remarks and suggestions considerably contributed to the improvement of this work.

To my thesis committee members Xavier Bekaert, Christian Duval, Stam Nicolis and Jan-Willem Van Holten, I wish to express my deep gratefulness. I am especially indebted to Jan-Willem Van Holten for his invaluable support in the course of this project.

I would also like to thank the people who have participated most directly in my formation and initiated me into the fascinating world of physics. I think about Claude Barrab\`es, Hector Giacomini, Amaury Mouchet, Jean-Claude Soret and Michael Volkov.

I make a friendly glance to my former and current officemates Tanaya Bhattacharya, Julien Garaud, Shuangwei Hu, J\'er\'emy Le Deunff, Elisa Meunier and Francesco Sardelli with whom I had interesting conversations about physics and much more.

I am also grateful to Nathalie Doris, Mokhtar Hassa\"ine, Frederik Scholtz and to the {\bf \textit{Laboratoire de Math\'ematiques et de Physique Th\'eorique}} of Tours University for their constant support.

Lastly, but by no means least importantly, I thank my loving and supporting family which always relies on me. In particular, I thank its female components so dear for my life: Mema, M'Angue, Caro, Suzy, V\'ero, Lydie, Prisca, Reine, Myjola, Ang\'ela, Kh\'elia, Oc\'eanne, Su, Princess, Swann and Mano\'e for their encouragements... I love you all.

I dedicate this thesis to my two princesses Isabelle-Fleur and Sol\'ena who share all the instants of my life. I love you both.

To my late brother Rich. I love you bro...

\cleardoublepage
\tableofcontents

\cleardoublepage

\section{Introduction}\label{chap:Intro}

The knowledge of the symmetries is essential in theoretical and condensed matter physics. Indeed, symmetries can be exploited to obtain valuable informations on the motion of a classical system or after quantization to generate the energy spectrum algebraically.

The usual classification provides us with discrete and continuous symmetry transformations. The discrete symmetries are described by finite groups while continuous symmetries, in which we are especially interested, are described by Lie groups.

A deep basis for the understanding of global conservation laws in modern physics was given by Emmy Noether in 1918 \cite{Noether}. She established that conservation laws directly follow from the symmetry properties of a physical system. See also \cite{Trautman1}. For instance, the invariance by time translation implies the conservation of the energy; the invariance by spatial translation yields the conserved momentum and the invariance under rotations provides us with the conserved angular momentum. 

In this thesis, we focus our attention on a novel way of deriving conserved quantities which has been put forward recently by van Holten \cite{vH}. In this formalism, invariants are constructed via Killing tensors which are, indeed, the main ingredients of this technique.

Our main endeavor will be to apply van Holten's covariant recipe to various physical systems.

\begin{enumerate}
\item
 
Firstly, we clarify the symmetries associated with isospin-Yang-Mills-Higgs field interactions.
To this end, we review, in the context of Kaluza-Klein theories, the classical equations describing the motion of an isospin-carrying particle evolving in a non-Abelian background.  Our presentation follows that of \cite{Kerner}, who first introduced these equations, using a ``Kaluza-Klein'' approach \cite{Kerner}. 
 
Next, we discuss the covariant van Holten formalism we use to investigate the symmetries of systems.  We note that the symmetry conditions of the van Holten formulation are the same as in the Forg\'acs-Manton-Jackiw (F-M-J) approach \cite{ForgacsManton,JackiwManton} to  symmetric gauge fields.
 
\item 

Most applications of the van Holten algorithm involve various (Abelian but also non-Abelian) monopoles and their symmetries.

In detail, for a ``naked'' Dirac monopole,
the angular momentum admits a celebrated radial term. It has been proved in turn that no 
globally defined conserved Runge-Lenz vector 
can exist \cite{Feher:1987}. It has, however, been found before by McIntosh and Cisneros, and by Zwanziger (MICZ) \cite{MIC,Zwanziger} that adding a suitable inverse-square potential can remove the obstruction such that the combined system can accommodate a conserved Runge-Lenz-type vector.

The archetype of non-Abelian monopoles corresponds to the one introduced in 1968 by Wu and Yang in pure Yang-Mills theory \cite{WuYang1}. One can wonder if a particle in the Wu-Yang field admits a Kepler-type dynamical symmetry. Generalizing the trick of McIntosh and Cisneros, and of Zwanziger, we find below the most general scalar potential such that the combined system admits a conserved Runge-Lenz vector. This result had to be expected, since Wu and Yang monopole is in fact an imbedded Dirac monopole.

\vskip2mm
Although no monopoles were ever seen in high-energy experiments, {\it monopole-like effective fields}
can arise in Condensed Matter Physics.
It has been noted by Moody, Shapere and Wilczek, 
for example, that an effective
non-Abelian field  arises  in a diatomic molecule
through Berry's phase due to nuclear motion \cite{MSW}. For some particular value of a certain parameter, it is just a Wu-Yang monopole field.
For  a full range of the parameter, however, the
effective field  becomes ``truly'' non-Abelian. Electric charge is not more conserved in this case. The system has still spherical symmetry, though, and Moody, Shapere and Wilczek do derive a conserved angular momentum -- but one which has an  ``unusual'' form. But they confess not  having a systematic way to obtain it. This goal has been achieved by Jackiw \cite{Jdiat} in the F-M-J framework mentioned above.

Here, after a short outline of Berry's phase, we re-derive the correct expression for the conserved angular momentum \cite{H-NGI}, using van Holten's algorithm. In addition, we constructed an ``unconventional'' conserved charge which reduces to the square of the electric charge in the Wu-Yang limit. 

\item
The next application of van Holten's approach concerns curved spaces of the Kaluza-Klein monopole type \cite{Sork,GrossPerry,GM,GRub
}. Mimicking what had been done for the MICZ system, we construct, on curved manifolds, conserved Runge-Lenz-type vectors along the geodesic motion. To this end, using the conservation of the ``vertical'' component of the momentum, we perform a dimensional reduction of our curved manifold. This allows us to find the conditions under which the dimensionally reduced manifold admits a Killing tensor field associated with a Kepler-type dynamical symmetry \cite{Ngome:2009pa}.
Our strategy is to lift $3D$ expressions to the extended Kaluza-Klein manifold.

Applied to a generalized Taub-NUT metric, we find the most general external potential which can be added such that the combined system exhibits a conserved Runge-Lenz-type vector. 

In the multi-center metric case \cite{GRub}, we show that,  under certain conditions,  a conserved scalar of Runge-Lenz-type does exist for  two-centers \cite{Ngome:2009pa}. For more than two centers no Runge-Lenz-type invariant does exist.

\item
Supersymmetries arise for fermions in a three-dimensional monopole background \cite{DV1,DJ-al,Pl,Leiva,Avery}. The Hamiltonian of the system then involves an additional spin-orbit coupling term, parametrized by the gyromagnetic ratio $g$. 

Below we construct the (super)invariants  using a SUSY extension of the van Holten algorithm. Our clue here is that the symmetry generators can be enlarged to Grassmann-algebra-valued Killing tensors \cite{Ngome:2010gg}. Conserved quantities are obtained for certain definite values of the gyromagnetic factor~:
$\N=1$ SUSY requires $g=2$ \cite{Spector}; a Kepler-type dynamical symmetry only arises, however,
for the anomalous values $g=0$ and $g=4$. The latter case has the additional bonus to contain an extra ``spin'' symmetry.

We find that the two contradictory conditions, namely that of having both  super and dynamical symmetry, can be conciliated by doubling the number of Grassmann variables. The anomalous systems with $\gyro=0$ and $\gyro=4$ will then become superpartners inside a unified $\N=2$ SUSY system.

For a planar fermion in any planar magnetic field, i.e. one perpendicular to the plane, an $\N=2$ SUSY arises without Grassmann variable doubling.

\item
We also construct a three-dimensional non-commutative oscillator with no kinetic term, but with a non-conventional momentum-dependent potential 
such that it admits a conserved Runge-Lenz-type vector. The latter is derived by adapting  van Holten's method to a ``dual'' description in momentum space \cite{ZHN}.

Our system, with monopole-type non-commutativity has the remarkable property to confine particle's motion to bounded trajectories, namely to (arcs of) ellipses. The best way to figure the motions followed by the particle is to think of them as generalizations of the familiar circular hodographs of the Kepler problem, to which they indeed reduce when the noncommutativity is turned off.


\end{enumerate}

\newpage

\section{Symmetries and conserved quantities in a non-Abelian  field theory}\label{chap:Sym}


{\normalsize
\textit{
The classical equations governing isospin-carrying particle motion in a non-Abelian background are derived using Kerner's Kaluza-Klein framework. The van Holten covariant method based on Killing tensors and the Forg\'acs-Manton-Jackiw approach based on the study of symmetric gauge fields are presented.
}}

\subsection{The ``Kaluza-Klein'' framework}

In this section, we deal with Kerner's extension of the Kaluza-Klein (KK) approach to a non-Abelian gauge
theory \cite{Kerner}. 

\hfil\break\underbar
{Abelian Kaluza-Klein theory}\hfil\break

First of all, let us recall that electromagnetism can be imbedded into general relativity (GR) by adding $\,U(1)\,$ local gauge invariance to the theory \cite{Kaluza,Klein}. See also \cite{Kplus1,Kerner3}. Indeed, let us consider the five-dimensional Einstein-Hilbert action given by
\beq
S_{5}=-\frac{1}{16\pi\tilde{G}_{5}}\int\! dx^{5}\sqrt{-g_5}\,\R_{5}\,,\label{KKdef}
\eeq
where $\tilde{G}_{5}$ is the coupling constant and $\,\R_5\,$ denotes the 5D scalar curvature.

Viewing the 5D manifold as a direct product of a 4D space-time with an unobservable space-like loop, and assuming that all components of the metric are independent of the extra coordinate, $\,y\,$, we get the most general  transformations allowed
\beq
x^{\mu}\longrightarrow x'^{\mu}(x^{\nu})\,,\quad y\longrightarrow y+f(x^{\mu})\,.\label{KKtransformation}
\eeq
 Putting $g_{44}=V$,
the 5D metric tensor reads therefore  as
\beq
g_{AB} =  \left(\begin{array}{cc} \g_{\mu\nu}+VA_{\mu}A_{\nu}  & \quad A_{\mu}V\\[4pt] A_{\nu}V  & V\end{array}\right),\quad\mu,\nu=0,\cdots,3\,.\label{KK5metric}
\eeq 
The transformations (\ref{KKtransformation}) imply that $\,A_{\mu}\,$ transforms as a gauge vector field,
\beq
g_{\mu4}\longrightarrow g_{\mu4}-V\p_{\mu}f\quad\Rightarrow\quad
A_{\mu}\longrightarrow A_{\mu}-\p_{\mu}f\,.
\eeq
The ``vertical'' translation yields hence a $\,U(1)$ gauge transformation for the vector field $A_{\mu}$ so that the theory (\ref{KK5metric}) is locally $U(1)$ gauge invariant. The Kaluza-Klein vector $A_{\mu}$ can thus be identify with the electromagnetic field.

Let us now embed the metric (\ref{KK5metric}) into the Einstein-Hilbert action defined in (\ref{KKdef}). We have $$\det(g_{AB})=\det(\g_{\mu\nu})V=g_4V\,,$$ and it is also useful to calculate the Christoffel connections. The 5D Ricci scalar $\R_5$ is expressed in terms of the 4D scalar curvature $\,\R_4$, the field strength $F_{\mu\nu}=\p_{\mu}A_{\nu}-\p_{\nu}A_{\mu}$ and the scalar field $\,V$,
\beq
\R_5=\R_4-\frac{1}{4}VF_{\mu\nu}F^{\mu\nu}-\frac{2}{\sqrt{V}}\Box\sqrt{V}\,.
\eeq
Substituting this $\R_5$ into the action in (\ref{KKdef}) and integrating with respect to the cyclic variable $\,y\,$, we obtain the 4D effective action,
\beq
S_{4}=-\frac{1}{16\pi G_5}\int_{\M_{4}}\!dx^{4}\sqrt{-g_4V}\big(\R_4-\frac{1}{4}VF_{\mu\nu}F^{\mu\nu}\big)+\frac{1}{8\pi G_5}\int_{\M_{4}}\!dx^{4}\sqrt{-g_4}\,\Box\sqrt{V}\,,\label{EffAction}
\eeq
where $G_5$ is the 5D Newton coupling constant. The second integral term in (\ref{EffAction}) can be dropped since it is a surface term and  does not affect therefore
the equations of the motion. Thus we end up with the following 4D action,
\beq
S_{4}=-\frac{1}{16\pi G_5}\int_{\M_{4}}\!dx^{4}\sqrt{-g_4V}\big(\R_4-\frac{1}{4}VF_{\mu\nu}F^{\mu\nu}\big)\,,
\eeq
wich involves GR and the Maxwell theory, coupled to an additional scalar field $\,V$.

Let us now study the dynamics of a classical point-like test particle of unit mass in our 5D space-time. Consider 5D geodesic motion,
\beq
\frac{d^{2}x^{A}}{d\tau^{2}}+\Gamma^{A}_{BC}\frac{dx^{B}}{d\tau}\frac{dx^{C}}{d\tau}=0\,,
\label{GeoEq}
\eeq
where $\tau$ denotes the proper time. Using the effective theory (\ref{KK5metric}) in (\ref{GeoEq}), a routine calculation yields the equations of the motion,
\beq\begin{array}{ll}\displaystyle
\frac{d}{d\tau}\big(VA_{\mu}\frac{dx^{\mu}}{d\tau}+V\frac{dy}{d\tau}\big)=\frac{dq}{d\tau}=0\,,\\[16pt]
\displaystyle
\frac{d^{2}x^{\mu}}{d\tau^2}+\Gamma^{\mu}_{\nu\lambda}\frac{dx^{\nu}}{d\tau}\frac{dx^{\lambda}}{d\tau}-qF^{\mu}_{\lambda}\frac{dx^{\lambda}}{d\tau}-\frac{q^2}{2}\frac{\p^{\mu}V}{V^2}=0\,.
\end{array}
\label{EqGeoM}
\eeq
The first equation in (\ref{EqGeoM}) tells us that the ``charge'',
\beq
q=V\left(A_{\mu}\frac{dx^{\mu}}{d\tau}+\frac{dy}{d\tau}\right)\,,
\eeq
is conserved along the 5D geodesics. The latter can also be viewed as being associated with translation, in the ``extra'' direction, generated by the Killing vector $\p_{y}$.  The second equation in (\ref{EqGeoM}) is a 4D geodesic equation involving in addition to the Lorentz force an interaction with the scalar field $V$. See \cite{Kerner2} for a point of view with $\,V=1\,$. See also \cite{Kibble,Trautman2}.

\hfil\break\underbar
{Non-Abelian generalization}.\hfil\break

The non-Abelian extension of the 5D KK approach was given by Kerner in \cite{Kerner}. 
First, we generalize our previous 5D manifold into a (4+d)-dimensional manifold noted as $\,
\M=\M^{4}\otimes \S^{d}\,$. The base $\M^{4}$ denotes the usual space-time with coordinates $x^{\mu}$, and $\S^{d}$ represents an unobservable $d$-dimensional extra space with the locally geodesic coordinates $\,y^{a},\,a,b=4,\cdots,(3+d)$. For definiteness, we fix $\,d=3\,$ so that $\S^3$, viewed as a Lie group, is isomorphic to the non-Abelian group $SU(2)$. Moreover, the compact manifold $\S^3$ admits the isometry generators, $\,\Xi_{j}=-i\xi^{b}_{j}(y)\p_{b}\,$, whose algebra reproduces the $SU(2)$ Lie algebra,
\beq
\left[\Xi_{j},\Xi_{k}\right]=i\,\varepsilon^{l}_{\,jk}\,\Xi_{l}\,,
\label{StructGroupEq}
\eeq 
and which imply the relation,
\beq
\xi^{b}_{k}(y)\,\p_{b}\xi^{a}_{j}(y)-\xi^{b}_{j}(y)\,\p_{b}\xi^{a}_{k}(y)=\varepsilon^{l}_{\,jk}\,\xi^{a}_{l}(y)\,.
\eeq
The anti-symmetric tensor $\,\varepsilon^{l}_{jk}\,$ denotes the structure constants of the  $SU(2)$ non-Abelian gauge group. In the KK approach the  7D diffeomorphism symmetry is broken into 4D infinitesimal coordinates transformations augmented with translations along the extra dimensions,
\beq
x^{\mu}\longrightarrow x^{\mu}+\d x^{\mu}\,,\quad y^{a}\longrightarrow y^{a}+f^{i}(x^{\nu})\xi_{i}^{a}(y)\,.\label{KKNAtransformation}
\eeq
Here the $f^{i}(x^{\nu})$ are functions. The 7D generalized metric, invariant under (\ref{KKNAtransformation}), then reads
\beq
\tilde{g}_{CD} =  \left(\begin{array}{cc} \g_{\mu\nu}+\k_{ab}B_{\mu}^aB_{\nu}^{b}  & \quad B_{\mu}^{b}\k_{ba}\\[4pt] \k_{ab}B_{\nu}^{a}  & \k_{ab}\end{array}\right),\quad C,D=0,\cdots,6\,,\label{KK7metric}
\eeq 
where $\,\k_{ab}\,$ is the $SU(2)$ invariant metric and 
\beq
\,B_{\mu}^{a}=A_{\mu}^{b}\,\xi^{a}_{b}\,.
\eeq
The $SU(2)$ Lie algebra-valued one-form 
 $\,A_{\mu}^{b}\,$ here will be identified with the
 Yang-Mills field. $\,A_{\mu}^{a}\,$ transforms indeed as a non-Abelian gauge field.
Under  (\ref{KKNAtransformation}) 
 the part $\,\k_{ab}\,$  of the metric (\ref{KK5metric}) 
 is preserved.
Using the formula $\xi'^{a}_{k}=\xi^{a}_{k}+\xi^{a}_{j}\varepsilon^{j}_{kl}f^{l}$ due to (\ref{KKNAtransformation}), the off-diagonal components  $\tilde{g}_{\mu b}$ of $\tilde{g}_{CD}$ change as
\beq
\tilde{A}^{a}_{\mu}=A^{a}_{\mu}(x)-\p_{\mu}f^{a}+\varepsilon_{bc}^{a}A^{b}_{\mu}f^{c}=
A^{a}_{\mu}(x)-D_{\mu}f^{a}
\,,
\label{NAgaugeTransf}
\eeq
where 
\beq
D_{\mu}f^{a}=\p_{\mu}f^{a}-\varepsilon_{bc}^{a}A^{b}_{\mu}f^{c}
\label{NACD}
\eeq
is the gauge-covariant derivative. The field strength of the potential $A_{\mu}^{b}\,$,
\beq
F^{a}_{\mu\nu}=\p_{\mu}A^{a}_{\nu}-\p_{\nu}A^{a}_{\mu}-\varepsilon_{bc}^{a}A_{\mu}^{b}A_{\nu}^{c}\,.
\label{KKYMF}
\eeq
changes in turn as
\beq
\tilde{F}^a_{\mu\nu}=F^a_{\mu\nu}-\varepsilon^a_{\;bc}f^bF^c_{\mu\nu}\,.
\eeq
For Abelian groups, the structure constants vanish so that the field strength is invariant and (\ref{NACD}) reduces to simple derivative.  

This is exactly how an infinitesimal gauge transformation, $ \d y^{a}=f^{i}(x^{\nu})\xi_{i}^{a}(y)$, acts on a non-Abelian gauge field. The result (\ref{NAgaugeTransf}) differs from the transformation law of Abelian gauge fields by the presence of the term $\varepsilon_{bc}^{a}A^{b}_{\mu}f^{c}$.

We now discuss the reduction of the dynamics starting from the 7D Einstein-Hilbert action,
\beq
S_{7}=-\frac{1}{16\pi\tilde{G}_{7}}\int_{\M_{4}}\!d^{4}xd^{3}y\sqrt{-g_7}\,\R_{7}\,.\label{KKNAdef}
\eeq
A tedious calculation provides us with the reduced scalar curvature so that the action (\ref{KKNAdef}) can be reduced as
\beq
S_{4}=-\frac{1}{16\pi G_{7}}\int_{\M_{4}}d^{4}x\!\sqrt{\g}\left(\R_{4}+\frac{1}{vol(\S^{3})}\int_{\S^{3}}\!d^{3}y\sqrt{\k}\,\R_{3}-\frac{1}{4}\k_{ab}F_{\mu\nu}^{a}F^{b\,\mu\nu}\right)\,.\label{KKNAdef2}
\eeq
The 7D Newton constant reads $G_{7}=\tilde{G}_{7}/vol(\S^{3})$ while $\R_{4}$ and $\R_{3}$ are the scalar curvatures associated with the metrics $\g_{\mu\nu}$ and $\k_{ab}$, respectively. The action (\ref{KKNAdef2}) describes an Einstein-like dynamics plus its coupling to the Yang-Mills fields. Note that the second term in (\ref{KKNAdef2}) is given by the curvature of the extra-space.

We focus our attention on the dynamics of a classical point-like test particle of unit mass in $(4+3)$-dimensional space-time. To this end, we consider the Lagrangian for geodesic motion in total space,
\beq
\L=\tilde{g}_{CD}\frac{dx^{C}}{d\tau}\frac{dx^{D}}{d\tau}\,.\label{LagrangianNAKK}
\eeq
For our metric (\ref{KK7metric}), the Lagrange function (\ref{LagrangianNAKK}) corresponds to
\beq
\L=\gamma_{\mu\nu}\frac{dx^{\mu}}{d\tau}\frac{dx^{\nu}}{d\tau}+\k_{ab}\big(\frac{dy^{a}}{d\tau}+A_{\mu}^a\frac{dx^{\mu}}{d\tau}\big)\big(\frac{dy^{b}}{d\tau}+A_{\mu}^b\frac{dx^{\mu}}{d\tau}\big)\,, 
\eeq 
and we evaluate the associated Euler-Lagrange equations,
\beqa\left\lbrace\begin{array}{ll}
\displaystyle
\frac{d}{d\tau}\left(\frac{\p\L}{\p\big(\frac{dx^{\a}}{d\tau}\big)}\right)-\frac{\p\L}{\p x^{\a}}=0\,,&\a=0,\cdots,3
\\[16pt]
\displaystyle
\frac{d}{d\tau}\left(\frac{\p\L}{\p\big(\frac{dy^{c}}{d\tau}\big)}\right)-\frac{\p\L}{\p y^{c}}=0\,,&c=4,5,6\,.
\end{array}
\right.
\label{ELeq}
\eeqa
The first equation in (\ref{ELeq}) yields the  motion projected  into real 4D space-time, whereas 
the second equation describes the motion
in  3D internal space.
In details, we have
\beq\begin{array}{ll}\displaystyle
\p_{c}\L=2\k_{ab}\left(\frac{dy^{b}}{d\tau}+A^{b}_{\nu}\frac{dx^{\nu}}{d\tau}\right)\varepsilon^a_{\;bc}A^b_{\mu}\frac{dx^{\mu}}{d\tau}\,,
\\[10pt]\displaystyle
\p_{\a}\L=\p_{\a}\gamma_{\mu\nu}\frac{dx^{\mu}}{d\tau}\frac{dx^{\mu}}{d\tau}+2\k_{ab}\left(\frac{dy^{a}}{d\tau}+A^{a}_{\mu}\frac{dx^{\mu}}{d\tau}\right)\p_{\a}A^b_{\nu}\frac{dx^{\nu}}{d\tau}\,.
\end{array}
\label{FirstTerms}
\eeq
Let us now identify the following quantity,
\beq
\I_a=\k_{ab}\left(\frac{dy^{b}}{d\tau}+A^{b}_{\nu}\frac{dx^{\nu}}{d\tau}\right)\,,
\eeq
as the classical isospin variable which describes
the motion in (non-Abelian) internal space.

Next, calculating the remaining terms in (\ref{ELeq}), we find
\beq\begin{array}{llr}\displaystyle
\frac{d}{d\tau}\left(\frac{\p\L}{\p\big(\frac{dy^{c}}{d\tau}\big)}\right)=2\frac{d\I_c}{d\tau}\,,\\[16pt]\displaystyle
\frac{d}{d\tau}\left(\frac{\p\L}{\p\big(\frac{dx^{\a}}{d\tau}\big)}\right)=\big(\p_{\mu}\gamma_{\a\nu}+\p_{\nu}\gamma_{\a\mu}\big)\frac{dx^{\mu}}{d\tau}\frac{dx^{\nu}}{d\tau}+2\gamma_{\a\nu}\frac{d^2x^{\nu}}{d\tau^2}\\[6pt]\displaystyle\qquad\qquad\qquad\qquad\qquad
+2\I_b\big(\p_{\b}A_{\a}^{b}\frac{dx^{\b}}{d\tau}+\varepsilon^b_{\;ca}A_{\a}^aA_{\mu}^c\frac{dx^{\mu}}{d\tau}\big)\,.
\end{array}
\label{Rest}
\eeq
Collecting the results (\ref{FirstTerms}) and (\ref{Rest}), we obtain the equations of motion of an isospin-carrying particle in a curved space plus a Yang-Mills field,
\beq
\left\lbrace\begin{array}{ll}\displaystyle
\frac{d^{2}x^{\b}}{d\tau^{2}}+\Gamma^{\b}_{\mu\nu}\frac{dx^{\mu}}{d\tau}\frac{dx^{\nu}}{d\tau}+\g^{\nu\b}F_{\mu\nu}^{b}\I_{b}\frac{dx^{\mu}}{d\tau}=0\,,\\[10pt]
\displaystyle
\frac{d\I_{c}}{d\tau}-\I_{a}\varepsilon_{\;bc}^{a}A^{b}_{\mu}\frac{dx^{\mu}}{d\tau}=0\,.
\end{array}\right.
\label{KKNAKerner-Wong}
\eeq

The first equation in (\ref{KKNAKerner-Wong}) describes the motion in  4D real space. Note here the  generalized Lorentz force 
\beq
\g^{\nu\b}F_{\mu\nu}^{b}\I_{b}\frac{dx^{\mu}}{d\tau}
\eeq
due to the Yang-Mills field, where the electric charge is replaced by the isospin  $\I^{a}$.
The derivation of the latter is analogous to that of the electric charge in the 5D KK theory, since it is also the contraction of the Killing vector field generating ``vertical'' translations with the direction field of the geodesic.

The second equation in (\ref{KKNAKerner-Wong}) says
that the isospin is parallel transported  in the internal space. 
Remark that the equations (\ref{KKNAKerner-Wong}) can also be obtained using the 7D geodesic equation,
\beq
\frac{d^{2}x^{C}}{d\tau^{2}}+\tilde{\Gamma}^{C}_{DE}\frac{dx^{D}}{d\tau}\frac{dx^{E}}{d\tau}\,,\quad C,D,E=0,\cdots,6.
\eeq

The equations (\ref{KKNAKerner-Wong}) are known as the Kerner-Wong equations. Indeed, some time after Kerner, Wong \cite{Wong}  obtained the same equations by ``dequantizing'' the Dirac equation. Later  Balachandran et al. \cite{Bal} also deduced the equations (\ref{KKNAKerner-Wong}) using a variational principle. Alternatively, they can be
studied using a symplectic approach,  \cite{Duval:1978,DHInt,Feher:1986*}.

\newpage

\subsection{van Holten's covariant Hamiltonian dynamics}\label{vHAl}

The standard approach to identify the constants of the motion associated with the symmetries of
a given mechanical system is achieved through Noether's theorem \cite{ForgacsManton,JackiwManton}, summarized in the next subsection.
More recently, however, an alternative approach has been put forward by van Holten \cite{vH}. To present his covariant Hamiltonian dynamics, let us consider a  non-relativistic charged isospin-carrying particle in three-dimensions with Hamiltonian
\beq
\H=\frac{\vpi^{2}}{2}+V(\vx,\,\I^{a})\,,\quad\vpi=\vp-e\vA\,.\label{H1}
\eeq 
Here $\vp\,$ and $\,\vpi\,$ define the canonical and the gauge-covariant momenta, respectively, and $V\,$ is an additional momentum-independent scalar potential. The gauge potential $\,\vA=\vA^{a}\,\I^{a}\,$, with the internal index $\,a=1,2,3\,$ referring to the non-Abelian $\su(2)$ Lie algebra, describes a static non-Abelian gauge field. Note that all dynamical variables here are gauge invariant.

Identifying the  $\su(2)$ Lie algebra of the non-Abelian variable with ${\IR}^3$, we consider the covariant phase space $\,\big(\vx,\,\vpi,\,\vec{\I}\,\big)\,$, where the dynamics, $$\dot{f}=\{f,H\}\,,$$ is defined by the covariant Poisson brackets,
\beq
\big\{f,g\big\}=D_jf\frac{\p g}{\p \pi_j}-\frac{\p f}{\p \pi_j}D_jg 
+e\,\I^aF_{jk}^a\frac{\p f}{\p \pi_j}\frac{\p g}{\p \pi_k}
-\epsilon_{abc}\frac{\p f}{\p \I^a}\frac{\p g}{\p \I^b}\I^c\,.
\label{PBracket}
\eeq
The field strength and the gauge covariant derivative read
\beq\begin{array}{ll}\displaystyle
F_{jk}
= \p_jA_k-\partial_kA_j - e\,\epsilon_{abc}\I^{a}A_j^bA_k^c\,,\\[10pt]\displaystyle
D_j=\partial_j-e\,\epsilon_{abc}\I^{a}A_j^b\frac{\p}{\p\I_c}\,,
\label{vHCovDerivative}
\end{array}
\eeq
respectively. The commutator of the covariant derivatives is recorded as
\beq
[D_i,D_j]=-\epsilon_{abc}\I^a{F}_{ij}^{b}\,\frac{\p}{\p\I^c}\ . 
\eeq
It is straightforward to obtain the non-vanishing fundamental Poisson-brackets,
\beq
\{x_i,\pi_j\}=\delta_{ij},
\qquad
\{\pi_i,\pi_j\}=e\,\I^{a}F^{a}_{ij},
\qquad
 \{\I^{a},\I^{b}\}=-\epsilon_{abc}\I^{c}.
\eeq
Let us remark that from the Jacobi identities we can derive the electromagnetic field equation,
\beq
\{\pi_{i},\{\pi_j,\pi_k\}\}+\{\pi_{j},\{\pi_k,\pi_i\}\}+\{\pi_{k},\{\pi_i,\pi_j\}\}=0\quad\Leftrightarrow\quad D_{i}\big(\I^{a}F^{a}_{ij}\big)=0\,.
\eeq
We can now derive the Kerner-Wong equations of motion [cf. (\ref{KKNAKerner-Wong})],
\beq\left\{
\begin{array}{lll}\displaystyle
\frac{d^2x_i}{dt^2}-e\,\I^aF^a_{ij}\frac{dx^j}{dt}+D_iV&=&0\,,
\\[6pt]\displaystyle
\frac{d\I^a}{dt}-\epsilon_{abc}\I^b\left(\frac{\p V}{\p\I^c}-e\,A^c_{j}\,\frac{dx^j}{dt}\right)&=&0\,.
\end{array}\right.
\label{Wongeq}
\eeq
To construct the dynamical quantities $\Q\big(\vx,\vpi,\vec{\I}\big)$ which are conserved along the motion, we use the covariant van Holten recipe \cite{vH}. The clue here is to expand constants of the motion in powers series of the covariant momenta,
\beq
\Q\big(\vx,\vpi,\vec{\I}\big)= C(\vx,\vec{\I})+C_i(\vx,\vec{\I})\pi_i+\frac{1}{2!}C_{ij}(\vx,\vec{\I})\pi_i\pi_j+
\cdots\label{constexp}
\eeq
Requiring $\Q$ to Poisson-commute with the Hamiltonian,
\beq
\{\Q,\H\}=0\,,
\eeq
leads us with the set of constraints, 
\beq\begin{array}{llll}
\displaystyle{C_iD_iV+\epsilon_{abc}\I^a\frac{\p C}{\partial\I^b}\frac{\p V}{\p\I^c}}=0,& o(0)
\\[8pt]
\displaystyle{D_iC=e\I^aF^a_{ij}C_j+C_{ij}D_jV+\epsilon_{abc}\I^a\frac{\p C_i}{\partial \I^b}\frac{\p V}{\p \I^c}},&
o(1)
\\[8pt]
\displaystyle{D_iC_j+D_jC_i=e\I^a(F^a_{ik}C_{kj}+F^a_{jk}C_{ki})+C_{ijk}D_kV+\epsilon_{abc}\I^a\frac{\p C_{ij}}{\partial \I^b}\frac{\p V}{\p \I^c}},& o(2)
\\[8pt]
\displaystyle{D_iC_{jk}+D_jC_{ki}+D_kC_{ij}=e\I^a(F^a_{il}C_{ljk}+F^a_{jl}C_{lki}+F^a_{kl}C_{lij})+C_{ijkl}D_lV}\\[10pt]\displaystyle{\qquad\qquad\qquad\qquad\qquad\qquad\qquad+\epsilon_{abc}\I^a\frac{\p C_{ijk}}{\partial \I^b}\frac{\p V}{\p \I^c}},& o(3)
\\
\vdots\qquad\qquad\qquad\qquad\qquad\vdots&\vdots
\end{array}\label{constraints}
\eeq
The series of constraints (\ref{constraints}) is a priori infinite since the expansion (\ref{constexp}) is also infinite. But for conserved quantities which admit finite expansion in covariant momenta, the series of constraints (\ref{constraints}) can be truncated at a finite order-(\textit{n+1}) provided we search for order-\textit{n} constants of motion. Hence, we can set $\,C_{i_1\dots i_{n+1}\dots}=0\,$, such that the higher-order constraint of (\ref{constraints}) becomes the covariant Killing equation,
\beq
D_{(i_1}C_{i_2\dots i_{n+1})}=0\,,\quad n\in\IN^{\star}.\label{KillingEq}
\eeq
It is worth noting that apart from the zeroth-order constants of the motion, i.e., which does not depend on the  covariant momentum, all order-\textit{n} invariants are deduced from the systematic method (\ref{constraints}) implying rank-\textit{n} Killing tensors, each Killing tensor solving the equation (\ref{KillingEq}). These Killing tensors also represent the higher-order coefficient of the expansion (\ref{constexp}) and, thus, can generate conserved quantities. The intermediate-order constraints of (\ref{constraints}) determine the other coefficient-terms of the invariant whereas the zeroth-order equation can be interpreted as a consistency condition between the scalar potential and the conserved quantity constructed.

To determine the conserved quantities of the system, we can consider the point of view consisting on solving first the equation (\ref{KillingEq}) in order to deduce Killing tensors. This task is extremely difficult due to the fact that the gauge covariant derivatives do not commute. But, in the particular case where
\beq
C_{i_2\dots i_{n+1}}(\vx,\vec{\I})\equiv C_{i_2\dots i_{n+1}}(\vx)\,,\label{SimpleCond}
\eeq
we can easily solve (\ref{KillingEq}) for $\,n=1\,$. We find the general form of the Killing vectors,
\beq
C_{i}=\C_{i\,Y_{1}\cdots Y_{m}}\,\xi^{Y_{1}\cdots Y_{m}}=a_{ij}x^{j}+b_{i}\,,\quad a_{ij}=-a_{ji}\,,\label{KV1}
\eeq
where $\,b_{j}\,$ and the anti-symmetric $\,a_{ij}\,$ denote constant tensors. It is worth mentioning that from the Killing vectors (\ref{KV1}), one can define the associated Killing tensors of Yano-type, $\,\C_{i\,Y_{1}\cdots Y_{m}}(\vx)\,$, which satisfy the Killing equation
\beq
D_{i}\C_{j\,Y_{1}\cdots Y_{m}}(\vx)+D_{j}\C_{i\,Y_{1}\cdots Y_{m}}(\vx)=0\,.
\eeq
Note that the Killing-Yano tensors are completely anti-symmetric differential forms in their indices.

To solve (\ref{KillingEq}) for $\,n=2\,$, even under the condition (\ref{SimpleCond}), remains, however, an awkward task. The trick is to construct the rank-$2$ Killing tensors $\,C_{ ij}(\vx)\,$ as a symmetrized product \cite{GRub} of Yano-type Killing tensors,
\beq
C_{ij}(\vx)=\C_{i\,Y_{1}\cdots Y_{m}}\;\widetilde{\C}_{j\,Y_{1}\cdots Y_{m}}+\widetilde{\C}_{i\,Y_{1}\cdots Y_{m}}\;\C_{j\,Y_{1}\cdots Y_{m}}\,.\label{Coef1}
\eeq
As an illustration, consider the two Killing-Yano tensors,
\beq
\left\lbrace\begin{array}{ll}
\C_{i\,Y}=\epsilon_{iYl}\,n^{l}\,,\quad\mbox{(with $\,\vn\,$ a constant unit vector)}\,,\\[10pt]
\widetilde{\C}_{j\,Y}=\epsilon_{jYk}\,x^{k}\,,
\end{array}\right.\label{KYT1}
\eeq
extracted from (\ref{KV1}). The symmetrized  product (\ref{Coef1}) of the Killing-Yano tensors (\ref{KYT1}) provides us with the rank-$2$ Killing tensor generating Kepler-type dynamical symmetry \cite{Ngome:2010gg},
\beq
C_{ij}(\vx)=2\d_{ij}\big(\vn\cdot\vx\big)-n_{i}x_{j}-n_{j}x_{i}\,.
\eeq
In the following, we focus our attention at given Killing tensors.

$\bullet$ For $n=1$, (\ref{KillingEq}) provides us with \underline{Killing vectors}. For example, we have, for any unit vector $\vn$, the generator of rotations around the axis $\vn$,\beq
\vC=\vn\times\vx\,
\label{rotKilling}
\eeq
leading to the conserved angular momentum.

The generator of space translations along an axis $\vn$,
\beq
\vC=\a\vn\,,\quad\a\in\IR\,,
\label{transKilling}
\eeq
implies a conserved quantity which is identified with the ``magnetic translations''.

$\bullet$ For $n=2$, then (\ref{KillingEq}) yields rank-2 \underline{Killing tensors}.
Similarly, for any unit vector $\vn$,
\begin{equation}
C_{ij}=2\delta_{ij}\, \vn\cdot\vx-(n_ix_j+n_jx_i)
\label{RLKilling}
\end{equation}
is a Killing tensor of rank $2$ associated with the conserved Laplace-Runge-Lenz vector.

The rank-2 Killing tensor implying the conservation of energy reads
\beq
C_{ij}=\delta_{ij}\,.
\eeq

The constant rank-2 Killing tensor generating the conserved Fradkin tensor, associated with the three-dimensional $\,SU(3)\,$ oscillator symmetry, is
\beq
C_{ij}=\a_{ij}\,,\quad\a_{ij}=\const.
\eeq

$\bullet$ For $n\geq3$, the equation (\ref{KillingEq}) provides us with higher-rank Killing tensors which, in general, generate product of already known constants of motion. Thus, no new conserved quantities and therefore no new symmetries arise in general from these higher-order Killing tensors.\\

In this section, we outlined the van Holten procedure (\ref{constraints}) to derive the symmetries of particle in flat space. This recipe can conveniently be extended to curved space provided the partial derivative is replaced by the metric covariant derivative, $\,\p_{i}\rightarrow\nabla_{i}\,$. Moreover, the van Holten algorithm is practical and efficient to derive linear and higher-order invariants in the momenta since the only requirement is to have a Killing tensor corresponding to a symmetry transformation.    

\newpage

\subsection{The Forg\'acs-Manton-Jackiw approach}\label{FMJapproach}

The invariants of a system can also be sought using the Forg\'acs-Manton-Jackiw approach based on the study of symmetric gauge fields \cite{ForgacsManton,JackiwManton,Jackiw}. Let us first consider indeed, the evolution of a free matter system, in the absence of a gauge field. In this case, the dynamics is characterized by several conserved quantities associated with spacetime diffeomorphisms.
For instance, invariance under temporal translation generates the conserved energy whereas the space rotational invariance generates conserved angular momentum.

Let us now assume that the matter field interacts with an external gauge field, $\,A_{\a}\,$. In general, the symmetry of the system is broken by this gauge-matter field interaction. However, when $\,A_{\a}\,$ and the matter field  are both invariant under the same three-dimensional space infinitesimal diffeomorphisms
\beq
\d x^{\a}=\omega^{\a}\,,\quad\a=1,2,3\,,\label{Diffeo}
\eeq
then, the constants of the motion, namely $\,C^{\,\omega}\,$, wins an extra term and can therefore be decomposed into two contributing parts,
\beq
C^{\,\omega}= C^{\,\omega}_{matter}+C^{\,\omega}_{gauge}\,.\label{CM}
\eeq
The first term on the right hand side of (\ref{CM}), which was the total conserved quantity in the absence of gauge field, corresponds, in the presence of an external gauge field, to the matter contribution augmented with that of the gauge field-matter interaction into the constant of the motion,
\beq
C^{\,\omega}_{matter}=\omega^{\a}\pi_{\a}\,,\label{MatterInv}
\eeq
where $\pi_{\a}$ represents the gauge covariant momentum. 

The second term of (\ref{CM}), namely $\,C^{\,\omega}_{gauge}\,$, is the gauge field's additional contribution restoring the full symmetry (\ref{Diffeo}) of the system. The Forg\'acs-Manton-Jackiw approach developped here is thus a systematic method to construct the gauge field additional contribution $\,C^{\,\omega}_{gauge}\,$ into the constant of the motion.

Let us first define the Lie derivative of a tensor dragged along the flow, $\,\C$, described by the vector field $\omega^{\a}$.
\begin{figure}[!h]
\begin{center}
\includegraphics[scale=.35]{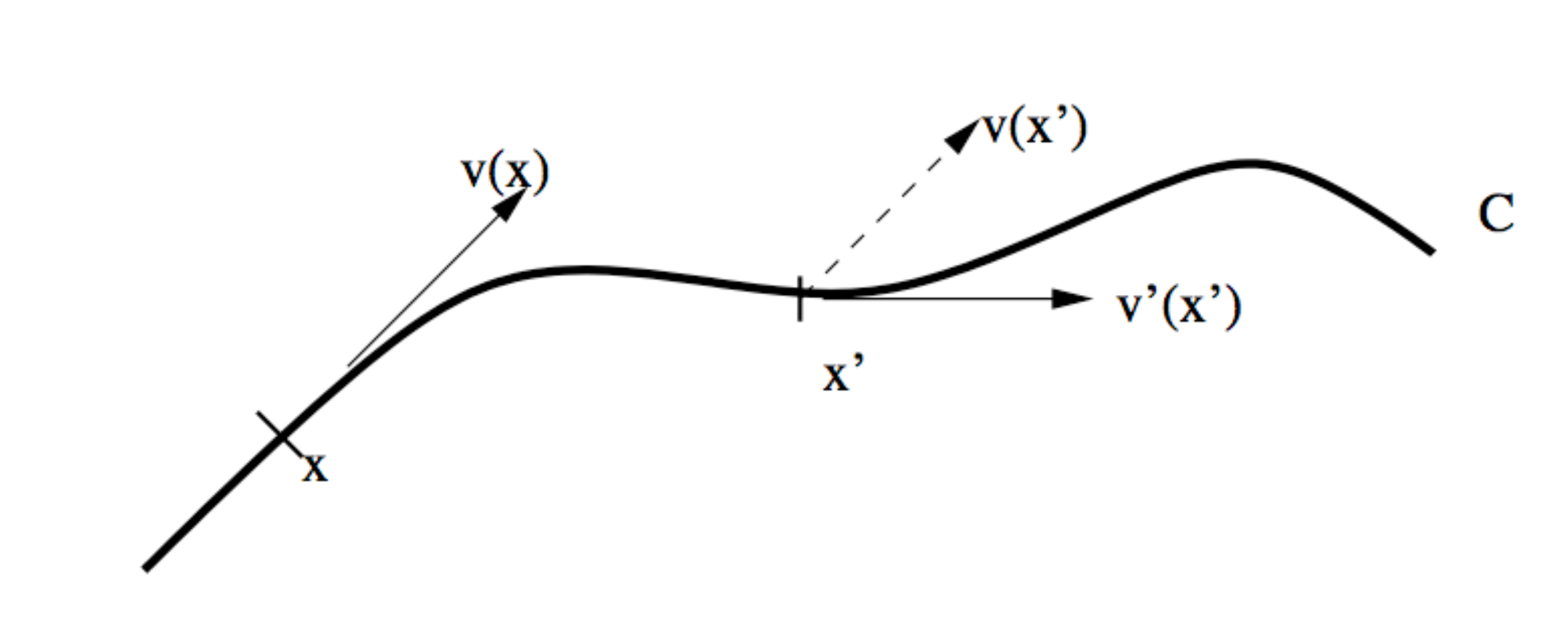}
\caption[]{
}
\end{center}
\end{figure}
\noindent
 For this, we consider the infinitesimal coordinate transformation,
\beq
x^{\a}\longrightarrow x'^{\a} = x^{\a} + \delta t \,\omega^{\a}\,,\quad \d t \ll 1\,,\label{transf}
\eeq
associated with the diffeomorphisms generate by the vector flow $\,\omega^{\a}\,$. See (\ref{Diffeo}). The formula (\ref{transf}) thus implies the tensor field transformations,
\beq\displaystyle
\upsilon^{\b}\left(x^{\a} \right)\longrightarrow \upsilon'^{\b}\left(x'^{\a} \right)\,,\quad
A_{\b}\left(x^{\a} \right)\longrightarrow A'_{\b}\left(x'^{\a} \right)\,.
\eeq
The Lie derivative along $\C$ takes the standard form
\beq\left\{\begin{array}{lll}
L_{\omega}\upsilon^{\a}=\d\upsilon^{\a}=\omega^{\mu}\p_{\mu}\upsilon^{\a}- \upsilon^{\mu}\p_{\mu}\omega^{\a}\\[7pt]
L_{\omega}A_{\b}=\d A_{\a}= \omega^{\mu}\p_{\mu}A_{\b}+ A_{\mu}\p_{\b}\omega^{\mu}\\[7pt]
L_{\omega} f = \omega^{\beta}\partial_{\beta}f\,,
\end{array}\right.
\eeq
and can be generalize to a $(p,q)$-tensor field by
\beq\begin{array}{rr}
\displaystyle{L_{\omega}T^{m_1,...,m_p}_{n_1,...,n_q}  = \omega^{\mu}\partial_{\mu}T^{m_1,...,m_p}_{n_1,...,n_q} + T^{m_1,...,m_p}_{\mu,n_2,...,n_q}\partial_{n_1}\omega^{\mu}+... +  T^{m_1,...,m_p}_{n_1,...,n_{q-1},\mu}\partial_{n_q}\omega^{\mu}}\\[8pt]\displaystyle{ - T^{\mu,m_2,...,m_p}_{n_1,...,n_q}\partial_{\mu}\omega^{m_1}-... - T^{m_1,...,m_{p-1},\mu}_{n_1,...,n_q}\partial_{\mu}\omega^{m_p}}\;.\end{array}
\eeq
Following Forg\'acs, Jackiw and Manton, we write the symmetry condition of the gauge field, $\,A_{\b}\,$, along the vector flow $\,\omega^{\b}\,$ as
\beq
L_{\omega} A_{\a} = D_{\alpha}{Q}^{\,\omega}\,,\label{SymCondGaugeField}
\eeq
where $\,{Q}^{\,\omega}\,$ is a differentiable Lie algebra-valued scalar function. Geometrically, the condition (\ref{SymCondGaugeField}) refers to the action of infinitesimal automorphism of the principal bundle. Note that the effect of the gauge freedom on $\,A_{\b}\,$,
\beq
A_{\a}\longrightarrow\widetilde{A}_{\a} = A_{\a} + \partial_{\a}\Lambda\,,
\eeq
does not affect the symmetry condition (\ref{SymCondGaugeField}), but only shifts the differentiable scalar field so that
\beq
L_{\omega}\widetilde{A}_{\a} = D_{\alpha}\widetilde{Q}^{\,\omega}\quad\hbox{with}\quad\widetilde{Q}^{\,\omega}={Q}^{\,\omega}+\omega^{\mu}\p_{\mu}\Lambda\,.\label{SymCondGaugeFieldTilde}
\eeq
Consequently, the gauge potential $\,\widetilde{A}_{\b}\,$ also remains invariant under the symmetry transformation generated by $\,\omega^{b}\,$.
Thus, as expected, we can conclude that the symmetry condition defined in (\ref{SymCondGaugeField}) is gauge invariant.
 
An equivalent way to describe the symmetry condition of a gauge field and therefore to obtain the gauge field contribution to the constant of the motion is to express the Lie derivative in term of the field strength $\,F_{\mu\nu}\,$,
\beq
L_{\omega} A_{\b} =\omega^{\mu}F_{\mu\b}+D_{\b}\big(\omega^{\mu}A_{\mu}\big)\,.
\eeq
Injecting this result into (\ref{SymCondGaugeField}), it is straightforward to obtain the following equivalent symmetry condition implying the gauge field contribution, discussed by Jackiw \cite{Jackiw}, 
\beq
F_{\b\mu}\,\omega^{\mu}=D_{\b}C^{\,\omega}_{gauge}\quad\hbox{with}\quad C^{\,\omega}_{gauge}=\omega^{\mu}A_{\mu}-Q^{\,\omega}\,.\label{FJMCond}
\eeq
Here, the gauge field contribution to the constant of the motion is a differentiable scalar function which can be determined by an integration of the equation (\ref{FJMCond}). 

The physics status of the term $\,C^{\,\omega}_{gauge}\,$ is now clear. Indeed, it represents the response of the external (symmetric) gauge field to a spacetime diffeomorphism. It restores the symmetry of the system and appears as a Lie algebra-valued scalar field contribution to the constant of motion. The complete constant of motion reads therefore as
\beq
C^{\,\omega}=\omega^{\nu}\pi_{\nu}+\int\omega^{\a}(x)F_{\a\b}(x)\,dx^{\b}\,.
\eeq
Let us remark that identifying the Lie algebra of the $SU(2)$ gauge group with $\,\IR^{3}\,$, the usual gauge covariant derivative, which we use in this section, becomes the gauge covariant derivative defined as (\ref{vHCovDerivative}) in the previous section. The rule is simply to replace the generators of the Lie algebra, $\,\tau^{a}\;(a=1,2,3)\,$, by the components of the isospin vector, $\,\I^{a}\,$. Under this transformation, the symmetry condition (\ref{FJMCond}) becomes precisely (with no scalar potential) the first-order condition in (\ref{constraints}) that a linear, in the covariant momentum, conserved quantity has to satisfy.

Thus, the Forg\'acs-Jackiw-Manton approach is, in fact, equivalent to the van Holten procedure for linear invariants. To generalize the first-cited method to higher-order contants of the motion, we require the symmetric gauge field to admitting higher-order Killing tensors. Then, as in the case of linear conserved quantities, the invariants can, in that event, be splitted into the two contributing parts (\ref{CM}),
\beq
C^{\,\omega}= C^{\,\omega}_{matter}+C^{\,\omega}_{gauge}\,.\nn
\eeq
In that event, the matter plus matter-gauge fields contributions give rise to the term
\beq
C^{\,\omega}_{matter}=\frac{1}{n!}\omega^{\mu_{1}\cdots\mu_{n}}\pi_{\mu_{1}}\cdots\pi_{\mu_{n}}\,,
\eeq
where $\,\omega^{\mu_{1}\cdots\mu_{n}}\,$ denotes the Killing tensor field generating the symmetry. The external gauge field carries, however, the contribution $\,C^{\,\omega}_{gauge}\,$ satisfying the constraints, 
\beq
D^{\left(\mu_{1}\right.}C^{\,\omega,\;\left.\mu_{2}\cdots\mu_{n-1}\right)}_{gauge}=F_{\b}^{\left(\mu_{1}\right.}\;\omega^{\,\omega,\;\left.\mu_{2}\cdots\mu_{n}\right)}_{\b}\,.
\eeq
We still here recognize the series of constraints of the van Holten algorithm (\ref{constraints}) for particle evolving in an external gauge field in the absence of an additional scalar potential.

\newpage

\section{Abelian monopoles}\label{chap:AM}

{\normalsize
\textit{
Dirac's quantization of magnetic monopole strength is obtained from the associativity of operators multiplication. (Dynamical) symmetries of the generalized Taub-NUT metric and its multi-center extension  are investigated.
}}

\subsection{Dirac monopole}\label{DiracMonop}

The concept of magnetic monopole is one of the most influential idea in modern theoretical physics. The hypothesis of particles carrying magnetic charge, $\,g\,$, was first made by Dirac \cite{Dirac}, who observed that the phase unobservability in quantum mechanics allows singularities as sources of magnetic fields, just as point electric monopoles are sources of electric fields. These singularities define the celebrated ``Dirac string'' whose position is not detectable. This implies the so-called Dirac quantization condition,
\beq
eg=\hbar c\frac{N}{2}\,,\quad N\in\IZ^{\star}\,.\label{DiracCondIntro}
\eeq
Consequently, the existence of a single magnetic monopole in the universe would explain the quantization of electric charge, for which there is no alternative explanation till this day.

In work preceding Dirac by over fifty years, Maxwell established the equations describing the electromagnetism. A surprising asymmetry inside these Maxwell's equations made Poincar\'e and J. J. Thompson to infer that a magnetic charge has to be introduced in the theory. The Maxwell equations with this assumption then read
\beq
\left\lbrace
\begin{array}{ll}
\displaystyle
\vnabla\cdot\vE=4\pi\rho_e\,,&\vnabla\cdot\vB=4\pi\rho_m\\[8pt]
\displaystyle
\vnabla\times\vB=\frac{1}{c}\frac{\p \vE}{\p t}+\frac{4\pi}{c}\vj_e\,,&\displaystyle\vnabla\times\vE=-\frac{1}{c}\frac{\p \vB}{\p t}-\frac{4\pi}{c}\vj_m\,,
\end{array}
\right.\label{MaxwellEq}
\eeq
where $\,\rho_{e,m}\,$ and $\,\vj_{e,m}\,$ denote the electric/magnetic charge and current density, respectively. But this introduction responded to a mathematical convenience and had, at that time, no physical reality; although, at the same period, P. Curie raised the possibility of the existence of free magnetic poles \cite{Curie}.

However, studying the motion of a charged particle in the field of an hypothetic isolated magnetic monopole, Poincar\'e \cite{Poinc1} observed that, as the particle is no longer deal with central forces, the angular momentum is no longer conserved and the motion is no longer necessarily planar. However, a certain amount of angular momentum resides in the magnetic field, and that a total angular momentum does exist,
\beq
\vJ=\vL-q\hx\,,\quad\vL=\vx\times\vpi\,.\label{PoincMom}
\eeq
Here $\,\vL\,$ denotes the mechanical angular momentum and the term $\,\big(-q\vx/r\big)\,$ represents the Poincar\'e momentum with $\,q\,$ denoting the magnetic pole strength. The total angular momentum (\ref{PoincMom}) is conserved along the motion.

Later, Wu and Yang \cite{NABAWY} showed that the Dirac string, which was introduced as a mathematical artifact, can be totally avoided using  two different choices of vector potential compatible with the monopole field strength. These two patches read
\beq
A_{r}=A_{\theta}=0\,,
A_{\phi}=
\left\lbrace
\begin{array}{ll}
\displaystyle
\frac{g}{r\sin\theta}\big(1-\cos\theta\big)\quad\mbox{for}\quad0\leq\theta\leq\frac{\pi}{2}+\d\,,\\[8pt]
\displaystyle
\frac{-g}{r\sin\theta}\big(1+\cos\theta\big)\quad\mbox{for}\quad\frac{\pi}{2}-\d\leq\theta\leq\pi\,,
\end{array}
\right.
\eeq
for any arbitrary $\,\d\,$ in the range $\,0<\d <\pi/2\,$. Each region contains a singularity if we try to extend them over the entire region around the monopole as Dirac did, but is regular in its  restricted hemisphere. In the overlapping region 
$$
\,\pi/2-\d\leq\theta\leq\pi/2+\d\,,
$$
 the two
patches are related by a gauge transformation, 
$$\,\vA_{N} = \vA_{S}-\vnabla\big(2g\Lambda(\vx)\big)\,.$$ The latter transformation changes the particle wave functions as
$$
\Psi(\vx)\longrightarrow\exp\big(2ieg\Lambda(\vx)\big)\Psi(\vx)\,,
$$
so that requiring the exponential to be single valued everywhere leads to the Dirac quantization condition \cite{NABAWY}, [cf. \ref{DiracCondIntro}].

From now on, we discuss the Dirac magnetic monopole without reference to singular patches or vector potential \cite{JackiwQuant}. To this end, we define the Hamiltonian of the monopole system as
\beq
\H=\frac{\pi^{2}}{2m}\,,\quad\pi_{j}=p_{j}-\frac{e}{c}A_{j}\,,\quad p_{j}=-i\hbar\p_{j}\,,\label{HMonop}
\eeq
and the following fundamental commutation rules are posited
\beqa
\left[x^{i},x^{j}\right]=0,
\quad
\left[x^{i},\pi_j\right]=i\hbar\delta^{i}_{j},
\quad
\left[\pi_i,\pi_j\right]=ie\frac{\hbar}{c}\epsilon_{ijk}B^{k}\,.\label{BracketsHmonop}
\eeqa 
Taking into account (\ref{HMonop}) and (\ref{BracketsHmonop}), we derive the gauge-invariant Lorentz-Heisenberg equations specifying the motion of a massive charged particle in the external monopole field $\,\vB\,$,
\beq\left\{\begin{array}{ll}
\displaystyle\dot{\vx}=\frac{i}{\hbar}\left[\H,\vx\right]=\frac{\vpi}{m}\\[8pt]
\displaystyle\dot{\vpi}=\frac{i}{\hbar}\left[\H,\vpi\right]=\frac{e}{2mc}\big(\vpi\times\vB-\vB\times\vpi\big)\,.
\end{array}\right.\label{EqMDirac}
\eeq
A priori no constraints on the monopole field $\,\vB\,$ are required in the previous equations of the motion. Indeed, equations (\ref{EqMDirac}) make sense both when $\,\vB\,$ is source-free, $\,\vnabla\cdot\vB=0\,$, or not, $\,\vnabla\cdot\vB\neq0\,$. However, when we look the Jacobi identities for the commutators of the momenta $\,\vpi\,$, we find
\beq
\displaystyle\epsilon_{ijk}\left[\pi_i,\left[\pi_j,\pi_k\right]\right]=2e\frac{\hbar^2}{c}\vnabla\cdot\vB\,,
\eeq
which vanishes, as it should, for a source-free magnetic fields, $\,\vB=\vnabla\times\vA\,$.

In order to obtain the exact form of $\,\vB\,$, we study now the Lie algebra associated with the O(3) symmetry of the monopole system. We first remark that the usual angular momentum operator, $\,\vL=\vx\times\vpi\,$, does not satisfy the o(3) Lie algebra, since we get an obstruction term inside of the commutator,
\beq
\left[L_{i},L_{j}\right]=i\hbar\epsilon_{ijk}L^{k}+ie\frac{\hbar}{c}\epsilon_{ijk}x^{k}\big(\vx\cdot\vB\big)\,.
\eeq
Following Jackiw \cite{Jackiw}, we restore the spherical symmetry of the system by adding a gauge field contribution, $\,\vC(\vx)\,$, into the angular momentum, $\,\vL\,$,
\beq
\vJ=\vL+\vC\,,
\eeq
so that we obtain the modified angular momentum algebra,
\beqa\begin{array}{lll}
\displaystyle\left[x^{i},J_{j}\right]=i\hbar\epsilon^{i}_{\;jk}x^{k}\,\\[8pt]
\displaystyle\left[\pi_{i},J_{j}\right]=i\hbar\epsilon^{\;\;\;k}_{ij}\pi_{k}+ie\frac{\hbar}{c}\left(x_{i}B_{j}-\d_{ij}\big(\vx\cdot\vB\big)\right)-i\hbar\p_{i}C_{j}\,\\[8pt]
\displaystyle\left[J_{i},J_{j}\right]=i\hbar\epsilon_{ijk}L^{k}+ie\frac{\hbar}{c}\epsilon_{ijk}x^{k}\big(\vx\cdot\vB\big)+i\hbar\epsilon_{ijk}x^{m}\big(\epsilon_{\;pl}^{k}\epsilon_{\;mp}^{n}\p_{n}C_{l}\big)\,.
\end{array}\label{AngMalgebra}
\eeqa
It is now clear that the contribution $\vC(\vx)\,$ restores the standard angular momentum algebra,
\beq\displaystyle\left[x^{i},J_{j}\right]=i\hbar\epsilon^{i}_{\;jk}x^{k}\,,\quad
\displaystyle\left[\pi_{i},J_{j}\right]=i\hbar\epsilon_{ijk}\pi^{k}\,,\quad
\displaystyle\left[J_{i},J_{j}\right]=i\hbar\epsilon_{ijk}J^{k}\,,
\eeq
provided that the following constraints are verified,
\beq\left\{\begin{array}{ll}
\displaystyle\p_{i}C^{j}= \frac{e}{c}\left(x_{i}B^{j}-\d_{i}^{j}\big(\vx\cdot\vB\big)\right)\,\\[8pt]
\displaystyle C^{k}=\frac{e}{c}x^{k}\big(\vx\cdot\vB\big)-x^{j}\big(\epsilon_{il}^{\;\;k}\epsilon_{ij}^{\;\;m}\p_{m}C^{l}\big)
\end{array}\right.\Longrightarrow\quad C^{k}+\frac{e}{c}x^{k}\big(\vx\cdot\vB\big)=0\,.\label{O(3)AlgbraCond}
\eeq
Consequently, the conserved generalized angular momentum along the motion becomes
\beq
\vJ=\vx\times\vpi-\frac{e}{c}\big(\vx\cdot\vB\big)\vx\,.\label{AngMomField}
\eeq
Moreover, the integrability condition coming from the equations in the left hand side of (\ref{O(3)AlgbraCond}),
\beq
\left[\p_{i},\p_{k}\right]C^{j}=0\,,
\eeq
imposes that the field $\,\vB\,$ satisfies the structural equation,
\beq
x_{k}\p_{i}B^{j}-x_{i}\p_{k}B^{j}+\d_{i}^{j}\big(B_{k}+x_{m}\p_{k}B^{m}\big)-\d_{k}^{j}\big(B_{i}+x_{m}\p_{i}B^{m}\big)=0\,,
\eeq
which can conveniently be solved with the magnetic monopole field,
\beq
\vB=g\frac{\vx}{r^{3}}\,,\label{MonopForm}
\eeq 
where $g\,$ represents the magnetic charge centered at the origin. In fact, the expression (\ref{MonopForm}) is the only spherically symmetric possibility consistent with the Jacobi identity (except in the origin),
\beq
\vnabla\cdot\vB=4\pi g\d^{3}(\vx)\,.\label{JacobiJackiw}
\eeq 
Note that the obstruction occurs only at the isolated location of the magnetic source, at origin, which has to be excluded.
 
Following Jackiw \cite{JackiwQuant}, the violation, at the origin, of the Jacobi identity (\ref{JacobiJackiw}) can be better understood by examining the unitary operator,
\beq
U(\va)=\exp\left(-\frac{i}{\hbar}\va\cdot\vpi\right)\,,
\eeq
which according to (\ref{BracketsHmonop}) implements finite translations by $\,\va\,$ on $\,\vx\,$,
\beq
U^{-1}(\va)\,\vx\,U(\va)=\vx+\va\,.
\eeq
As the momenta operators do not commute according to (\ref{BracketsHmonop}), we obtain \footnote{we use the Baker-Campbell-Hausdorff formula: $\,e^{A}e^{B}=e^{A+B+\frac{1}{2}\left[A,B\right]+\cdots}\,$.}
\beq
U(\va)U(\vb)=\exp\left(-\frac{ie}{\hbar c}\Phi\big(\vx,\va,\vb\big)\right)U(\va+\vb)\label{CompFormula}\,.
\eeq
Here $$\Phi\big(\vx,\va,\vb\big)=\frac{1}{2}\big(\va\times\vb\big)\cdot\vB\,$$ 
represents the flux of the magnetic source through the triangle defined by the three vertices located at $\,\vx\,$, $\,\vx+\va\,$ and $\,\vx+\va+\vb\,$.
\begin{figure}[!h]
\begin{center}
\includegraphics[scale=0.5]{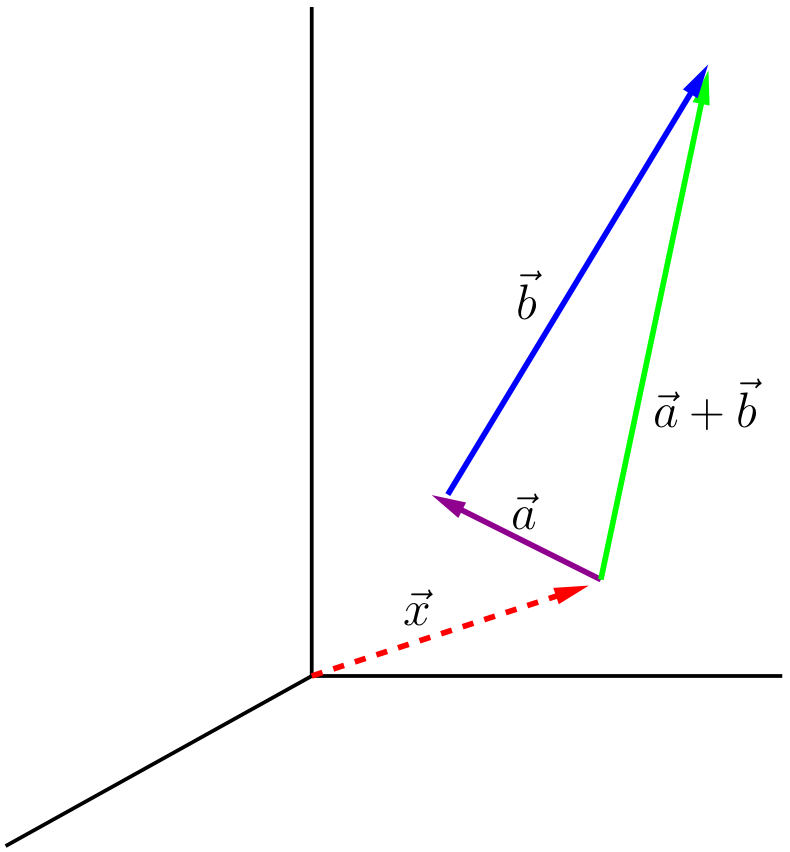}
\caption[]{Magnetic flux through the triangle.}\label{NG3.jpg}
\end{center}
\end{figure}

\noindent
Using (\ref{CompFormula}) it is straightforward to derive the following expressions,
\beq
\begin{array}{ll}\displaystyle
\left(U(\va)U(\vb)\right)U(\vc)=\exp\left(-\frac{ie}{\hbar c}\Phi\big(\vx,\va,\vb\big)\right)\exp\left(-\frac{ie}{\hbar c}\Phi\big(\vx,\va+\vb,\vc\big)\right)U(\va+\vb+\vc)\,,
\\[18pt]\displaystyle
U(\va)\left(U(\vb)U(\vc)\right)=\exp\left(-\frac{ie}{\hbar c}\Phi\big(\vx-\va,\vb,\vc\big)\right)\exp\left(-\frac{ie}{\hbar c}\Phi\big(\vx,\va,\vb+\vc\big)\right)U(\va+\vb+\vc)\,.
\end{array}\nn
\eeq
Combining the two previous formulas, we obtain 
\beq
\left(U(\va)U(\vb)\right)U(\vc)=\exp\left(-\frac{ie}{\hbar c}\Omega(\vx,\va,\vb,\vc\big)\right)U(\va)\left(U(\vb)U(\vc)\right)\,,\label{Associativity}
\eeq
where the first term on the right-hand side of (\ref{Associativity}) reads
\beq
e^{-\frac{ie}{\hbar c}\Omega(\vx,\,\va,\,\vb,\,\vc\big)}=e^{-\frac{ie}{\hbar c}\Phi\big(\vx,\,\va,\,\vb\big)}e^{-\frac{ie}{\hbar c}\Phi\big(\vx,\,\va+\vb,\,\vc\big)}e^{\frac{ie}{\hbar c}\Phi\big(\vx,\,\va,\,\vb+\vc\big)}e^{\frac{ie}{\hbar c}\Phi\big(\vx-\va,\,\vb,\,\vc\big)}\,.
\eeq
Here $\,\Omega(\vx,\va,\vb,\vc\big)\,$ can be interpreted as the total magnetic flux,
\beq
\Omega(\vx,\va,\vb,\vc\big)=\int d\vS\cdot\vB=\int d\vx\;\vnabla\cdot\vB\,,
\eeq
emerging out from the tetrahedron constructed with the three vectors $\,\va\,$, $\,\vb\,$, $\,\vc\,$ with one vertex at $\,\vx\,$. See Figure \ref{NG2.jpg} below.
\begin{figure}[!h]
\begin{center}
\includegraphics[scale=0.45]{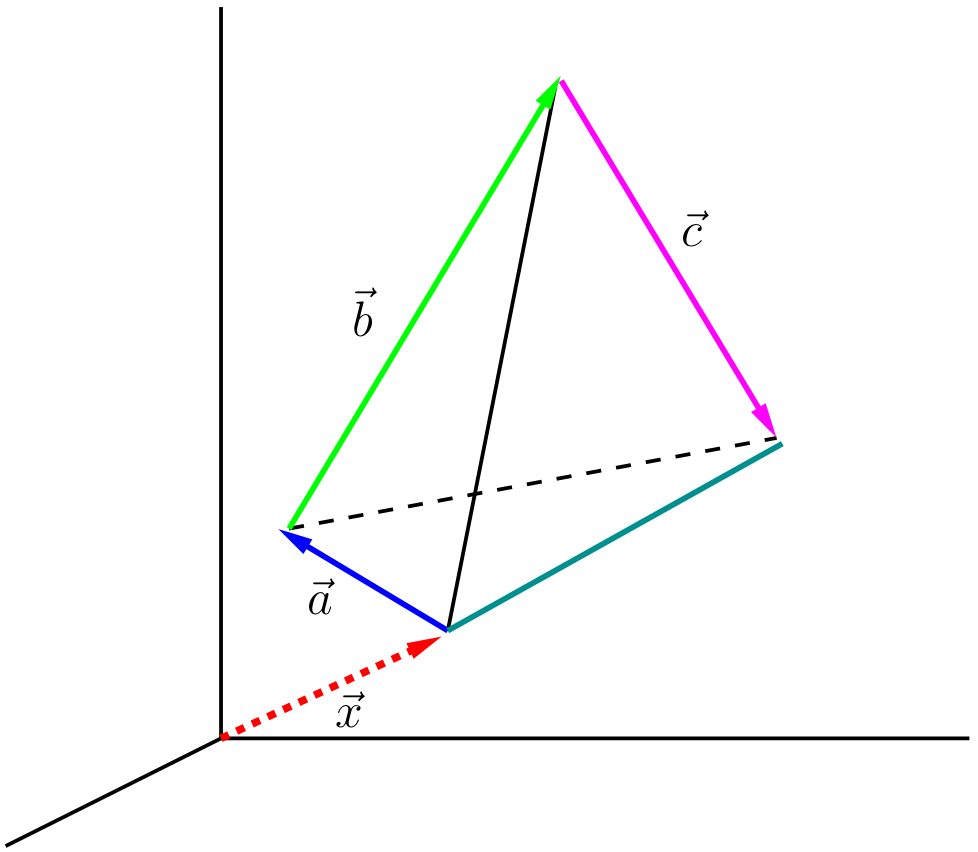}
\caption[]{Magnetic flux through the tetrahedron.}\label{NG2.jpg}
\end{center}
\end{figure}

\noindent
\noindent Positing the following axiom of Quantum mechanics;
\begin{Axiom}:
Quantum mechanics realized by linear operators acting on a Hilbert space requires  operator multiplication to be associative.
\end{Axiom}

We therefore obtain that the phase factor on the right hand side of (\ref{Associativity}) has to be unobservable so that the flux is quantized for arbitrary $\,\va\,$, $\,\vb\,$ and $\,\vc\,$,
\beq
\exp\left(-\frac{ie}{\hbar c}\Omega(\vx,\va,\vb,\vc\big)\right)=\exp\left(-i2\pi N\right)=1\,,\quad N\in\IZ\,.
\eeq
Consequently we obtain
\beq
\int d\vx\;\vnabla\cdot\vB=2\pi\frac{\hbar c}{e}N\quad\hbox{with}\quad\vnabla\cdot\vB=4\pi g\d^{3}(\vx)\neq0\,,\label{JackiwEq}
\eeq
which provides us with the Dirac's quantization relation \cite{Dirac},
\beq
\frac{eg}{\hbar c}=\frac{N}{2}\,,\quad N\in\IZ^{\star}\,.\label{QRelation}
\eeq
Note that the equation (\ref{JackiwEq}) saves the associativity of operators acting on Hilbert space and thus implies the quantization of the magnetic charge. The only requirement here is that the magnetic field must be a point source or a set of point sources in order to conserve the integrality of the left hand side of (\ref{JackiwEq}).

Let us now investigate the classical dynamics of a particle evolving in a magnetic monopole field, augmented with a scalar potential $\,V\,$. We inquire, in particular, about Kepler-type dynamical symmetries. Our investigations lead us to the well-known Mcintosh-Cisneros-Zwanziger system \cite{MIC,Zwanziger}.
 
  We start with searching conserved quantities associated with the system. A relevant recipe to search for constants of the motion is the van Holten algorithm, presented in Section \ref{vHAl}. From now on we fix $\,\hbar=c=1\,$ and we expand the constant of motions in terms of covariant momenta,
\beq
\Q\big(\vx,\vpi\big)= C(\vx)+C_i(\vx)\pi_i+\frac{1}{2!}C_{ij}(\vx)\pi_i\pi_j+\cdots
\eeq
and we require $\,\Q\,$ to Poisson-commute with the Hamiltonian of the system. This therefore implies to solve the series of constraints,
\beq\left\{\begin{array}{llll}
C_i\p_iV=0,& o(0)
\\[8pt]
\p_iC=e\,F_{ij}C_j+C_{ij}\p_jV\,&
o(1)
\\[8pt]
\p_iC_j+\p_jC_i=e\,(F_{ik}C_{kj}+F_{jk}C_{ki})+C_{ijk}\p_kV\,& o(2)
\\[8pt]
\p_iC_{jk}+\p_jC_{ki}+\p_kC_{ij}=e(F_{il}C_{ljk}+F_{jl}C_{lki}+F_{kl}C_{lij})+C_{ijkl}\p_lV\,& o(3)
\\
\vdots\qquad\qquad\qquad\qquad\qquad\vdots&\vdots
\end{array}\right.\nn
\eeq 
Searching for conserved quantity linear in the momentum, we recall that introducing the Killing vector generating space rotations,
\beq
\vC=\vn\times\vx\,,
\eeq
we directly get the associated generalized angular momentum [see (\ref{AngMomField})],
\beq
\vJ=\vx\times\vpi-eg\hx\,.\label{AngMomDirac}
\eeq
Considering quadratic conserved quantities, we first obtain that the rank-$2$ Killing tensor,
\beq
C_{ij}=\d_{ij}\,,
\eeq
generates the conserved energy of the system,
\beq
\E=\frac{\vpi^{2}}{2}+V(r)\,.
\eeq
On the other hand, inserting into the algorithm the rank-$2$ Killing tensor generating the Kepler-type dynamical symmetry,
\beq
C_{ij}=2\d_{ij}\big(\vn\cdot\vx\big)-n_{i}x_{j}-n_{j}x_{i}\,,\label{RLDirac1}
\eeq
we solve the second-order constraint with,
\beq
\vC=eg\frac{\vn\times\vx}{r}\,.\label{RLDirac2}
\eeq
Next, we insert (\ref{RLDirac1}) and (\ref{RLDirac2}) into the first-order constraint of the algorithm and we investigate the integrability condition of this equation by requiring the vanishing of the commutator,
\begin{eqnarray}\displaystyle{
\left[\,\partial_i\,,\,\partial_j\, \right]C=0\quad\Longrightarrow\quad
\Delta\left(V(r)- \frac{e^2g^2}{2r^2}\right)=0 }\,.
\label{goodpotential}
\end{eqnarray}
Thus, the bracketed quantity must satisfy a \emph{Laplace equation} so that a Runge-Lenz-type vector does exist only for radial effective potential of the form,
\beq
V(r)=\frac{e^2g^2}{2r^2}+\frac{\beta}{r}+\gamma\quad\hbox{with}\quad\b,\gamma\;\in\;\IR\,.
\label{DiracGooodPot}
\eeq
Consequently, the zeroth-order constraint is identically satisfied and the solution of the first-order constraint reads,
\beq
C=\b\frac{\vn\cdot\vx}{r}\,.\label{RLDirac3}
\eeq
Collecting the results (\ref{RLDirac1}), (\ref{RLDirac2}) and (\ref{RLDirac3}), we get the Runge-Lenz vector conserved along the particle's motion,
\beq
\vK=\frac{1}{2}\big(\vpi\times\vJ-\vJ\times\vpi\big)+\b\hx\,.\label{RungeLenzDirac}
\eeq
Note that the presence of the fine-tuned inverse-square term in (\ref{DiracGooodPot}) is \textit{mandatory} for canceling the effect of the monopole. For a ``naked'' monopole, $V\equiv0$, in particular, no conserved Runge-Lenz vector does exist \cite{Feher:1986*}.

Now we can give a complete description of the classical motion of a charged particle in the Dirac monopole field, augmented with the potential (\ref{DiracGooodPot}). A MICZ system in what follows. Firstly, from the angular momentum (\ref{AngMomDirac}) we obtain
\beq
\vJ\cdot\frac{\vx(t)}{r}=-eg\,,\label{CondOne}
\eeq
so that the trajectory followed by the particle lies on a cone with axis $\,\vec{J}\,$ and fix opening angle $\,\theta\,$ defined by
\beq\theta=\arccos\left(\frac{-eg}{J}\right)\,.
\eeq
Secondly, the conservation of the Runge-Lenz vector (\ref{RungeLenzDirac}) allows us to construct the new conserved vector,
\beq
\vN=\frac{\b}{eg}\vJ+\vK\qquad\hbox{such that}\qquad\vN\cdot\vx(t)=J^{2}-e^2g^2=\const\,.\label{SeconCond}
\eeq
The result (\ref{SeconCond}) implies that the particle motion also lies in the oblique plan perpendicular to $\,\vN\,$.
Consequently, combining (\ref{CondOne}) and (\ref{SeconCond}) the particle motion is viewed to be confined onto conic sections \cite{GM,FHKK}. 
\begin{figure}[!h]
\begin{center}
\includegraphics[scale=.45]{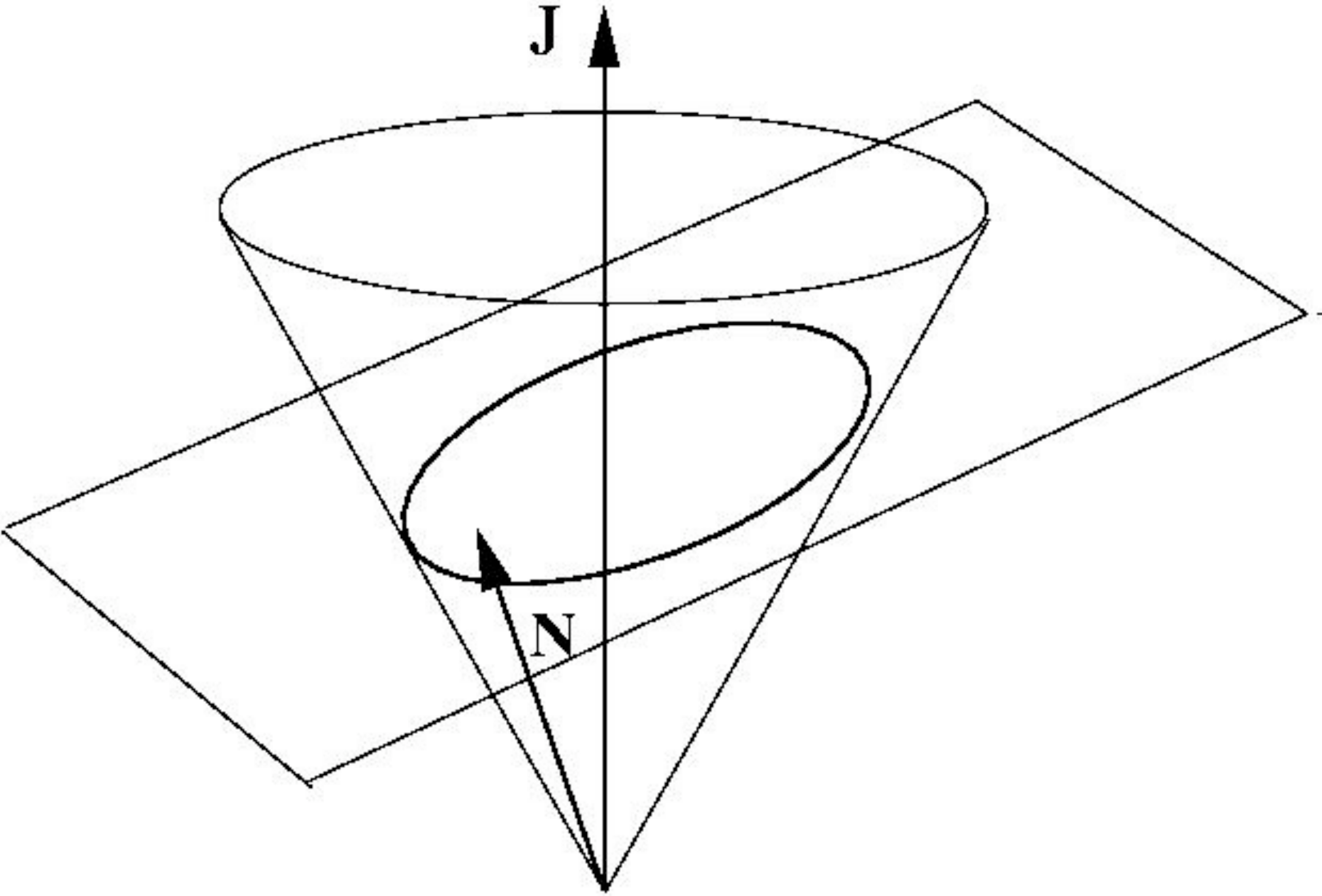}
\caption[]{The motion lies on the conic section  obtained by intersecting  the cone, due to conserved angular momentum $\vJ$, with  the oblique plane determined by the additional conserved quantity $\vN\,$.}
\end{center}
\end{figure}

The particular form of the conic section depends only on the angle $\,\b\;(\hbox{mod}\left[\pi\right])$ given by 
\beq
\cos\beta=\frac{\vJ\cdot\vN}{J\,N}\,,
\eeq
and which determines the inclination of the oblique plane in comparison to the angular momentum vector.

We thus obtain the following properties :
\beq\hbox{For}\quad\left\{
\begin{array}{lll}
\displaystyle\b\in\left[0,\,\frac{\pi}{2}-\a\right[\\[8pt]
\displaystyle\b=\frac{\pi}{2}-\a\,,\\[8pt]
\displaystyle\b\in\left]\frac{\pi}{2}-\a,\,\frac{\pi}{2}\right]
\end{array}
\right.\hbox{the trajectories lie on}\quad
\left\{
\begin{array}{lll}
\hbox{ellipses}\\[8pt]
\hbox{parabolae}\\[8pt]
\hbox{hyperbolae}\,.
\end{array}
\right.
\eeq
It is worth noting that the momentum-trajectories called the hodographs are also confined to a plane perpendicular to the conserved vector $\,\vN\,$ since
\beq
\vN\cdot\vpi(t)=0\,.
\eeq
But in some interesting cases, the momentum trajectories can be completely determined. For example, in the Kepler problem the $\vpi$-trajectories are known to be (arcs) of circle. In the context of non-commutative oscillator mechanics (see later), we prove that the hodographs of the MICZ-system lie on (arcs) of ellipses \cite{ZHN}.

Another illustration of using the symmetry (\ref{RungeLenzDirac}) is to derive the energy spectrum from the dynamical symmetry \cite{FH,Horvathy:1990zz,FHKK}. To this end, we return to quantum mechanics and consider the vectors $\,\vJ\,$ and $\,\vK\,$ defined in (\ref{AngMomDirac}) and (\ref{RungeLenzDirac}), respectively, as operators of the Hilbert space satisfying the quantized commutation relations,
\beqa
\left[J_{i},J_{j}\right]=i\epsilon_{ijk}J^{k},
\quad
\left[J_{i},K_j\right]=i\epsilon_{ijk}K^{k},
\quad
\left[K_i,K_j\right]=i\big(2\gamma-2\H\big)\epsilon_{ijk}J^{k}\,.\eeqa
Let us define, on the fixed-energy eigenspace $\,\H\Psi=\E\Psi\,$, the rescaled Runge-Lenz operator,\beqa
\vec{\widetilde{K}}=\left\{\begin{array}{lll}
\displaystyle\big(2\gamma-2\E\big)^{-\frac{1}{2}}\;\vK\quad\hbox{for}\quad\E<\gamma\\[8pt]
\vK\quad\hbox{for}\quad\E=\gamma\\[8pt]
\displaystyle\big(2\E-2\gamma\big)^{-\frac{1}{2}}\;\vK\quad\hbox{for}\quad\E>\gamma\,
\end{array}\right..
\eeqa
We therefore obtain the commutation relations between $\,\vJ\,$ and $\,\vec{\widetilde{K}}\,$,
\beqa
\left[J_{i},J_{j}\right]=i\epsilon_{ijk}J^{k},
\quad
\left[J_{i},\widetilde{K}_j\right]=i\epsilon_{ijk}\widetilde{K}^{k},
\quad
\left[\widetilde{K}_i,\widetilde{K}_j\right]=\left\{\begin{array}{lll}i\epsilon_{ijk}J^{k}\quad\hbox{for}\quad\E<\gamma\\[8pt]
0\quad\hbox{for}\quad\E=\gamma\\[8pt]
-i\epsilon_{ijk}J^{k}\quad\hbox{for}\quad\E>\gamma\,
\end{array}
\right.
\eeqa
Thus,
\beqa
\hbox{for}\quad\left\{\begin{array}{lll}
\E<\gamma\\[8pt]
\E=\gamma\\[8pt]
\E>\gamma
\end{array}\right.\,,\quad
\vec{\widetilde{K}}\;\hbox{and}\;\vJ\;\,\hbox{generate the}
\;\left\{\begin{array}{lll}o(4)\;\hbox{Lie algebra}\\[8pt]
o(3)\otimes\IR^{3}=e(3)\\[8pt]
o(3,1)\;\hbox{Lie algebra}\,
\end{array}\right.
\eeqa
For a fixed value of the energy, $\,\E<\gamma\,$, we consider the more convenient commuting operators 
\beqa
\vA=\frac{1}{2}\big(\vJ+\vec{\widetilde{K}}\big)\quad\hbox{and}\quad\vB=\frac{1}{2}\big(\vJ-\vec{\widetilde{K}}\big)\,,
\eeqa
verifying the following relations
\beqa
\left[A_{i},A_{j}\right]=i\epsilon_{ijk}A^{k},
\quad
\left[B_{i},B_j\right]=i\epsilon_{ijk}B^{k},
\quad
\left[A_i,B_j\right]=0\,.
\eeqa
Then, the operators $\,\vA\,$ and $\,\vB\,$ extend the manifest $\,o(3)\,$ symmetry into a dynamical $\,o(3)\oplus o(3)=o(4)\,$ Lie algebra. The common eigenvector $\,\Psi\,$ of the commuting operators, $\,\H,\,\vA^{\,2},\,\vB^{\,2}\,$ satisfies,
\beq
\vA^{\,2}\Psi=a\big(a+1\big)\Psi\,,\quad\vB^{\,2}\Psi=b\big(b+1\big)\Psi\,,\quad\H\Psi=\E\Psi\,,
\eeq
where $\,a\,$ and $\,b\,$ are half-integers. Considering the so far non-negative number,
\beq
n=-\frac{\b}{\sqrt{2\gamma-2\E}}\,,\label{BSEnergySpectrum}
\eeq
we use the Casimir operators,
\beq
\vec{\widetilde{K}}^{\,2}=-\vJ^{\,2}+e^2g^2-1+\frac{\b^2}{2\gamma-2\E}\quad\mbox{and}\quad
\vJ\cdot\vec{\widetilde{K}}=-\frac{eg\b}{\sqrt{2\gamma-2\E}}\,,
\eeq
to obtain the equalities,
\beqa
\left\{\begin{array}{ll}
\displaystyle a\big(a+1\big)+b\big(b+1\big)=\frac{1}{2}\big(e^2g^{2}-1+n^{2}\big),
\\[8pt]
a\big(a+1\big)-b\big(b+1\big)=\big(eg\big)n\,.
\end{array}\right.\label{EqSpectrum}
\eeqa
Solving the equations (\ref{EqSpectrum}) provide us with the relations,
\beqa\left\{\begin{array}{llll}
2a+1=\pm\big(n+eg\big)\\[8pt]
2b+1=\pm\big(n-eg\big)\\[8pt]
a-b=\pm eg\\[8pt]
a+b+1=n\,.
\end{array}\label{QRelation1}\right.
\eeqa
Let us recall that from equation (\ref{QRelation}), the product $\,\big(eg\big)\,$ is quantized in integers or half-integers [in units $\,\hbar=c=1\,$]. Consequently the first relation in (\ref{QRelation1}) implies that $\,n\,$ is integer or half-integer depending on the value of $\,\big(eg\big)\,$ being integer or half-integer.

We can now derive from (\ref{BSEnergySpectrum}) the bound-state energy spectrum,
\beq
\E_{n}=\gamma-\frac{\b^{2}}{2n^2}\,,\quad n=\pm eg+1\,,\pm eg+2\,\cdots\,,\label{SpectrumDirac}
\eeq
with the integer value of the degeneracy
\beq
n^2-e^2g^2=\big(n-eg\big)\big(n+eg\big)\,,
\eeq
since $\,n\,$ and $\,eg\,$ are simultaneously integers or half-integers.

\newpage

\subsection{Kaluza-Klein-type monopoles}

Kaluza-Klein theory is one of the oldest ideas attempting to unify gravitation and gauge theory \cite{Kaluza,Klein}. The physical assumption, in this framework, is that the world contains four space-time dimensions, plus an extra cyclic dimension so small that it can not be observed. Thus, the ordinary general relativity in five dimensions is considered to possess a local $\,U(1)\,$ gauge symmetry arising from the isometry transformation of the hidden extra dimension.

Later, Sorkin \cite{Sork}, and Gross and Perry \cite{GrossPerry}, introduced the Kaluza-Klein monopole which is obtained by imbedding the Taub-NUT gravitational instanton into Kaluza-Klein theory. The global stationary metric obtained,
\beq
\begin{array}{cc}
ds^2=-dt^2+f(r)\left(dr^2+r^2\big(d\theta^2+\sin^{2}\theta d\phi^2\big)\right)+f^{-1}(r)\big(dx^{4}+A_{\phi} d\phi\big)^2\,,\\[8pt]
\mbox{with}\quad\theta\in\left[0,\pi\right]\,,\quad\phi\in\left[0,2\pi\right]\,,\quad A_{\phi}\equiv\mbox{Dirac potential}\,,
\end{array}
\eeq
has lead to exact solution of the equations of the four-dimensional Euclidean gravity, approaching the vacuum solution at spatial infinity.

In 1986, Gibbons and Manton studied the hidden symmetry of Kaluza-Klein-type metrics and found, in the context of monopole scattering \cite{GM,GRub}, that the geodesic motion in the Taub-NUT metric admits a Kepler-type dynamical symmetry \cite{FH,GR,CFH1,CFH3}. (See \cite{FHKK} for a review).

A better understanding of such hidden symmetries of Kaluza-Klein-type monopoles was achieved by various generalizations \cite{VVHH1,Japs1,Japs,VVHH,VVHH2,VVHH3,Visi2,Visi3,Visi1,Krivonos,Ballesteros, Ballesteros1,K-L,Visi,K-L-S,Nerses,Visinescu:2011,Marquette2}.

More recently, Gibbons and Warnick considered geodesic motion on hyperbolic space \cite{GW} and found a large class of systems admitting such a dynamical symmetry. 

Our aim, in this section, is to present a systematic analysis of Kaluza-Klein-type metrics admitting a conserved Runge-Lenz-type conserved quantity. To this end, we consider the stationary family of metrics,
\begin{eqnarray}\displaystyle dS^2=f(\vx)\,\delta_{ij}\,dx^i\,dx^j+h(\vx)\,\big(dx^4+A_k\,dx^k\big)^2\,\label{metricTN}\,.
\end{eqnarray}
In these metrics, $\,f(\vx)\,$ and $\,h(\vx)\,$ are real functions and the $1$-form $\,A_k\,$ is the gauge potential of a charged Dirac monopole.

Inspired by Kaluza's hypothesis, as the fourth dimension here is considered to be cyclic, we use the conservation of the ``vertical'' component of the momentum to reduce the four-dimensional problem to one in three dimensions, where we have strong candidates for the way these symmetries act \cite{Ngome:2009pa}. Then, the lifting problem can be conveniently solved using the Van Holten technique [see section \ref{vHAl}].\\

Let us first investigate the four-dimensional geodesic motion of a classical point-like test scalar particle with unit mass. The Lagrangian of geodesic motion  on the  $4$-manifold endowed with the metric (\ref{metricTN}) is  
\begin{eqnarray}\displaystyle{
\mathcal{L}=\frac{1}{2}\,f(\vx)\,\delta_{ij}\,\frac{d x^i}{dt}\,\frac{d x^j}{dt}+\frac{1}{2}\,h(\vx)\,\big(\,\frac{d x^4}{dt}+A_k\,\frac{d x^k}{dt}\,\big)^2-U(\vx)}\label{LagrangTN}\,,
\end{eqnarray}
where we also added an external scalar potential, namely $U(\vx)$, for later convenience.
The canonical momenta conjugate to the coordinates $(x^j,\,x^4)$ read as
\begin{eqnarray}\begin{array}{ll}\displaystyle{
p_j= \frac{\partial \mathcal{L}}{\displaystyle{\partial\big({d x^j}/{dt}\big)}}  =f(\vx)\,\delta_{ij}\,\frac{d x^i}{dt}+h(\vx)\,\big(\,\frac{d x^4}{dt}+A_k\,\frac{d x^k}{dt}\,\big)\,A_j\;,}\ 
\\[14pt]\displaystyle{
p_4= \frac{\partial \mathcal{L}}{\displaystyle{\partial\big({d x^4}/{dt}\big)}} =h(\vx)\,\big(\,\frac{d x^4}{dt}+A_k\,\frac{d x^k}{dt}\,\big)=q}\;.
\end{array}
\end{eqnarray}
 The ``vertical'' momentum, $\,p_4=q\,$, associated with the periodic variable, $\,x^4\,$, is conserved and can be interpreted as conserved electric charge. Thus, we can introduce the covariant momentum,
\begin{eqnarray}\displaystyle{
\Pi_j=f(\vx)\,\delta_{ij}\,\frac{d x^i}{dt}=p_j-q\,A_j}\;.\label{DimReduc}
\end{eqnarray}
The geodesic motion on the 4-manifold  projects therefore onto the curved 3-manifold with metric $\,g_{ij}(\vx)=\,f(\vx)\,\delta_{ij}\,$, augmented with a scalar potential. The Hamiltonian reads as
\beq
\mathcal{H}= \frac{1}{2}\,g^{ij}(\vx)\Pi_{i}\,\Pi_{j}+V(\vx)\quad\hbox{with}\quad V(\vx)=\frac{q^{2}}{2h(\vx)}+U(\vx)\,.
\label{HamVanHolten}
\eeq
For a particle without spin, the covariant Poisson brackets are given by \cite{Souriau1970}
\begin{eqnarray}\displaystyle{
\left\lbrace B,D \right\rbrace = \partial_kB\,\frac{\partial D}{\partial \Pi_k}-\frac{\partial B}{\partial \Pi_k}\,\partial_kD+qF_{kl}\,\frac{\partial B}{\partial \Pi_k}\,\frac{\partial D}{\partial \Pi_l}}\,,
\end{eqnarray}
where $\,\displaystyle{F_{kl}=\partial_kA_l - \partial_lA_k}\,$
is the monopole field strength. Then, the nonvanishing fundamental brackets are
\beq
\displaystyle{
\left\lbrace x^i,\;\Pi_j\right\rbrace=\delta^i_{j}\ ,
\quad\left\lbrace \Pi_i,\;\Pi_j\right\rbrace =q\,F_{ij} }\,.
\eeq
We can now deduce the Hamilton equations yielding the geodesic motion of the scalar particle on the $\,3\,$-manifold,
\begin{eqnarray}
\dot{x}^i&=&\left\lbrace x^i,\,\mathcal{H}\right\rbrace=g^{ij}(\vx)\,\Pi_j,
\\[5pt]
\dot{\Pi}_{i}&=&\left\lbrace \Pi_i,\,\mathcal{H}\right\rbrace=\displaystyle{
q\,F_{ij} \,\dot{x}^j-\partial_i\displaystyle{V} +\Gamma^k_{ij}\,\Pi_k\,\dot{x}^j}\;. \label{Lorentzbis}
\end{eqnarray}
Note that the Lorentz equation (\ref{Lorentzbis}) involves also in addition to the monopole and potential terms a curvature term, typical for motion in curved space, which is quadratic in the velocity.\\

We  now inquire about the symmetries of the system. For our investigation, we recall that constants of the motion, noted as $\,Q\,$, which are polynomial in the momenta, can be derived following van Holten's algorithm \cite{vH}. The clue in this technique is to expand $\,Q\,$ into a power series of the covariant momentum,
\begin{eqnarray}\displaystyle{
Q= C+C^i\,\Pi_i+\frac{1}{2!}\,C^{ij}\,\Pi_i\Pi_j+\frac{1}{3!}\,C^{ijl}\,\Pi_i\Pi_j\Pi_l+\cdots}\,,
\label{devQbis}
\end{eqnarray}
and to require $Q$ to Poisson-commute with the Hamiltonian augmented with an effective potential, 
$\,\displaystyle{\mathcal{H}=\frac{1}{2}\,\vec{\Pi}^{2}+V(\vx)
}\,$. This yields the series of constraints,
\beqa
\left\lbrace
\begin{array}{lllll} 
C^m\;\partial_m\,V(\vx)=0& \hbox{o(0)} & \\[9pt]
\partial_n C=q\,F_{nm}\,C^m+C_n^{\;m}\partial_m\,V(\vx) & \hbox{o(1)} &\\[9pt]
\mathcal{D}_{i}C_{l}+\mathcal{D}_{l}C_i=q\,\left( F_{im}\,C_l^{\;m}+F_{lm}\,C_i^{\;m} \right)+C_{il}^{\;\;k}\partial_k\,V(\vx)& \hbox{o(2)} &\\[9pt]
\mathcal{D}_iC_{lj}+\mathcal{D}_jC_{il}+\mathcal{D}_lC_{ij}=q\,\left( F_{im}\,C_{lj}^{\;\;m}+F_{jm}\,C_{il}^{\;\;m}+ F_{lm}\,C_{ij}^{\;\;m}\right)\\
\qquad\qquad\qquad\qquad\qquad\quad+\quad C_{ijl}^{\;\;\;m}\partial_m\,V(\vx)& \hbox{o(3)} &
\\
\cdots\cdots\,.
\end{array}\right.\label{ConsTraints}
\eeqa
which have to be solved. Here the zeroth-order constraint can be interpreted as a consistency condition for the effective potential. It is worth noting that the expansion can be truncated at a finite order provided some higher-order constraint reduces to a Killing equation,  
\beq
\mathcal{D}_{\left(i_1\right.}C_{\left. i_2\;\cdots\;i_n \right)}=0\,,\label{KillingNAbis}
\eeq
where the covariant derivative is constructed with the Levi-Civita connection so that
\beq
\displaystyle{\mathcal{D}_iC^j =\partial_iC^j +\Gamma^j_{\;ik}\,C^k}\,.
\eeq
Then, $\,\displaystyle{C_{i_1\cdots i_p}=0}\,$ for all $\,p\,\geqslant\,n\,$ and the constant of motion takes the polynomial form,
\beq
Q=\sum_{k=0}^{p-1}\,\frac{1}{k!}\,C^{i_1\cdots i_k}\,\Pi_{i_1}\cdots\Pi_{i_k}\,.
\eeq\\

The previously presented van Holten recipe is based on Killing tensors of the $\,3\,$-manifold. Indeed, the conserved angular momentum is associated with a rank-1 Killing tensor (i.e. Killing vector), which generates spatial rotations. Rank-$\,2\,$ Killing tensors lead to conserved quantities quadratic in covariant momenta $\,\vec{\Pi}\,$'s, etc.
Note that Killing tensors has been advocated by Carter in the context of the Kerr metric \cite{Carter}. 

Let us discuss two particular Killing tensors on the $\,3\,$-manifold which carries the metric,
\beq g_{ij}(\vx)=f(\vx)\,\delta_{ij}\,.
\eeq
Our strategy is to find conditions for lifting  the Killing tensors, which generate the conserved angular momentum and the Runge-Lenz vector of planetary motion in flat space, respectively, to the ``Kaluza-Klein'' 
$4$-space.\\

$\bullet\,$ First, we search for a rank-1 Killing tensor generating ordinary spatial rotations as
\beqa\displaystyle{C_i=g_{ij}(\vx)\,\epsilon^j_{\;\;kl}\;n^k\,x^l}\,.
\eeqa
We require $\,C_{i}\,$ to satisfy the Killing equation $\,\displaystyle{
\mathcal{D}_{\left(i\right.}C_{\left. j \right)}=0}\,$, so that we obtain the following theorem \cite{Ngome:2009pa}:
\begin{prop}
On the curved $3$-manifold carrying the metric $\,g_{ij}(\vx)=f(\vx)\,\delta_{ij}$, the rank-1 tensor
\beq
C_i=g_{ij}(\vx)\,\epsilon^j_{\;\;kl}\;n^k\,x^l\,\nn
\eeq
is a Killing tensor generating spatial rotations around the fixed unit vector $\,\vn\,$, provided
\beq
\left(\vx\times\vnabla\,f(\vx)\right)\cdot\vn=0\,.\label{CondTh1}
\eeq
\label{angularM}
\end{prop}\vspace*{-12mm}
Note that Theorem \ref{angularM} can be satisfied for some, but not all $\vn$'s. In the two-center metric case, for example, it only holds for $\vn$ parallel to the axis of the two centers (see the next section).

 An important case to consider is when the metric is radial,
\beq
f(\vx)=f(r)\,,\label{RadialMs}
\eeq
including the Taub-NUT metrics. In that event, the gradient is parallel to $\,\vx\,$ and (\ref{CondTh1}) holds for all $\vn$'s. Thus, Theorem \ref{angularM} is always satisfied for radial metrics.\\

$\bullet\,$ Next, inspired by the known flat-space expression, we consider the rank-2 Killing tensor associated with the Runge-Lenz-type conserved quantity
\beq
C_{ij}=2\,g_{ij}(\vx)\,n_k\,x^k-g_{ik}(\vx)\,n_j\,x^k-g_{jk}(\vx)\,n_i\,x^k\,.
\label{KillT2}
\eeq
In order to deduce conditions on the metrics admitting a Kepler-type dynamical symmetry, we impose $\,\displaystyle{ \mathcal{D}_{\left( i\right.}C_{\left. j l\right)}}\,$ to vanish. A tedious calculation provides us with
\begin{eqnarray}
\begin{array}{lr}\displaystyle{ \mathcal{D}_{\left( i\right.}C_{\left. j l\right)}      =2\,n_{k}\,x^{m}\,\left(\, g_{ij}(\vx)\,\Gamma^{k}_{lm}+g_{il}(\vx)\,\Gamma^{k}_{jm}+g_{jl}(\vx)\,\Gamma^{k}_{im}\,
\right)}-n_{i}\displaystyle{x^{m}\partial_{m} g_{jl}(\vx)}\\[12pt]\quad\quad\quad\quad\quad\quad\quad\quad\quad\quad\quad\quad\quad\displaystyle{-\;n_{j}\,x^{m}\partial_{m}g_{il}(\vx)-n_{l}\,x^{m}\partial_{m}g_{ij}(\vx)\;.
}\end{array}\label{F1}
\end{eqnarray}
Let us now calculate each term on the right hand side of (\ref{F1}). We first obtain
\begin{eqnarray}\begin{array}{ll}\displaystyle{
n_{i}\,x^{m}\,\partial_{m}g_{jl}(\vx)=f^{-1}(\vx)\,n_{i}\,g_{jl}(\vx)\,x^{m}\,\partial_{m}f(\vx)}
\\[10pt]\displaystyle{
n_{j}\,x^{m}\,\partial_{m}g_{il}(\vx)=f^{-1}(\vx)\,n_{j}\,g_{il}(\vx)\,x^{m}\,\partial_{m}f(\vx)}
\\[10pt]\displaystyle{
n_{l}\,x^{m}\,\partial_{m}g_{ij}(\vx)=f^{-1}(\vx)\,n_{l}\,g_{ij}(\vx)\,x^{m}\,\partial_{m}f(\vx)}\,,\end{array}\label{F2}
\end{eqnarray}
and next the curvature terms yield,
\begin{eqnarray*}\begin{array}{cc}\displaystyle{
2\,g_{jl}(\vx)\,n^{k}\,x^{m}\,\Gamma_{im}^{k}=f^{-1}(\vx)\,n_{i}\,g_{jl}(\vx)\,x^{m}\partial_{m}f(\vx)+f^{-1}(\vx)\,n_{m}\,x^{m}\,g_{jl}(\vx)\,\partial_{i}f(\vx)}\\[10pt]\displaystyle{\quad\quad\quad+\;f^{-1}(\vx)\,n_{k}\,g_{jl}(\vx)\,g^{nk}(\vx)\,g_{im}(\vx)\,x^{m}\,\partial_{n}f(\vx)}\,,
\end{array}\label{F3}
\end{eqnarray*}
\begin{eqnarray*}\begin{array}{cc}\displaystyle{
2\,g_{il}(\vx)\,n^{k}\,x^{m}\,\Gamma_{jm}^{k}=f^{-1}(\vx)\,n_{j}\,g_{il}(\vx)\,x^{m}\partial_{m}f(\vx)+f^{-1}(\vx)\,n_{m}\,x^{m}\,g_{il}(\vx)\,\partial_{j}f(\vx)}\\[10pt]\displaystyle{\quad\quad\quad+\;f^{-1}(\vx)\,n_{k}\,g_{il}(\vx)\,g^{nk}(\vx)\,g_{jm}(\vx)\,x^{m}\,\partial_{n}f(\vx)}\,,
\end{array}\label{F4}
\end{eqnarray*}
\begin{eqnarray*}\begin{array}{cc}\displaystyle{
2\,g_{ij}(\vx)\,n^{k}\,x^{m}\,\Gamma_{im}^{k}=f^{-1}(\vx)\,n_{l}\,g_{ij}(\vx)\,x^{m}\partial_{m}f(\vx)+f^{-1}(\vx)\,n_{m}\,x^{m}\,g_{ij}(\vx)\,\partial_{l}f(\vx)}\\[10pt]\displaystyle{\quad\quad\quad+\;f^{-1}(\vx)\,n_{k}\,g_{ij}(\vx)\,g^{nk}(\vx)\,g_{lm}(\vx)\,x^{m}\,\partial_{n}f(\vx)}\,.
\end{array}\label{F5}
\end{eqnarray*}
Inserting (\ref{F2}) and the previous curvature terms into (\ref{F1}), we get
\beqa
\mathcal{D}_{\left( i\right.}C_{\left. j l \right)}  
=f^{-1}(\vx)\bigg( g_{\left(ij\right.}\partial_{\left.l\right)}f(\vx)\,n_{m}\,x^{m}- g_{\left(ij\right.}x_{\left.l\right)}\,n^{m}\partial_{m}f(\vx)\bigg)
\,.\nn
\eeqa
Requiring $\,\displaystyle{
\mathcal{D}_{\left( i\right.}C_{\left.j l \right)}=0} \,$ yields the following theorem \cite{Ngome:2009pa}:
\begin{prop}\label{RLenzCond}
On the curved $3$-manifold carrying the metric $\,g_{ij}(\vx)=f(\vx)\,\delta_{ij}$, the tensor
\beq
C_{ij}=2\,g_{ij}(\vx)\,n_k\,x^k-g_{ik}(\vx)\,n_j\,x^k-g_{jk}(\vx)\,n_i\,x^k\nn
\eeq
is a symmetrical rank-$\,2\,$ Killing tensor, associated with the Runge-Lenz-type vector, provided
\beq
\vn\times\left(\vx\times\vnabla\,f(\vx)\right)=0\,.
\eeq
\end{prop}
Note that the radial metrics (\ref{RadialMs}) satisfy again the Theorem \ref{RLenzCond} so that, in addition to the rotational symmetry, they also admit a Kepler-type dynamical symmetry.\\

We can also remark that taking into account the compatibility condition of the metric tensor,
\beq
\mathcal{D}_{k}\,g_{ij}(\vx)=0\,,
\eeq
the $\,g_{ij}(\vx)\,$ always verifies the order-$\,2\,$ Killing equation $\,\displaystyle{
\mathcal{D}_{\left( k\right.}g_{\left.ij\right)}=0}\,$. Hence, the metric tensor is itself a symmetrical rank-2 Killing tensor and the associated conserved quantity is the Hamiltonian \cite{GR,vH}.\\

Having determined the generators of the symmetry  which were previously the object of our considerations, we can construct the associated constants of the geodesic motion using the algorithm (\ref{ConsTraints}). We investigate the radially symmetric generalized TAUB-NUT metric so that (\ref{metricTN}) becomes,
\beq
dS^2=f(r)\,\delta_{ij}\,dx^i\,dx^j+h(r)\,\big(dx^4+A_k\,dx^k\big)^2\,.\label{radmetricTN}
\eeq
Then, the Lagrangian (\ref{LagrangTN})  takes the form,
\begin{eqnarray}\displaystyle{
\mathcal{L}=\frac{1}{2}\,f(r)\,\dot{\vx}^{\,2}+\frac{1}{2}h(r)\,\big(\,\frac{d x^4}{dt}+A_k\,\frac{d x^k}{dt}\,\big)^2-U(r)}\,,\label{LagrangTNG}
\end{eqnarray}
where the scalar potential $\,U(r)\,$ is necessary to furnish Killing $2$-tensors.
Respectively associated with the cyclic variables $\,x^{4}\,$ and time $\,t\,$, the conserved electric charge and the energy read  
\beq
\displaystyle{
q=h(r)\,\big(\,\frac{d x^4}{dt}+A_k\,\frac{d x^k}{dt}\,\big)}\;,\quad\displaystyle{\mathcal{E}=  \frac{\vec{\Pi}^{2}}{2\,f(r)}+\frac{q^2}{2\,h(r)}+U(r)}\,.\label{ConsEC}
\eeq
Using the relations (\ref{ConsEC}), we can rearrange the dimensionally reduced Hamiltonian as
\beq
\displaystyle{
\mathcal{H}= \frac{1}{2}\,\vec{\Pi}^{2}+f(r)\,W(r)}
\displaystyle{\quad\hbox{with}\quad
\,W(r)=U(r)+\frac{q^{2} }{2\,h(r)}+ \frac{\mathcal{E} }{f(r)}-\mathcal{E}}\,,\label{FormPot1}
\eeq
which we can now use to derive conserved quantities via the algorithm (\ref{ConsTraints}).

{\bf{1)}} First, we look for conserved angular momentum which is linear in the covariant momentum since the $3$-metric now satisfies Theorem \ref{angularM}. Hence, $\,C_{ij}=C_{ijk}=\cdots =0\,$ so that (\ref{ConsTraints}) reduces to
\beqa\left\lbrace\begin{array}{lll} 
 C^m\;\partial_m\,\big(f(r)\,W(r)\big)=0&\hbox{o(0)}& 
\\[10pt]
\partial_n C=q\,F_{nm}\,C^m &\hbox{o(1)}& 
\\[10pt]
\mathcal{D}_{i}C_{l}+\mathcal{D}_{l}C_i=0\,.&\hbox{o(2)}&
\label{SystOrder1}
\end{array}\right.
\eeqa

$\bullet$ The second- and the first-order constraints yield
\beq
\displaystyle{
C_i=g_{im}(r)\,\epsilon^m_{\;\;\;nk}\;n^n\;x^k }
\quad\hbox{and}\quad\displaystyle{
C=-qg\,n_k\,\frac{x^k}{r}}\,,
\eeq
respectively.
The zeroth-order consistency condition in (\ref{SystOrder1})  is satisfied for an arbitrary radial effective potential, providing us with the conserved angular momentum,
\begin{eqnarray}
\displaystyle{\vJ=\vx\times\vPi-qg\,\hx}\,,
\label{ANGTN}
\end{eqnarray}
involving the  typical monopole term.

{\bf{2)}} Let us now turn to quadratic conserved quantities. In that event, we have $\,C_{ijk}=\nobreak\cdots =0\,$ which implies the series of constraints,
\begin{eqnarray}\left\lbrace \begin{array}{llll} 
C^m\;\partial_m\,\big(f(r)\,W(r)\big)=0&\hbox{o(0)}& 
\\[10pt]
\partial_n C=q\,F_{nm}\,C^m+C_{n}^{\;m}\,\partial_m\,\big(f(r)\,W(r)\big)&\hbox{o(1)}& 
\\[10pt]
\mathcal{D}_{i}C_{l}+\mathcal{D}_{l}C_i=q\,\left( F_{im}\,C_l^{\;m}+F_{lm}\,C_i^{\;m} \right)&\hbox{o(2)}&
\\[10pt]
\mathcal{D}_iC_{lj}+\mathcal{D}_jC_{il}+\mathcal{D}_lC_{ij}=0\,.&\hbox{o(3)}&
\end{array}\right.\label{SystOrder2}
\end{eqnarray}

$\bullet$ Taking $\,C_{ij}=g_{ij}(r)\,$ as a rank-2 Killing tensor, we deduce from the second-order equation of (\ref{SystOrder2}) that $\,C_i =0\,$. As expected, the first-order and the zeroth-order consistency relation are both satisfied by any radial effective potential $\,C=f(r)\,W(r)\,$.  The conserved energy 
associated, therefore, read as
\begin{eqnarray}
\mathcal{\E}= \frac{1}{2}\vec{\Pi}^{2}+f(r)\,W(r)\,.\label{FixE}
\end{eqnarray}

$\bullet$ Next, we search for a Runge-Lenz-type vector generating the Kepler-type dynamical symmetry of the system. Since Theorem \ref{RLenzCond} is satisfied by the considered radial $\,3\,$-metric, we have to solve the constraints (\ref{SystOrder2}) using the rank-2 Killing tensor 
\beq\displaystyle{ C_{ij}= 2\,g_{ij}(r)\,n_k\,x^k-g_{ik}(r)\,n_j\,x^k-g_{jk}(r)\,n_i\,x^k}
\label{RL1}
\eeq
inspired by its form in the Kepler problem. We solve the second-order constraint of (\ref{SystOrder2}), and we get
\begin{eqnarray}\displaystyle{ C_{i}= \frac{q\,g}{r}\,g_{im}(r)\,\epsilon^{m}_{\;\;\;jk}\,n^j\,x^k}\,.\label{RL2}
\end{eqnarray} 
Next, inserting (\ref{RL1}) and (\ref{RL2}) into the first-order constraint of (\ref{SystOrder2}), we obtain
\begin{eqnarray*}
\displaystyle{
\partial_j C=\left(\frac{\big(f(r)\,W(r)\big)'}{r}+\frac{q^2g^2}{r^4}\right)x_j\,n_k\,x^k-\left(r\big(f(r)\,W(r)\big)'+\frac{q^2g^2}{r^2} \right)n_j}\,.\end{eqnarray*}
It is now easy to analyze the integrability condition of the previous equation by requiring the vanishing of the commutator,
\begin{eqnarray}\displaystyle{
\left[\,\partial_i\,,\,\partial_j\, \right]C=0\quad\Longrightarrow\quad
\Delta\left(f(r)\,W(r)- \frac{q^2g^2}{2r^2}\right)=0 }\,.
\label{Laplacecond}
\end{eqnarray}
Thus, the bracketed quantity must satisfy the 
\emph{Laplace equation}.

The zeroth-order equation is identically satisfied. Consequently, a Runge-Lenz-type conserved vector does exist only when the radial effective potential is 
\begin{eqnarray}\displaystyle{
f(r)\,W(r)=\frac{q^2g^2}{2r^2}+\frac{\beta}{r}+\gamma}\quad\hbox{with}\quad\displaystyle{\beta, \gamma\;\in\;\IR }\,.\label{FormPot2}
\end{eqnarray}
Equivalent to the result of Gibbons and Warnick \cite{GW}, the formulas (\ref{FormPot1}) and (\ref{FormPot2}) allow us to announce the theorem \cite{Ngome:2009pa}:
\begin{prop}
For the generalized TAUB-NUT metric (\ref{radmetricTN}), the most general potentials $\,U(r)\,$ permitting the existence of a Runge-Lenz-type conserved vector are given by
\beq
U(r)=\left(\frac{q^2g^2}{2r^2}+\frac{\beta}{r}+\gamma
 \right)\frac{1}{f(r)} -\frac{q^2}{2\,h(r)}+\mathcal{E}\,,
 \label{pot37}
\eeq
\label{PotCond}
where $\,q\,$ and $\,g\,$ denote the particle and the monopole charge. And $\,\beta\,$, $\gamma$ are free constants and $\,\E\,$ is the fixed energy [cf. (\ref{FixE})].
\end{prop}
Inserting now (\ref{FormPot2}) into the first-order constraint of (\ref{SystOrder2}) provides us with 
\beq
\partial_n C=\frac{\beta}{r}\,n_n-\frac{\beta}{r^3}\,n_k\,x^k\,x_n\,,
\eeq
which is solved by
\beq
\displaystyle{C=\frac{\beta}{r}\,n_k\,x^k }\,.\label{RL3}
\eeq
Collecting the results (\ref{RL1}), (\ref{RL2}) and (\ref{RL3}) yield the conserved Runge-Lenz-type vector,
\begin{eqnarray}\displaystyle{
\vec{K}=\vec{\Pi}\times\vec{J}+\beta\,\frac{\vec{x}}{r}}\,.\label{RLTN}
\end{eqnarray}
Due to the simultaneous existence of the conserved angular momentum (\ref{ANGTN}) and the conserved Runge-Lenz vector (\ref{RLTN}), we obtain a complete description of the motion for generalized TAUB-NUT metric. Indeed, the motions of the particle are confined to conic sections \cite{FH}. Our class of metrics, which satisfy Theorem \ref{PotCond}, includes the following.

\begin{enumerate}
\item
The original TAUB-NUT case \cite{Sork,GrossPerry}  with vanishing external $\,U(r)=0$,
\beq
f(r)=\frac{1}{h(r)}
=1+\frac{4m}{r},
\label{TNcase}
\eeq
where $m$ is real \cite{FH,GR}. We note that the monopole scattering case corresponds to $\,m=-1/2\,$, see \cite{GM,FH,GRub}. We then obtain for
\beq\gamma=q^2/2-\mathcal{E}\quad\hbox{and charge}\quad g=\pm 4m\,,
\eeq
the conserved Runge-Lenz vector,
\beq
\vec{K}=\vPi\times\vec{J}-4m\left(\mathcal{E}-q^{2}\right)\hx\,.\label{KKRL}
\eeq
\item  Lee and Lee \cite{LeeLee} argued that for monopole
scattering with independent
components of the Higgs expectation values, the geodesic Lagrangian (\ref{LagrangTN}) should be replaced by $L\to L-U(r)$,
where the external potential reads
\beq
U(r)=\frac{1}{2}\,\frac{a^{\;\,2}_{0}}{1+\displaystyle\frac{4m}{r}}\ .
\label{LLpot}
\eeq
It is now easy to see that this addition merely shifts the
value in the brackets in (\ref{Laplacecond}) by a constant and corresponds to a shift of $a^{\;\,2}_{0}/2$ in the energy. Hence, the Laplace equation in (\ref{Laplacecond}) is still satisfied. So the previously found Runge-Lenz vector (\ref{KKRL}) is still valid.\vspace*{-1,7mm}
\item The metric associated with winding strings \cite{GR2} where
\beq
f(r)=1,
\qquad 
h(r)=\frac{1}{\big(1-\displaystyle\frac{1}{r}\big)^2}\,.
\eeq
For charge $\,g=\pm 1\,$, we deduce from Theorem \ref{PotCond},
\beq\displaystyle{\left(\beta+q^{2}\right)-r\,\left(U(r)-\gamma+\frac{q^{2}}{2}-\mathcal{E}\right)=0}\,,\nn
\eeq
so that for the fixed energy, $\,\mathcal{E}=q^{2}/2-\gamma+U(r)\,$, the conserved Runge-Lenz vector reads as
\beq\displaystyle{
\vec{K}=\dot{\vx}\times\vec{J}-q^{2}\,\hx}\,.
\eeq
\item The extended TAUB-NUT metric \cite{Japs1,Japs} where
\begin{eqnarray}\begin{array}{cc}\displaystyle{
f(r)= b+\frac{a}{r},
\quad 
h(r)= \frac{a\,r+b\,r^2}{1+d\,r+c\,r^2}}\,,
\end{array}
\label{japexp}
\end{eqnarray}
with the constants 
$\,(a,\, b,\, c,\, d\,) \in\IR\,$. With the choices $\,U(r)=0\,$ and charge $\,g=\pm 1\,$,  Theorem \ref{PotCond} requires
\begin{eqnarray*}\displaystyle{
-r\,f(r)\,\mathcal{E}+\frac{r\,f(r)}{h(r)}\,\frac{q^{2}}{2}-\frac{q^{2}}{2\,r}-\gamma\,r}=\beta=\const\,.
\end{eqnarray*}
Inserting here (\ref{japexp}) yields
$$
\Big(-a\,{\cal E}+\frac{1}{2}d\,q^2-\beta\Big)+
r\Big(-b\,{\cal E}+\frac{1}{2}c\,q^2-\gamma \Big)=0,
$$
which holds when
\beq
\beta=-a\,{\cal E}+\frac{1}{2}d\,q^2\quad\hbox{and}\quad\gamma=-b\,{\cal E}+\frac{1}{2}c\,q^2\,.
\eeq
Then, we get the conserved Runge-Lenz vector
\beq
\vK=\vPi\times\vec{J}-\left(a\,
\E-\frac{1}{2}\,d\,q^{2}\right)\hx\,.
\eeq

\item Considering the oscillator-type metric discussed by Iwai and Katayama \cite{Japs1,Japs}, the functions $\,f(r)\,$  and $\,h(r)\,$ take the form
\beq
\displaystyle
f(r)= b+ar^2\quad\mbox{and}\quad h(r)= \frac{ar^4+br^2}{1+cr^2+dr^4}\,.
\eeq
A direct calculation leads to the following Runge-Lenz-type vector \cite{Marquette1},
\beq
\vK=\left(b+ar^2\right)\,\dot{\vx}\times\vJ+\beta\,\frac{\vx}{r}\,.
\eeq
Which is conserved only for scalar potential of the form
\beq
\displaystyle
U(r)=\left( \frac{q^2g^2}{2r^2}+\frac{\beta}{r}+\gamma\right)\left( b+ar^2\right)^{-1} -q^2\left(  \frac{1+cr^2+dr^4}{ar^4+br^2}\right)\,.
\eeq
\end{enumerate}
Let us just conclude by outlining that the five examples treated above are shown to be particular cases deduced from the general expression (\ref{RLTN}), see \cite{Ngome:2009pa}. See also \cite{Igata} for applications on the Kerr metric. The case of SUSY of the Kerr metric is investigated in \cite{Galajinsky}.

\newpage

\subsection{Multi-center metrics}

The multi-center metrics family in which we are interested in this section are known to be Euclidean vacuum solutions of the Einstein equations, with self-dual curvature. The multi-center metrics can also be viewed as an extension of the TAUB-NUT metrics studied in the previous section \cite{GR}.

Let us begin by considering a scalar particle moving in the Gibbons-Hawking space \cite{GH}, which generalizes the TAUB-NUT space. The Lagrangian function associated with this dynamical system, as in (\ref{LagrangTN}), is given by
\beq
\mathcal{L}=\frac{1}{2}\,f(\vx)\,\dot{\vx}^{\,2}+\frac{1}{2}f^{-1}(\vx)\,\big(\,\frac{dx^4}{dt}+A_k\,\frac{dx^k}{dt}\,\big)^2-U(\vx)\,.\nn
\eeq
But here the functions $\,f(\vx)\,$ obey the following ``self-dual'' ANSATZ \cite{GH},
\beq
\vnabla\,f=\pm\vnabla\times\vA\,.
\eeq
Hence, $\,f(\vx)\,$ is an harmonic function,
\beq
\Delta\,f(\vx)=0\,.\label{HarmonicEq}
\eeq
The most general solution of (\ref{HarmonicEq}) is given by
\begin{eqnarray}\displaystyle{
f(\vx)=f_{0}+\sum_{i=1}^{N}\frac{m_{i}}{\vert\vx-\va_{i}\vert}}\quad\hbox{with}\quad(f_{0}\,,\;m_{i})\in \IR ^{N+1}\,.
\end{eqnarray} 
Thus, the multi-center metric admits multi-NUT singularities so that the \textit{jth} NUT singularity is characterized by the charge $\,m_{j}\,$ and is located at $\,\va_{j}\,$. However, we can remove these singularities provided all NUT charges are equal,
\beq
m_{1}=m_{2}=\cdots=m_{i}=\frac{g}{2}\,.
\eeq
In this case, the cyclic variable $\,x^{4}\,$ is periodic with the range,
\beq
0 \leq x^{4}\leq 4\pi\,\frac{g}{N}\,.
\eeq
We can now investigate the symmetries associated with the projection of the particle's motion onto the curved 3-manifold described by the multi-center metric tensor,
\begin{eqnarray}\displaystyle{
g_{jk}(\vx)=\left(
f_{0}+\sum_{i=1}^{N}\frac{m_{i}}{\vert\vx-\va_{i}\vert}\right)\,\delta_{jk}}\,.\label{MCmetric}
\end{eqnarray}
The projected Hamiltonian is given by
\beq
\mathcal{H}= \frac{1}{2}\vec{\Pi}^{2}+f(\vx)W(\vx)\quad\hbox{with}\quad
W(\vx)=U(\vx)+\frac{q^{2}}{2}f(\vx)+ \mathcal{E}f^{-1}(\vx)-\mathcal{E}\,.
\eeq
Let us first note that for multi-center metric (\ref{MCmetric}), it is straightforward to deduce from Theorem \ref{RLenzCond} the metric condition,
\beq
\displaystyle\sum_{i=1}^{N}\frac{\big(\vn\cdot\vx\big)\va_{i}-\big(\vn\cdot\va_{i}\big)\vx}{\vert\vx-\va_{i}\vert^{3}}=0\,,
\eeq
which can not hold for more than two centers. Thus, we state the following theorem \cite{Ngome:2009pa}:\vspace*{-1,5mm}
\begin{prop}\label{NmetricRL}
In the curved 3-manifold carrying the $\,N$-center metric,
\beq
g_{jk}(\vx)=\left(
f_{0}+\sum_{i=1}^{N}\frac{m_{i}}{\vert\vx-\va_{i}\vert}\right)\,\delta_{jk}\,,\nn
\eeq
no symmetry of the Kepler-type occurs for $N>2$.
\end{prop}\vspace*{-1mm}
For simplicity, from now on we limit ourselves to a discussion of the two-center metrics, 
\beq
f(\vx)=f_{0}+\frac{m_{1}}{\vert\vx-\va\vert}+\frac{m_{2}}{\vert\vx+\va\vert}\,,\quad\vert\vx\pm\va\vert\neq0\,,\label{2CenterM}
\eeq
which are relevant for diatomic molecule systems and which possess some interesting symmetry properties. These metrics include, as special regular cases, those listed in 
Table \ref{Table1}.

\begin{table}[!h]
\begin{center}
\begin{tabular}{|| l || c || r ||}
  \hline\hline
  $\,f_{0}\,$ & $\,N\,$ & Type of Metric \\
  \hline\hline
  0 & 1 & ($m_{1}$ or $m_{2}$ = 0) Flat space\;\\ \hline
  1 & 1 & ($m_{1}$ or $m_{2}$ = 0) TAUB-NUT\, \\ \hline
  0 & 2 & Eguchi-Hanson\, \\\hline
  1 & 2 & Double TAUB-NUT\,.\\
  \hline\hline 
  \end{tabular}\caption{Examples of two-center metrics.}    
  \label{Table1}
  \end{center}
\end{table}

Let us now apply the van Holten algorithm (\ref{ConsTraints}) to derive the symmetry of the two-center metric.

{\bf{1)}} First, finding conserved quantities linear in the covariant momentum require to solve the reduced series of constraints,
\beqa\left\lbrace\begin{array}{lll}
C^m\;\partial_m\,\big(f(\vx)\,W(\vx)\big)=0 &\hbox{o(0)}& 
\\[8pt]
\partial_n C=q\,F_{nm}\,C^m &\hbox{o(1)}& 
\\[8pt]
\mathcal{D}_{i}C_{l}+\mathcal{D}_{l}C_i=0\,. &\hbox{o(2)}& 
\label{SystMCOrder1}
\end{array}\right.
\eeqa

$\bullet$ From Theorem \ref{angularM}, we deduce  the rank-1 Killing tensor satisfying the second-order constraint of (\ref{SystMCOrder1}),
\begin{eqnarray}\displaystyle{
C_i=g_{im}(\vx)\,\epsilon^m_{\;\;\;l k}\;\frac{a^l}{a}\,x^k }\,,\quad a=\vert\vert\va\vert\vert\,.\label{Killing1MC}
\end{eqnarray}
The tensor (\ref{Killing1MC}) generates rotational symmetry around the axis through the two centers. Next, injecting both (\ref{Killing1MC}) and the magnetic field of the two centers,
\beq
\displaystyle{\vB=m_{1}\,\frac{\vx-\va}{\vert\vx-\va\vert^{3}}+m_{2}\,\frac{\vx+\va}{\vert\vx+\va\vert^{3}}}\,,\label{2CenterMagneticField}
\eeq
 into the first-order equation of (\ref{SystMCOrder1}) yield  
\beq\displaystyle{C=-q\,\left( m_1\,\frac{\vx-\va}{\vert \vx-\va\vert}+m_2\,\frac{\vx+\va}{\vert \vx+\va\vert}\right)\cdot\frac{\va}{a} }\;.
\eeq
Finally we obtain, as conserved quantity, the projection of the angular momentum onto the axis of the two centers,
\begin{eqnarray}\displaystyle\mathcal{J}_a=\mathcal{L}_{a}-q\left( m_1\,\frac{\vx-\va}{\vert \vx-\va\vert}+m_2\,\frac{\vx+\va}{\vert \vx+\va\vert}\right)\cdot\frac{\va}{a}}\quad\hbox{with}\quad\displaystyle{\mathcal{L}_{a}=\left(\vx\times\vPi\right)\cdot\frac{\va}{a}\,,
\end{eqnarray}
which is consistent with the axial symmetry of the two-center metric.

\vskip2mm
{\bf{2)}} Now we study quadratic conserved quantities,
$\,\displaystyle{Q=C+C^i\,\Pi_i+\frac{1}{2}\,C^{ij}\,\Pi_i\Pi_j}\,$. Putting $\,C_{ijk}=C_{ijkl}=\nobreak\cdots =0\,$, leaves us with,
\beqa\left\lbrace\begin{array}{llll}
C^m\;\partial_m\,\big(f(\vx)\,W(\vx)\big)=0&\hbox{o(0)}&\\[8pt]
\partial_n C=q\,F_{nm}\,C^m+C_{n}^{\;m}\,\partial_m\,\big(f(\vx)\,W(\vx)\big)&\hbox{o(1)}&
\\[8pt]
\mathcal{D}_{i}C_{l}+\mathcal{D}_{l}C_i=q\,\left( F_{im}\,C_l^{\;m}+F_{lm}\,C_i^{\;m} \right)&\hbox{o(2)}&
\\[8pt]
\mathcal{D}_iC_{lj}+\mathcal{D}_jC_{il}+\mathcal{D}_lC_{ij}=0\,.&\hbox{o(3)}&
\end{array}\right.\label{SystMCOrder2}
\end{eqnarray}

$\bullet$ We consider the reducible rank-2 Killing tensor,
\begin{eqnarray}\displaystyle{
C_{ij}=\frac{2}{a^2}\,g_{im}(\vx)\,g_{jn}(\vx)\,\epsilon^m_{\;\;\;l k}\,\epsilon^n_{\;\;\;p q}\,a^l\,a^p\,x^k\,x^q+\frac{2}{a^2}\,g_{il}(\vx)\,g_{jm}(\vx)\,a^l\,a^m}\,,\label{KillTensor}
\end{eqnarray}
which is a symmetrized product of Killing-Yano tensors. $\,\displaystyle{C_{i}=g_{im}(\vx)\,\epsilon^m_{\;\;\;l k}\;\frac{a^l}{a}\,x^k}\,$ generates rotations around the axis of the two centers and $\,\displaystyle{\widetilde{C}_j=g_{jm}(\vx)\,\frac{a^m}{a} }\,$ generates spatial translation along the axis of the two centers. Injecting (\ref{KillTensor}) into the second-order constraint of (\ref{SystMCOrder2}) yields
\begin{eqnarray}\displaystyle{ C_{i}= -\frac{2\,q}{a^{2}}\,g_{im}\,\epsilon^m_{\;\;\;jk}\,a^j\,x^k\,a_{l}\left(m_1\,\frac{x^{l}-a^{l}}{\vert \vx-\va\vert}+m_2\,\frac{x^{l}+a^{l}}{\vert \vx+\va\vert}\right)}\,.
\end{eqnarray} 
For vanishing effective potential, we solve the first-order constraint with
\beq
\displaystyle{
C=\frac{q^2}{a^{2}}\,\left(m_1\,\frac{\left(x^{l}-a^{l}\right)}{\vert \vx-\va\vert}\,a_{l}+m_2\,\frac{\left(x^{l}+a^{l}\right)}{\vert \vx+\va\vert}\,a_{l}\right)^{2}}
\eeq
so that we obtain the \emph{square of the projection of the angular momentum onto the axis of the two centers, plus 
a squared component along the axis of the two centers of the covariant momentum}, 
\begin{eqnarray}\displaystyle{
Q=\mathcal{J}_a^{2}+\Pi^2_a}\,.
\end{eqnarray}
As expected, this conserved quantity is not really a new constant of the motion \cite{GR,Valent1,Valent,DuvalValent}.

$\bullet$ Now we turn to the Kepler-type dynamical symmetry. Let us first check if a rank-$2$ Killing tensor associated with Runge-Lenz type conserved quantity does exist. To this end we apply Theorem \ref{RLenzCond} to the two-center metric,
\beq
g_{jk}(\vx)=f(\vx)\d_{jk}\,,\quad f(\vx)=\left(f_{0}+\frac{m_{1}}{\vert\vx-\va\vert}+\frac{m_{2}}{\vert\vx+\va\vert}\right).
\eeq
We obtain
\beq\displaystyle{
\vn\times\left(\vx\times\vnabla\,f(\vx)\right)=\left(\frac{m_2}{\vert\vx+\va\vert^3}-\frac{m_1}{\vert\vx-\va\vert^3}\right)\left(\vx\times\va \right)\times\vn}=0\,,
\eeq
according to Theorem \ref{RLenzCond}. Consequently we get
\begin{eqnarray}\displaystyle
\frac{m_2}{\vert\vx+\va\vert^3}-\frac{m_1}{\vert\vx-\va\vert^3}=0\qquad\mbox{or}\qquad\vx=k\,\va\,,\quad k=\const\,.\label{SphereMotion}
\end{eqnarray}
The right condition in (\ref{SphereMotion}) restricted to motions parallel to $\,\va\,$ and therefore implies no interesting case.
 
Considering the first case given by (\ref{SphereMotion}), we assume that both charges are positive $\,m_{1}>0\,$, $\,m_{2} > 0\,$, and we write $\,\va=\left(a_{1},\,a_{2},\,a_{3}\right)\,$. Thus, the left Equation in (\ref{SphereMotion}) becomes 
\begin{eqnarray}\begin{array}{cc}\displaystyle{
\left(x-a_{1}\,\rho\right)^{2}+\left(y-a_{2}\,\rho\right)^{2}+\left(z-a_{3}\,\rho\right)^{2}=    a^{2}\,\left(\rho^{2}-1\right)}\\[10pt]\displaystyle{
\hbox{with}\quad\rho = \frac{m_1^{2/3}+m_2^{2/3}}{m_2^{2/3}-m_1^{2/3}}}\;.
\end{array}\label{Star32}
\end{eqnarray}
We recognize here the equation of a 2-sphere of center $\,\va\,\rho\,$ and radius $\,\displaystyle{R = a\,\sqrt{\rho^{2}-1}}\,$, noted as $\,\mathcal{S}^{2}\,$. The latter shows that for two-center metric, a Kepler-type dynamical symmetry is only possible for motion confined onto the sphere, $\,\mathcal{S}^{2}\,$.

Before searching for the exact form of the associated Runge-Lenz conserved quantity, let us first check that the motions can be consistently confined onto this $2$-sphere, $\,\mathcal{S}^{2}\,$.

To this end, we assume that the initial velocity is tangent to $\,\mathcal{S}^{2}\,$ and, using the equations of motion, we verify that at time $\,\big(t+\delta t\big)\,$ the velocity remains tangent to $\,\mathcal{S}^{2}\,$. Thus we write
\beq
\vv(t_{0}+\delta t)= \vv_{0}+ \delta t\;\dot{\vv}_{0}\quad\hbox{with $ \vv_{0}=\vv(t_{0})$ tangent to $\,\mathcal{S}^{2}\,$}.\label{Acc}
\eeq
The equations of motion in the effective scalar potential chooses as (\ref{FormPot3}) have the shape
\beq
\dot{\vPi}=q\,\vv\times\vB-\vnabla\left(f(\vx)W(\vx)\right)-\frac{v^{2}}{2}\left(\frac{m_1}{\vert\vx-\va\vert^3}+\frac{m_2}{\vert\vx+\va\vert^3}\right)\vx\,.
\eeq
Injecting the expressions of the magnetic field of the two-center (\ref{2CenterMagneticField}) and the effective potential (\ref{FormPot3}), we obtain
\beq
\dot{\vPi}_{0}=\left(\frac{m_1}{\vert\vx-\va\vert^3}+\frac{m_2}{\vert\vx+\va\vert^3}\right)\left[q^{2}\left(\frac{m_1}{\vert\vx-\va\vert}+\frac{m_2}{\vert\vx+\va\vert}\right)-\frac{v^{\;2}_{0}}{2}+\beta\right]\vx_{0}=f(\vx)\,\dot{\vv}_{0}\,.\nn
\eeq
Thus  
$\vv(t_{0}+\delta t)$
in (\ref{Acc}) becomes
$$
 \vv_{0}+ \gamma\,\delta t\;\vx_{0}\,,
\quad
\gamma= f^{-1}(\vx)\left(\frac{m_1}{\vert\vx-\va\vert^3}+\frac{m_2}{\vert\vx+\va\vert^3}\right)\left(q^{2}\left(\frac{m_1}{\vert\vx-\va\vert}+\frac{m_2}{\vert\vx+\va\vert}\right)-\frac{v^{\;2}_{0}}{2}+\beta\right),
$$
 where $\vv_{0}$ and $\vx_{0}$ are tangent to the $2$-sphere $\,\mathcal{S}^{2}\,$. Hence, the velocity remains tangent to $\,\mathcal{S}^{2}\,$ along the motion.

Having shown the consistency of motions on the $\,2\,$-sphere $\,\mathcal{S}^{2}\,$, we can state the following theorem \cite{Ngome:2009pa}:
\begin{prop}\label{SphereTh}\vspace{2,0mm}
In the curved 3-manifold carrying the 2-center metric,
\beq
g_{jk}(\vx)=\left(
f_{0}+\frac{m_{1}}{\vert\vx-\va\vert}+\frac{m_{2}}{\vert\vx+\va\vert}\right)\,\delta_{jk}\,,
\eeq
a scalar Runge-Lenz-type conserved quantity does exist only for a particle moving along the axis of the two centers or for motions confined on the two-sphere of radius $\,\displaystyle{R = a\,\sqrt{\rho^{2}-1}}\,$ centered at $\,\va\,\rho\,$ $(\,m_{1},\,m_{2}>0)\,$. In the Eguchi-Hanson case $\,(m_{1}=m_{2})\,$, the $2$-sphere is replaced by the median plane of the two centers.
\end{prop}
Our method here is particular since instead searching for Kepler-type dynamical symmetry directly, we already look at the conditions of its existence. Knowing now, for the two-center metric, that only motions confined on $\,\mathcal{S}^{2}\,$ allow a Runge-Lenz-tpe conserved quantity, we can solve the second-order constraint of (\ref{SystMCOrder2}) using (\ref{Star32}). Thus, we obtain
\begin{eqnarray}\displaystyle{ C_{i}= \frac{q}{a}\,g_{im}\,\epsilon^m_{\;\;\;jk}\,a^j\,x^k\,\left(\frac{m_1}{\vert \vx-\va\vert}+\frac{m_2}{\vert \vx+\va\vert}\right)}\,,\label{RL2MC}
\end{eqnarray} 
where the only component of $\,\vn\,$ is along  the axis $\,\displaystyle{\va}/{a}\,$. The final step consists to solve the first-order constraint of (\ref{SystMCOrder2}). Indeed, following (\ref{FormPot2}) the clue is to choose
\beq
f(\vx)W(\vx)=
\frac{q^{2}}{2}\left(\frac{m_1}{\vert \vx-\va\vert}+\frac{m_2}{\vert \vx+\va\vert}\right)^{2}
+\beta\left(\frac{m_1}{\vert \vx-\va\vert}+\frac{m_2}{\vert \vx+\va\vert}\right)+\gamma\,,
\label{FormPot3}
\eeq
with $\,\beta,\,\gamma\;\in\;\IR\,$. Let us precise that this potential satisfies the consistency condition given by the zeroth-order constraint of (\ref{SystMCOrder2}). Moreover, the leading coefficient of the effective potential cancels the obstruction due to the magnetic field of the two centers, and the remaining part on the right-hand side of (\ref{FormPot3}) leads to
\begin{eqnarray}\displaystyle{C=\beta\,\left( m_1\,\frac{\vx-\va}{\vert \vx-\va\vert}+m_2\,\frac{\vx+\va}{\vert \vx+\va\vert}\right)\cdot\frac{\va}{a} 
}\,.\label{RL3MC}
\end{eqnarray}
Collecting our results (\ref{KillT2}), (\ref{RL2MC}) and (\ref{RL3MC}) provide us with the scalar,
\beq\displaystyle{
K_{a}=\left(\vPi\times\vJ\right)\cdot\frac{\va}{a}+\frac{\beta}{q}\,\left(\mathcal{L}_{a}-\mathcal{J}_{a}\right)
}\,,\label{RLScalar}
\eeq
which represents, in the case of two-center metrics (\ref{2CenterM}), a conserved Runge-Lenz-type scalar for particle moving on the 2-sphere of center positioned at $\,\va\,\rho\,$ and radius $\,\displaystyle{R = a\,\sqrt{\rho^{2}-1}}\,$, combined with the effective potential (\ref{FormPot3}).

\newpage

\subsection{Killing-St\"ackel Tensors on extended manifolds}\label{LiftedKillingTensor}

Having discussed, in the two previous sections, the conditions on Killing tensors which are related to the existence of constants of motion on the dimensionally reduced curved manifold. We can observe that the Killing tensor generating the Runge-Lenz-type quantity preserved by the geodesic motion can be lifted to an extended manifold.

Let us study this lifting problem in detail, by considering the geodesic motion of a particle before  dimensional reduction (\ref{DimReduc}). The particle evolves on the extended 4-manifold carrying the metric $\,g_{\mu\nu}(x)\,$ with $\,\mu,\nu=1,\cdots,4\,$. A rank-2 Killing-St\"ackel tensor on this curved 4-manifold is a symmetric tensor, $\,C_{\mu\nu}\,$, which satisfies
\beq\mathcal{D}_{\left( \lambda\right.}C_{\left.\mu\nu\right)} =0\,,\quad\lambda,\,\mu,\,\nu=1\,,\cdots,\,4\,.
\eeq
For the Killing-St\"ackel tensor generating the Runge-Lenz-type conserved quantity, the degree-2 polynomial function in the canonical momenta $\,p_{\mu}\,$ associated with the local coordinates $\,x^{\mu}\,$,
\beq
K=\frac{1}{2}\,C^{\mu\nu} \, p_{\mu} \, p_{\nu}\quad(\mu,\,\nu=1,\cdots,4)\,,
\eeq
is preserved along the geodesics. Then, the lifted Killing-St\"ackel tensor on the 4-manifold, which directly yields the Runge-Lenz-type conserved quantity is written as
\beq
\displaystyle{
C^{\mu\nu} =  \left(\begin{array}{cc} C^{i\,j}  & \quad C^{i\,4}\\[4pt]C^{4\,j}  & \quad C^{4\,4} \end{array}\right)\,,\quad i,\,j=1\,,2\,,3}\,.\label{LKT1}
\eeq
The tensor $\,C_{i\,j} \,$ is, therefore, a rank-2 Killing tensor on the dimensionally reduced curved 3-manifold carrying the metric $ \,g_{ij}(\vx)=\,f(\vx)\,\delta_{ij}\,$, which generates a Runge-Lenz-type quantity  conserved along the projection of the geodesic motion onto the curved 3-manifold. The off- and the diagonal contravariant components read
\beq
C^{i\,4}=C^{4\,i}=\frac{1}{q}C^{i}-C^{i}_{\;\;k}\,A^{k}\,,\quad
C^{4\,4}=\big({2}/{q^{2}}\big)C-\big({2}/{q}\big)C_{k}\,A^{k}+C_{jk}\,A^{j}A^{k}\,.
\eeq
The term $\,A^{k}\,$ represents the component of the vector potential of the magnetic field. In the case of the generalized TAUB-NUT metrics, the terms $\,C\,$ and $\,C_{k}\,$ are the results (\ref{RL3}) and (\ref{RL2}) of the first- and the second-order constraints of (\ref{SystOrder2}), respectively. In the case of the two-center metrics, $\,C\,$ and $\,C_{k}\,$ are given by the results (\ref{RL3MC}) and (\ref{RL2MC}), respectively.

\noindent$\bullet$ As an illustration, let us consider a particle in the gravitational potential, $\,\displaystyle{V(r)=-\frac{m_{0}\,G_{0}}{r}}$, described by the Lorentz metric \cite{Barg,DBKP,Balachandran,DGH},
\beq
dS^{2}=d\vx^{\,2}+2\,dx^{4}dx^{5}-2\,V(r)\left(dx^{5}\right)^{2}\,.\label{BargMetric}
\eeq
The variable $\,x^{5}=t\,$ is the non-relativistic time  and $\,x^{4}\,$ the vertical coordinate. Rotations, time translations and ``vertical'' [on the fourth direction] translations generate as conserved quantities the angular momentum $\,\vL\,$, the energy and the fixed mass $\,m\,$, respectively. The Runge-Lenz-type conserved quantity, along null geodesics of the 5-manifold described by the metric (\ref{BargMetric}),
\beq
K=\frac{1}{2}\,C^{ab} \, p_{a} \, p_{b}\quad\hbox{with}\quad a,\,b,\,c=1,\cdots, 5\
\eeq
 is derived from the trace-free rank-2 Killing-St\"ackel tensor \cite{DGH},
\beq
C^{ab}=\left(\frac{\hat{\eta}}{g^{c}_{\;\;c}}\right)g^{ab}-\eta^{ab}\quad\hbox{with}\quad\hat{\eta}=\eta^{ab}g_{ab}\,.\label{LlightKillingTensor}
\eeq
For some $\,\vn\,\in\IR^{3}\,$, the nonvanishing contravariant components of $\,\eta\,$ are given by
\beq
\eta^{ij}=n^{i}\,x^{j}+n^{j}\,x^{i}-\hat{\eta}\,\delta^{ij}\quad\hbox{and}\quad\eta^{45}=\eta^{54}=\hat{\eta}=n_{i}\,x^{i}\,,\label{KTReducedM}
\eeq
where we recognize, in the left hand side of (\ref{KTReducedM}), the generator of Kepler-type symmetry in the dimensionally reduced 3-manifold. 

A calculation of each matrix element of the Killing tensor (\ref{LlightKillingTensor}) leads to $\,C^{ab}\,$ whose only nonvanishing components
are,
\beq
\displaystyle{
C^{ij} =  2\,\hat{\eta}\,\delta^{ij}-n^{i}\,x^{j}-n^{j}\,x^{i}\quad\hbox{and}\quad C^{44} =2\,\hat{\eta}\,V(r)\,.}\label{LlightKillingTensor2}
\eeq
Consequently, the associated Runge-Lenz-type conserved quantity reads as
\beq
\vK\cdot\vn=\left(\vp\times\vL+m^{2}\,V(r)\,\vx\right)\cdot\vn\,.
\eeq
Note, in the previous expression, that the mass ``$\,m\,$'' is preserved by the ``vertical'' reduction. In the original Kepler case, we thus deduce on the dimensionally reduced flat 3-manifold that the symmetric tensor
\beq
C^{ij} =  2\,\delta^{ij}\,n_{k}\,x^{k}-n^{i}\,x^{j}-n^{j}\,x^{i}
\eeq
is a Killing-St\"ackel tensor generating the Runge-Lenz-type conserved quantity along the projection of the null geodesic of the 5-manifold onto the 3-manifold carrying the flat Euclidean metric.

\section{Non-Abelian gauge fields and the Berry phase}\label{chap:NAM}


{\normalsize
\textit{
Conserved quantities of an isospin-carrying particle in non-Abelian monopole-like fields are investigated. In the effective non-Abelian field for nuclear motion, obtained through the Berry phase in a diatomic molecule, due to Wilczek et al., an unusual conserved charge and angular momentum are constructed. 
}}

\subsection{The Wu-Yang monopole}\label{WYsection}

In 1968, T.T. Wu and C.N. Yang found the first ``monopole-like'' classical solution of the Yang-Mills field equations \cite{WuYang1}. Such a solution can also be viewed as an extension of the Abelian Dirac monopole solution when usual electrodynamics, with $\,U(1)\,$ symmetry, is considered as a part of a larger theory. The generator of the electromagnetism $\,U(1)\,$ subgroup should be embedded into the non-Abelian $\,SU(2)\,$ gauge group. 

In order to investigate the Wu-Yang monopole solution, we consider here a Yang-Mills theory described by the Lagrangian density,
\beq
\L=-\frac{1}{4}F^a_{\mu\nu}\,F^{a\,\mu\nu}\,,\quad \mu,\,\nu=0,1,\,2,\,3\,.\label{LagDensity}
\eeq
where $\,F^a_{\mu\nu}\,$ represents the antisymmetric Yang-Mills field strength tensor taking values in the 
Lie algebra of the gauge group $\,SU(2)\,$,
\beq
F^a_{\mu\nu}=\partial_{\mu}A^a_{\nu}-\partial_{\nu}A_{\mu}^a - e\,\varepsilon_{abc}A_{\mu}^bA_{\nu}^c\,,\quad a=1,\,2,\,3\,.
\eeq
Here $\,e\,$ and the antisymmetric tensor $\,\varepsilon_{abc}\,$ denote the gauge coupling constant and the structure constant of the gauge group, respectively. As expected, 
\beq
A_{\mu}=A_{\mu}^a\,\tau^a\,,
\eeq
takes value in the $\,su(2)\,$ Lie algebra. The Hermitian and traceless infinitesimal generators of the $\,SU(2)\,$ gauge group verify the commutation relation,
\beq
\left[\tau^a,\tau^b\right]=i\varepsilon_{abc}\tau^c\,,\quad\tau^{a}=\frac{1}{2}\sigma^{a}\,,
\eeq
where the $\,\sigma^{a}\,$ are Pauli matrices. Our approach now is to find the equations of the motion using a variational principle with the Yang-Mills Lagrangian density (\ref{LagDensity}). We obtain the classical source-free Yang-Mills (YM) equations
\beq
\partial_{\b}F^{d\,\a\b}-e\,\varepsilon^{adc}A^c_{\b}\,F^{a\,\a\b}=0\,,
\eeq
which can be written in a more compact way as
\beq
D_{\b}F^{d\,\a\b}=0
\,.
\label{eqYM}
\eeq
Searching for solutions of the equations of the motion (\ref{eqYM}) in the ``temporal'' gauge,
\beq
A^{a}_{0}=0\,,
\eeq
only time-independent gauge field and local gauge invariance are permitted. We can now posit the spherically symmetric Wu-Yang ANSATZ \cite{WuYang1},
\begin{eqnarray}\begin{array}{ll}\displaystyle{
A^a_{i}=g\,\varepsilon_{iaj}\,x^{j}\,\frac{\left(1-\Phi\left( r\right)\right) }{r^2}}\,,\\[12pt]
\displaystyle{r^2=x_i\,x^i\quad\hbox{with}\quad i,\,j=1,2,3}\,,
\end{array}\label{ansatz}
\end{eqnarray}
where $\,a\,$, $\,(i,j)\,$ and $\,g\,$ represent the color index, the space indices and the quantized Wu-Yang monopole charge, respectively. Also remark that $\,\Phi\,$ is a radial real function which is to be determined. As expected, a direct calculation implies that all time-dependent components of the field strength vanish,
\beq
F_{0\mu}^a=0\,,\quad\mu=0,1,\cdots,4\,,
\eeq
whereas the spatial components of the $2$-form curvature reads
\beq\begin{array}{ll}
\displaystyle F_{ij}^a=g\,\varepsilon_{ijk}\left\lbrace 2\d^{ak}\frac{\big(1- \Phi\big)}{r^{2}}-eg\,x^{k}x^{a}\left(\frac{1- \Phi}{r^{2}}\right)^2\right\rbrace\\[13pt]\displaystyle\quad\quad\,\,+\frac{g}{r}\frac{d}{dr}\left(\frac{1- \Phi}{r^2} \right)\left\lbrace\varepsilon_{jal}\,x^i\,x^l-\varepsilon_{ial}\,x^j\,x^l\right\rbrace \,.\label{cfYM}
\end{array}
\eeq
The right hand side bracket of (\ref{cfYM}) can be rewritten by using the relation,
\beq
\big(\varepsilon_{jal}\,x^i\,x^l-\varepsilon_{ial}\,x^j\,x^l\big)\tau^{a}=\varepsilon_{ijk}\big(r^{2}\d^{ak}-x^{a}x^{k}\big)\tau^{a}\,,
\eeq
so that the ``Wu-Yang'' field strength reduces to
\beq
\displaystyle
F^a_{ij}=g\,\varepsilon_{ijk}\left[-\frac{\delta^{ak}}{r}\,\frac{d\Phi}{dr}+\frac{x^a\,x^k}{r^3}\left(\frac{d\Phi}{dr}-eg\frac{\Phi^2}{r}+2\big(eg-1\big)\frac{ \Phi}{r}+\frac{2-eg}{r}\right)\right]\,.\label{ReducField}
\eeq
Injecting the relations (\ref{ansatz}) and (\ref{ReducField}) into the Yang-Mills field equations (\ref{eqYM}) provide us with the non-linear Wu-Yang equation,
\beq
r^2\,\frac{d^2\Phi}{dr^2}-eg\big(1-eg+eg\Phi\big)\big(\Phi-1\big)\left(\Phi+\frac{2-eg}{eg}\right)=0\,.\label{YMNLEq}
\eeq
Note that due to the non-linear nature of the YM equations, to search analytical solutions of (\ref{YMNLEq}) is an unconquerable task.

 Hence, we investigate numerically the non-linear equation (\ref{YMNLEq}) viewed as a dynamical system. To this end, without loss of generality, we reduce ourselves to the case where $\,eg=1\,$ \footnote{See the formula (\ref{QRelation}) in section \ref{DiracMonop}.}. Thus, the equation (\ref{YMNLEq}) takes the simple form,
\beq
r^2\,\frac{d^2 \Phi}{dr^2}- \Phi\big(\Phi-1\big)\big(\Phi+1\big)=0\,.\label{nonlineq}
\eeq
We first posit the variable change
\beq
r=\exp(\tau)\,,\quad\tau\in\IR\,,\;r\in\IR^+\,,
\eeq
where $\,\tau\,$ is viewed as an evolution parameter. Next, we multiply the resulting $\,\tau\,$-dependent equation (\ref{nonlineq}) by $\,\displaystyle\frac{d\Phi(\tau)}{d\tau}\,$ so that we get
\beq\displaystyle
 \frac{d}{d\tau}\left\lbrace\frac{1}{2}\left(\frac{d\Phi(\tau)}{d\tau}\right)^2-\frac{1}{4}\left(\Phi^2(\tau)-1\right)^2\right\rbrace =\left(\frac{d\Phi(\tau)}{d\tau}\right)^2\,.\label{nonlineq2}
\eeq
Then, the equation (\ref{nonlineq2}) can be interpreted as the equation of motion of a unit mass particle with non-conserved Hamiltonian \footnote{Here the dot means derivative w.r.t. the evolution parameter $\,\displaystyle\frac{d}{d\tau}\,$.},
\beq
\displaystyle{\H=\frac{1}{2}\dot{\Phi}^2+\V\quad\hbox{with}\quad\V=-\frac{1}{4}\left(\Phi^2-1\right)^2}\,.
\eeq
Here $\,\V\,$ represents an Higgs potential in which the particle evolves. Let us remark that the kinetic frictional force,
\beq
F_{+}=\dot{\Phi}\,,
\eeq
exerted on the particle has a positive friction coefficient and makes the energy to grow, since
\beq
\frac{dE}{d\tau}\geq 0\,.
\eeq 
The system receives energy from the exterior so that the Rayleigh function, $\,\R\,$, is negative and reads 
\beq
\R=-\int_{0}^{\dot{\Phi}} F_{+}d\dot{\Phi}=-\frac{1}{2}\,\dot{\Phi}^2\,.
\eeq
When positing the conjugate momentum as $\,\Psi=\dot{\Phi}\,$, we can construct the canonical phase-space $\,\big(\Phi,\,\Psi\big)\,$ in which we define an extension of the equations of the motion of our non-conservative system (\ref{nonlineq2}) as
\beq
\left\lbrace\begin{array}{ll}
\displaystyle\dot{\Phi}=\frac{\p\H}{\p\Psi}=\Psi\,,\\[10pt]
\displaystyle\dot{\Psi}=-\frac{\p\H}{\p\Phi}-\frac{\p\R}{\p\Psi}=\Psi+ \Phi\big(\Phi^2-1\big)\,.
\end{array}\right.\label{veqYM}
\eeq
Then, we can now describe the curve solutions of this Hamiltonian system (\ref{veqYM}) \cite{Protog,Breitenlohner:1993es}. To this end, we draw in the phase-plane $\,\big(\Phi,\,\Psi \big)\,$, the vector field $\,\big(\dot{\Phi},\,\dot{\Psi}\big)\,$, representing the velocity of each phase-point. Hence, the orbit solutions of the dynamical system lie on curves tangent to the velocity vector field. However, we restrict our investigation to finite orbits which are the only solutions physically consistent. We first search for critical points, which can be considered as orbits degenerated to a point, by solving the constraints
\beq
\left\lbrace\begin{array}{ll}
\Psi=0\,,\\[10pt]
\Psi+ \Phi\big(\Phi^2-1\big)=0\,.
\end{array}\right.
\eeq 
A simple algebra leads to the three critical points $\;\big(\Phi,\,\Psi\big)=\left\lbrace\big(0,\,0\big)\,;
\big(\pm1,\,0\big)\right\rbrace\,$, to which we characterize the equilibrium by analysing the eigenvalues and the eigenvectors of the stability matrix,
\beq
\Delta =-\left(\begin{array}{cc}\displaystyle
-\frac{\p^2 H}{\p\Phi\,\p\Psi}\quad&\quad\displaystyle-\frac{\p^2\H}{\p\Psi^2}\\[10pt]
\displaystyle\frac{\p^2\H}{\p\Phi^2}+\frac{\p^2\R}{\p\Phi\,d\Psi}\quad&\quad\displaystyle\frac{\p^2\H}{\p\Phi\,\p\Psi}+\frac{\p^2\R}{\p\Psi^2}
\end{array}\right)=\left(\begin{array}{cc}
0\quad&\quad1\\[10pt]
3\Phi^2-1\quad&\quad1
\end{array}\right).
\eeq
$\bullet$ For the fixed point $\,\big(\Phi,\,\Psi\big)= \big(0,\,0 \big)\,$, the eigenvalues of the stability matrix read, $$\,\displaystyle
\lambda_{\pm}={1}/{2}\pm{i\sqrt{3}}/{2}\,.$$
As the complex conjugated eigenvalues $\,\lambda_{\pm}\,$ have both a positive real part, then the orbits are spiraling out with respect to the focus $\big(0,\,0 \big)\,$.
Hence, the critical point $\,\big(0,\,0 \big)\,$ is unstable and is considered to be a negative attractor.\vspace*{3mm} 

\noindent
$\bullet$ For the fixed points $\,\big(\Phi,\,\Psi\big) =\big(\pm1,\,0\big)\,$, the eigenvalues of the stability matrix $\,\Delta\,$ read $$\lambda_{+}=2\quad\hbox{and}\quad\lambda_{-}=-1\,.$$ The fixed points are saddle points with the stable direction given by $\,\lambda_{-}\,$ and the unstable direction given by $\,\lambda_{+}\,$.\vspace*{3mm} 

\noindent
$\bullet$ We also consider the two bounded solutions, noted $\,\Phi_{\mp}\,$, represented by the curves joining the negative attractor $(0,\,0)$ to the saddle points $(\mp1,\,0)$. The corresponding curves solution are represented in the phase portrait below by $\,\Phi_{-}\,$ in pink line and $\,\Phi_{+}\,$ in cyan.
\begin{figure}[!h]
\begin{center}
\includegraphics[scale=0.7]{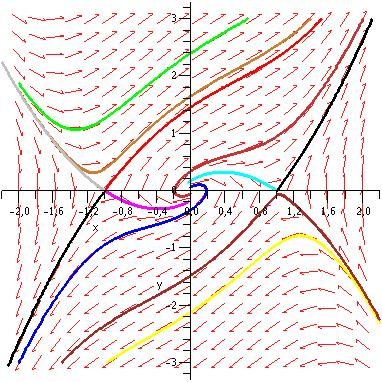}
\caption[]{Phase portrait of the Hamiltonian system (\ref{veqYM}).}\label{PPhase1}
\end{center}
\end{figure}

Let us now discuss the bounded solutions of the non-linear differential equation (\ref{nonlineq}). For $\,r\in\IR^+\,$, we have the five orbits solution,
\beq
\Phi(r)=\left\lbrace-1,\,1,\,0,\,\Phi_{-},\,\Phi_{+}\right\rbrace,
\eeq
that we introduce in the Wu-Yang forms (\ref{ansatz}) and (\ref{ReducField}) to obtain the shape of the gauge field.

\noindent
$\bullet$ The degenerate orbit solution $\,\Phi=-1\,$ leads to the pure gauge field
\beq
A^a_{i}=2g\varepsilon_{iaj}\,\frac{x^{j}}{r^2}\quad\hbox{since}\quad F_{0i}^a=F_{jk}^a=0\,.
\eeq
$\bullet$ The choice $\,\Phi=1\,$ corresponds to the null gauge potential,
\beq
A^a_{i}=0\quad\hbox{with}\quad F_{0i}^a=F_{jk}^a=0\,.
\eeq
It is worth noting that the two previous gauge potentials, with both vanishing field strength, can be transformed the one by the other using a suitable gauge transformation. They are therefore gauge equivalent.

\noindent
$\bullet$ We consider the two curves solution $\,\Phi_{-}\,$ and $\,\Phi_{+}\,$, admitting the asymptotic limits,
\beq
\Phi_{\pm}=\left\lbrace\begin{array}{ll}\Phi_{\pm}^{\infty}\quad\hbox{when}\quad r\gg1\\[8pt]
\Phi_{\pm}^{0}\quad\hbox{when}\quad r\ll1\end{array}\right.
\qquad\hbox{with}\qquad
\left\lbrace\begin{array}{ll}
\displaystyle\lim_{r\rightarrow\infty}\Phi_{\pm}^{\infty}=\pm1\\[8pt]
\displaystyle\lim_{r\rightarrow0}\Phi_{\pm}^0=0\,.\end{array}\right.
\eeq
Taking into account the behavior of the curves solution in the neighborhood of the origin, we can neglect the cubic term of the non-linear equation (\ref{nonlineq}). Thus, we deduce that $\,\Phi_{\pm}^0\,$ satisfy the differential equation,
\beq
r^2\frac{d^2\Phi_{\pm}^0(r)}{dr^2}+\Phi_{\pm}^0(r)=0\,,
\eeq
which provides us with the non-analytic solution,
\beq
\displaystyle
\Phi_{\pm}^0 =\pm\a\sqrt{r}\cos\left(\frac{\sqrt{3}}{2}\ln\frac{r}{r_0} \right)\,,\quad\hbox{with}\quad\big(\a,\,r_0\ll1\big) \,\in\,\IR^{\star}\times\IR^{+\star}\,.\label{FirstZBound}
\eeq
In the case where $\,r\gg1\,$, the equation (\ref{nonlineq}) reduces to the simple form
\beq
\frac{d^2\Phi_{\pm}^{\infty}(r)}{dr^2}=\frac{\const}{r^3}+O\left(\frac{1}{r^4}\right)\,,
\eeq

so, we derive the behavior of the radial functions $\,\Phi_{\pm}\,$ at infinity as
\beq
\displaystyle
\Phi_{\pm}^{\infty}(r)=\pm 1\,\mp\frac{\gamma}{r}+O\left(\frac{1}{r^2}\right)\,,\quad\hbox{where}\quad\gamma > 0\,.
\eeq
After the complete description of the asymptotic behavior of the functions $\,\Phi_{\pm}\,$, our business now is to fill the gap between these two limit cases. Here we investigate the ``solution'' $\,K_{+}\,$ but the case $\,K_{-}\,$ is not more complicated. Thus, we integrate numerically the non-linear equation (\ref{nonlineq}) from a point $\,r_0\,$, located in the neighborhood of the origin so that $\,\Phi_{+}\,$ is approximated by $\,\Phi_{+}^{0}\,$, till a sufficiently great value of $\,r\,$ so that$\,\Phi_{+}\,$ can be approximated by $\,\Phi_{+}^{\infty}\,$. Following the usual procedure, let us begin the numerical integration by adding some analytic correction terms, $\,a_i\,$, into the early expression (\ref{FirstZBound}) of $\,\Phi_{+}^{0}\,$. Thus, we express the numerical lower bound as
\beq
\tilde{\Phi}_{+}^0(r)= \pm\a\sqrt{r}\cos\left(\frac{\sqrt{3}}{2}\ln\frac{r}{r_0} \right)+\sum_{i=0}^4\,a_i\big(\a,\,r_0\big)\,r^i\,,\label{solraccordement}
\eeq
where the coefficients $\,a_i\,$ depend on the values of $\,r_0\ll1\,$ and the fixed parameter $\,\a\,$. For the fixed values of the integration parameters,
\beq
r_0=0.007873997658\,,\qquad
\a=-0.8873554901\,,
\eeq
and with the initial conditions of the numerical integration given by
$$\,
\left(
\tilde{\Phi}_{\pm}^0(r_0),\;
\displaystyle\frac{d}{dr}\tilde{\Phi}_{\pm}^0(r_0)
\right)\,,
$$ we obtain, for $\,r\in\left[r_0,14\right]\,$, the curve solution $\,\Phi_{+}(r)\,$ of the non-linear equation (\ref{nonlineq}). For $e\!=\!g\!=\!1$, we draw the field strength intensity of the usual $SU(2)$ Dirac monopole (in circling dashes) together with the intensity of the field strength solution (\ref{ReducField}), namely $
\displaystyle B_{\pm}(r)=\frac{g}{r^2}\big(1-\Phi_{+}^2\big)\,$, carrying the branch $\,\Phi_{+}\,$ (in heavy line).
\begin{figure}[!h]
\begin{center}
\includegraphics[scale=0.55]{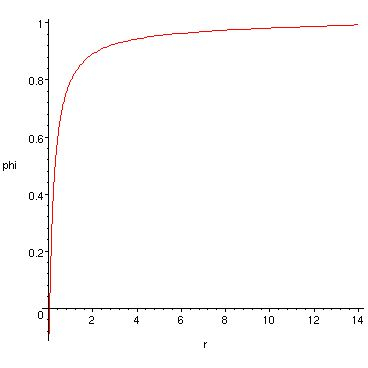}\qquad\qquad\qquad
\includegraphics[scale=0.55]{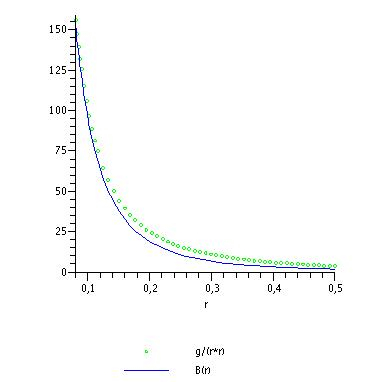}
\caption[]{In the top side we plot the curve solution $\,\Phi_{+}\,$; and in the bottom side we compare the field strength intensities of $B_+$ with the imbedded Dirac monopole.}\label{Fig5}
\end{center}
\end{figure}

It is worth mentioning that, in the asymptotic limits, the two curves coincide. The latter is due to the fact that when the radius tends to zero, $\,B_{\pm}(r)\,$ corresponds to the length of a Dirac monopole field; and when the radius tends to infinity, we get the null length of the vacuum. In the case of intermediary values of the radius, the two curves become quite different. This is a consequence of the radial function $\,\Phi_{+}\,$ which takes non-quantized but continuous values between zero and one. (See Figure \ref{Fig5}).

$\bullet$ Let us now investigate the last bounded solution, $\,\Phi_{\pm}=0\,$.
The comparison of the intensities of field strength made above provides us with strong assumption on the nature of the solution when $\,\Phi_{\pm}=0\,$. Let us analyze further by injecting the solution $\,\Phi(r)=0\,$ into the Wu-Yang ansatz (\ref{ansatz}). In that event, we obtain the gauge field
\beq
\displaystyle
A^{a\,WY}_{i}=g\varepsilon_{iaj}\frac{x^{j}}{r^2}\,,\label{WYMonop}
\eeq
which implies that the field strength (\ref{ReducField}) reduces to that of a Wu-Yang monopole,
\beq
\displaystyle
F_{ij}^{a\,WY}=g\varepsilon_{ijk}\frac{x^kx^a}{r^4}\,.\label{WYMonop1}
\eeq
The energy density,
\beq
\displaystyle
\E(r)=\frac{1}{4}\,F_{i j}^{a\,WY}\,F^{a\, i j\,WY}=\frac{g^2}{2r^4}\,,
\eeq
is singular at the origin $\,r=0\,$, so the Wu-Yang monopole possesses an infinite magnetic field energy.

Moreover we can prove, as suggested by the figure in the bottom of (\ref{Fig5}), that the non-Abelian Wu-Yang monopole can be viewed as an imbedded Dirac monopole. Indeed, we consider the trivial imbedding of Dirac monopole field $\,A^{D\,U(1)}_{\mu}\,$ into $\,SU(2)\,$,
\beq
A_{\mu}^{D\,U(1)}\longrightarrow A_{\mu}^{D\,SU(2)}=A_{\mu}^{a\,D}\,\tau_a\,,\quad \big(\tau_{a}=\sigma_{a}/2\big)\,,
\eeq
where the Abelian gauge potential reads
\beq
A_{\mu}^{a\,D}=\left\lbrace\begin{array}{ll}
0\quad\hbox{for}\quad a=1,2\,,\\[8pt]
A_{\mu}^{3\,D}=\pm\,g\left(1\mp\cos\theta\right)\p_{\mu}\phi\,.
\end{array}\right.
\eeq
Thus, a short algebra yields the imbedded gauge potential,
\beq
A_{\mu}^{D\,SU(2)}=\left\lbrace
\begin{array}{ll}
A_{0}^{D\,SU(2)}=0\quad\hbox{(temporal gauge)}\,,\\[8pt]
\displaystyle
A_{i}^{D\,SU(2)}=\pm\,g\,\frac{\left(1\mp\cos\theta\right)}{r\sin\theta}\,\left(-\sin\phi,\,\cos\phi,\,0\right)\,\tau_3\,.
\end{array} 
\right.
\eeq
In order to avoid the Dirac string singularity, we apply the singular ``hedgehog'' gauge transformation,
\beq
\displaystyle
A_{\mu}^{D\,SU(2)}\longrightarrow U\left(A_{\mu}^{D\,SU(2)}+\frac{i}{e}\p_{\mu} \right)U^{-1}=A_{\mu}^{WY}\,,\label{GaugeTransf}
\eeq
which rotates the unit vector on the sphere $\,S^2\,$ to the third axis in isospace. Thus, $\,U\,$ is charaterized by the unitary matrix,
\beq
U\big(\theta,\phi\big)= \left(\begin{array}{cc}\displaystyle\cos\frac{\theta}{2} &\displaystyle-\sin\frac{\theta}{2}\,\exp\left(-in\phi\right)\\[8pt]
\displaystyle\sin\frac{\theta}{2}\exp\left(in\phi\right) &\displaystyle\cos\frac{\theta}{2}\end{array}\right)\,,\quad n\in\IN^{\star}.\label{UMatrix}
\eeq
Applying (\ref{GaugeTransf}) with (\ref{UMatrix}), it is straightforward to derive the gauge equivalent potential,
\beq
\left\lbrace\begin{array}{llll}
A_0^{a\,WY}=0\qquad \hbox{(temporal gauge)}\,,\\[8pt]\displaystyle
A_i^{1\,WY}=\frac{n}{e}\cos\theta\sin\theta\cos\big(n\phi\big)\p_i\phi+\frac{1}{e}\sin\big(n\phi\big)\p_i\theta\,,\\[8pt]\displaystyle
A_i^{2\,WY}=\frac{n}{e}\cos\theta\sin\theta\sin\big(n\phi\big)\p_i\phi-\frac{1}{e}\cos\big(n\phi\big)\p_i\theta\,,\\[8pt]\displaystyle
A_i^{3\,WY}=-\frac{n}{e}\sin^2\theta\,\p_i\phi\,.
\end{array}\right.\label{PotSpheric}
\eeq
When taking into account the Dirac quantization relation \footnote{See the formula (\ref{QRelation}) in section \ref{DiracMonop}.}, $\,eg=n=1\,$, and the following algebraic relations,
\beq
\displaystyle
\p_{i}\theta=\frac{1}{r^2}\left(z\cos\phi,\,z\sin\phi,\,-\frac{x}{\cos\phi}\right)\quad\hbox{and}\quad
\p_{i}\phi=\frac{1}{r^2\sin^2\theta}\left(-y,\,x,\,0\right)\,,
\eeq
the gauge potential (\ref{PotSpheric}) transforms into the cartesian form as
\beq
\left\lbrace
\begin{array}{llll}
A_0^{a\,WY}=0\qquad\hbox{(temporal gauge)}\,,\\[8pt]\displaystyle
A_i^{1\,WY}=\frac{g}{r^2}\big(0,\,z,\,-y\big)\,,\\[8pt]\displaystyle
A_i^{2\,WY}=\frac{g}{r^2}\big(-z,\,0,\,x\big)\,,\\[8pt]\displaystyle
A_i^{3\,WY}=\frac{g}{r^2}\big(y,\,-x,\,0\big)\,.
\end{array}\right.\label{ToBeCompact}
\eeq
By compacting (\ref{ToBeCompact}), we hence recover the exact expression [see (\ref{WYMonop}) and (\ref{WYMonop1})] of the non-Abelian Wu-Yang gauge potential with a ``hedgehog'' magnetic field.
It is now clear that the non-Abelian Wu-Yang monopole field can be obtained by imbedding the Abelian Dirac monopole field into an $\,SU(2)\,$ gauge theory.

Let us now inquire about conserved quantities. To do this, we first Identify the  $\su(2)$ Lie algebra of the non-Abelian generator, $\,\tau_{a}\,$, with ${\IR}^3$. Thus, we make the replacement,
\beq
\tau_{a}\longrightarrow\I_{a}\,,\;\;\hbox{with the internal index}\;\; a=1,2,3\,.
\eeq
Here the non-Abelian variable $\,\I_{a}\,$ represents the isospin vector which satisfies the Poisson-bracket algebra,
\beq
 \{\I^{a},\I^{b}\}=-\epsilon^{abc}\I_{c}\,.
\eeq
Hence, we consider an isospin-carrying particle \cite{Bal,Duval:1978,Duval:1980,DHInt} moving in a Wu-Yang monopole field \cite{Schechter,Boulware:1976tv,Stern:1977,Wipf:1986}, augmented by a scalar potential \cite{Schonfeld,Feher:1984ik}, described by the gauge covariant Hamiltonian,
\beq
\H=\frac{\vpi^{2}}{2}+V(\vx,\,\I^{a})\,,\quad\vpi=\vp-e\I^{a}\vA^{a\;WY}\,.\label{HWY}
\eeq 
We define the covariant Poisson-brackets as 
\beq
\big\{f,g\big\}=D_jf\frac{\p g}{\p \pi_j}-\frac{\p f}{\p \pi_j}D_jg 
+e\,\I^aF_{jk}^{a\;WY}\frac{\p f}{\p \pi_j}\frac{\p g}{\p \pi_k}
-\epsilon_{abc}\frac{\p f}{\p \I^a}\frac{\p g}{\p \I^b}\I^c\,,
\label{PBracketbis}
\eeq
where $D_j$ is the covariant derivative,
\beq
D_jf=\p_jf-e\epsilon_{abc}\I^aA_j^{b\;WY}\frac{\p f}{\p\I^c}.
\label{covder}
\eeq
Thus, the commutator of the covariant derivatives is recorded as
\begin{equation}
[D_i,D_j]=-\epsilon_{abc}\I^aF_{ij}^{b\;WY}\frac{\p}{\p\I^c}\ . 
\label{covdercomm}
\end{equation}
Following van Holten's recipe \cite{vH}, conserved quantities $\Q\big(\vx,\vec{\I},\vpi\big)$ can conveniently be sought for in the form of an expansion into powers of the covariant momentum,
\beq
\Q\big(\vx,\vec{\I},\vpi\big)= C(\vx,\vec{\I})+C_i(\vx,\vec{\I})\pi_i+\frac{1}{2!}C_{ij}(\vx,\vec{\I})\pi_i\pi_j+
\cdots\label{constexp1}
\eeq
Requiring $\Q$ to Poisson-commute with the Hamiltonian,
$\{\Q,\H\}=0\,$,
provides us with the set of constraints to be satisfied, 
\beq\begin{array}{llll}
\displaystyle{C_iD_iV+\epsilon_{abc}\I^a\frac{\p C}{\p\I^b}\frac{\p V}{\p\I^c}}=0,& o(0)
\\[8pt]
\displaystyle{D_iC=e\I^aF^{a\;WY}_{ij}C_j+C_{ij}D_jV+\epsilon_{abc}\I^a\frac{\p C_i}{\p\I^b}\frac{\p V}{\p\I^c}},&
o(1)
\\[8pt]
\displaystyle
D_iC_j+D_jC_i=e\I^a(F^{a\;WY}_{ik}C_{kj}+F^{a\;WY}_{jk}C_{ki})+C_{ijk}D_kV+\epsilon_{abc}\I^a\frac{\p C_{ij}}{\p\I^b}\frac{\p V}{\p\I^c},& o(2)
\\[8pt]
\displaystyle
D_iC_{jk}+D_jC_{ki}+D_kC_{ij}=e\I^a(F^{a\;WY}_{il}C_{ljk}+F^{a\;WY}_{jl}C_{lki}+F^{a\;WY}_{kl}C_{lij})\\[8pt]\displaystyle{\qquad\qquad\qquad\qquad\qquad\qquad+C_{ijkl}D_lV+\epsilon_{abc}\I^a\frac{\p C_{ijk}}{\p\I^b}\frac{\p V}{\p\I^c}},& o(3)
\\
\vdots\qquad\qquad\qquad\qquad\qquad\vdots&\vdots
\end{array}\label{ConstraintsWY}
\eeq
To start, we search for zeroth-order conserved quantity. Thus $C_i=C_{ij}=\dots=0\,$, so that the series of constraints (\ref{ConstraintsWY}) reduces to
\beqa
\left\lbrace
\begin{array}{ll} 
\displaystyle\epsilon_{abc}\I^a\frac{\p C}{\p\I^b}\frac{\p V}{\p\I^c}=0 &\hbox{o(0)} 
\\[8pt]
D_iC=0\,. &\hbox{o(1)}
\end{array}\label{CC}
\right.
\eeqa
The zeroth-order equation of (\ref{CC}) is identically satisfied for rotationally invariant potentials with respect to $\,\vx\,$ and $\,\vec{\I}\,$. Applying for the consistency condition,
\beq
\left[D_i,D_j\right]C=-\epsilon_{abc}\I^aF_{ij}^{b\;WY}\frac{\p C}{\p\I^c}=0\,,
\eeq
we get the shape of the derivative of $\,C\,$ along the isospin variable,
\beq
\frac{\p C}{\p\I^c}=f\big(r,\I\big)x_{c}+h\big(r,\I\big)\I_{c}\,.\label{ShapeC}
\eeq
Injecting (\ref{ShapeC}) into the first-order constraint of (\ref{CC}) implies, for an arbitrary function $\,h\big(r,\I\big)\,$, that
\beq
\p_{i}C=f\big(r,\I\big)\left(\I_{i}-\frac{\vx\cdot\vec{\I}}{r^2}x_{i}\right)\,,\quad[\hbox{with}\;\;eg=1]\,.\label{Derf}
\eeq 
From the commutation rule, 
\beq
\displaystyle
\left[\,\partial_i\,,\,\partial_j\, \right]C=0\quad\Longrightarrow\quad
\frac{1}{r}\left(\frac{df}{dr}+\frac{f}{r}\right)=0\,,
\eeq
we derive the exact form of the function $\,f(r,\I)\,$,
\beq 
f\big(r,\I\big)=\frac{\b}{r}\,,\quad\b=\const\in\IR\,.\label{Funcf}
\eeq
Taking into account the result (\ref{Funcf}), the equations (\ref{Derf}) and (\ref{ShapeC}) lead to the system of equtions,
\beq
\left\lbrace
\begin{array}{ll}\displaystyle
\p_{i}C=\b\left(\frac{\I_{i}}{r}-\frac{\vx\cdot\vec{\I}}{r^3}x_{i}\right)\,,\\[10pt] \displaystyle
\frac{\p C}{\p\I^a}=\b\frac{x_{a}}{r}+h\big(r,\I\big)\I_{a}\,,
\end{array}
\right.
\eeq
which is solved uniquely by the covariantly constant charge,
\beq
\Q(\vx,\vec{\I}\big)=\b\,\frac{\vx\cdot\vec{\I}}{r}+\gamma\,\Q\big(\I\big)\,,\quad \gamma=\const\in\IR\,.\label{ConsCharge}
\eeq
This charge can be viewed as a linear combination of two quantities separately conserved along the particle's motion since $\,\b\,$ and $\,\gamma\,$ are arbitrary real numbers. Note that the first term on the right-hand side of (\ref{ConsCharge}), namely,
\beq
\displaystyle
\Q_{0}=\frac{\vx\cdot\vec{\I}}{r}\,,
\eeq
can be seen as a conserved electric charge; and its conservation admits a nice interpretation in terms of fiber bundles \cite{HRaw1,HRaw2}. The $\,SU(2)\,$ gauge field is a connection form defined on a  bundle over the $3$-dimensional space so that for Wu-Yang monopole, the $\,\su(2)\,$ connection living on the (trivial) bundle reduces to the $U(1)$ Dirac monopole bundle. This is the reason why the electric charge is conserved in the Wu-Yang case: the latter is, as already seen, an imbedded Abelian Dirac monopole.

Next, we study conserved quantities which are linear in $\pi_i$, $C_{ij}=C_{ijk}=\dots=0$. We therefore have to solve the constraints,
\beqa
\left\lbrace
\begin{array}{lll} 
\displaystyle
C_iD_iV+\epsilon_{abc}\I^a\frac{\p C}{\p\I^b}\frac{\p V}{\p\I^c}=0\,,& o(0)\\[8pt]
\displaystyle
D_iC=e\I^aF^{a\;WY}_{ij}C_j+\epsilon_{abc}\I^a\frac{\p C_i}{\p\I^b}\frac{\p V}{\p\I^c}\,,&
o(1)\\[8pt]
D_iC_j+D_jC_i=0\,.&
o(2)
\end{array}\label{C2}
\right.
\eeqa
 When the potential is invariant with respect to joint rotation of $\vx$ and $\vec{\I}$, inserting the Killing vector generating the spatial rotations,
\beq
\vC=\vn\times\vx\,,
\eeq
into the series of constraints (\ref{C2}) yields
\beq
C=-\Q_{0}\frac{\vn\cdot\vx}{r}\,.
\eeq
Collecting the two previous results provides us with the conserved angular momentum,
\beq
\vJ=\vx\times\vpi-\Q_{0}\hx\,.\label{WYAngMom}
\eeq
Let us now turn to quadratic conserved quantities, $C_{ijk}=C_{ijkl}=\dots=0$. The set of constraints (\ref{ConstraintsWY}) reduces to
\beqa
\left\lbrace
\begin{array}{llll}
\displaystyle{C_iD_iV+\epsilon_{abc}\I^a\frac{\p C}{\p\I^b}\frac{\p V}{\p\I^c}}=0,& o(0)
\\[8pt]
\displaystyle{D_iC=e\I^aF^{a\;WY}_{ij}C_j+C_{ij}D_jV+\epsilon_{abc}\I^a\frac{\p C_i}{\p\I^b}\frac{\p V}{\p\I^c}},&
o(1)
\\[8pt]
\displaystyle
D_iC_j+D_jC_i=e\I^a(F^{a\;WY}_{ik}C_{kj}+F^{a\;WY}_{jk}C_{ki})+\epsilon_{abc}\I^a\frac{\p C_{ij}}{\p\I^b}\frac{\p V}{\p\I^c},& o(2)
\\[8pt]
\displaystyle
D_iC_{jk}+D_jC_{ki}+D_kC_{ij}=0\,.& o(3)
\end{array}\label{CWY}\right.
\eeqa
We observe that the rank-$2$ Killing tensor generating the Kepler-type dynamical symmetry has the property,
\beq
C_{ij}=2\delta_{ij}\, \vn\cdot\vx-(n_ix_j+n_jx_i)\,.\label{RLKilling}
\eeq
Inserting (\ref{RLKilling}) into (\ref{CWY}), from the 2nd-order equation we find, therefore,
\beq
\vC=\Q_{0}\frac{\vec{n}\times{\vx}}{r}\,. 
\label{WYvC}
\eeq
The first-order equation requires in turn
\beq
D_iC = \left\lbrace\Q_{0}^2\,\big(\vn\cdot\vx\big)\frac{x_{i}}{r^4}-\Q_{0}^2\,\frac{n_{i}}{r^2}\right\rbrace+2\,\vn\cdot\vx\,D_{i}V-n_{i}\,\vx\cdot\vD V-x_{i}\,\vn\cdot\vD V\,.
\label{WYCeq}
\eeq
Restricting ourselves to potentials falling off at infinity and invariant by rotations with respect to $\,\vx\,$ and $\,\vec{\I}\,$, we hence obtain
\beq
V=\frac{\Q_{0}^2}{2r^2}+\frac{\a}{r}+\b
\quad\hbox{and}\quad
C=\a\,\frac{\vec{n}\cdot{\vx}}{r},
\label{goodpot}
\eeq
where $\a$ and $\b$ are arbitrary constants. Note that the inverse-square term in the previous potential is fixed by the requirement of canceling the bracketed term on the right-hand side of (\ref{WYCeq}). Collecting our results yields,
\beq
\vK=
\vpi\times{\vJ}+\a{\hx}\,,
\label{WYRL}
\eeq
which is indeed a conserved  Runge-Lenz vector for an isospin-carrying particle in the Wu-Yang monopole field, combined with the potential (\ref{goodpot}) \cite{HWY}. Let us emphasize that the fine-tuned inverse-square term is necessary to overcome the obstruction in solving the constraint equation \cite{H-NGI}; without it, no Runge-Lenz vector would exist. The expression (\ref{WYRL}) is actually not surprising since the Wu-Yang monopole field, as we proved it, corresponds to an imbedded Dirac monopole field.

The importance of the conserved quantities $\vJ$ and $\vK$ is understood by noting that they determine the trajectory~: multiplying the conserved angular momentum by the position, $\vx$, yields
\begin{equation}
\vJ\cdot{\hx}=-\Q_{0},
\label{cone}
\end{equation}
so that the particle moves, as always in the presence of a monopole,  on the surface of a cone of half opening angle,
\beq
\varphi={\rm arccos}(|\Q_{0}|/J)\quad\hbox{where}\quad J=|\vJ|\,.
\eeq
On the other hand, the projection of the position onto the vector
$\vN$, given by,
\begin{equation}
\vN=\vK+(\a/\Q_{0})\vJ,\qquad
\vN\cdot\vx=J^2-\Q_{0}^2=\const,
\label{Nvector}
\end{equation}
implying that the trajectory lies on a plane perpendicular to $\vN$. The motion is, therefore, a \emph{conic section}. Careful analysis would show that the trajectory is an ellipse, a parabola, or a hyperbola, depending on the energy, being smaller, equal or larger than $\b$ \cite{Feher:1984ik,Feher:1984xc,Feher:1986,Feher:1987,CFH3}. In particular, for sufficiently low energies, the motion remains bounded.

The conserved vectors $\vJ$ and $\vK$ satisfy, furthermore,
the commutation relations
\beq
\big\{J_i,J_j\big\}=\epsilon_{ijk}J_k\,,\quad
\big\{J_i,K_j\big\}=\epsilon_{ijk}K_k\,,\quad
\big\{K_i,K_j\big\}=2(\b-\H)\epsilon_{ijk}J_k\,,
\label{JKcommrel}
\eeq
with the following Casimir relations,
\begin{equation}
\vJ\cdot\vK=-\a\,\Q_{0},
\qquad
K^2=2(\H-\b)(J^2-\Q_{0}^2)+\a^2. 
\label{casimirs}
\end{equation}
Normalizing $\,\vK\,$ by $\,\left[2(\b-\H)\right]^{1/2}\,$, we get, therefore an $\,\SO(3)/ {\rm E(3)}/ \SO(3,1)\,$ dynamical symmetry, depending on the energy being smaller/equal/larger than $\,\b\,$ \footnote{The dynamical symmetry actually extends to an isospin-dependent representation of $\,\SO(4,2)\,$
\cite{HWY}.}.

We remark that although our investigations have been purely classical,
there would be no difficulty to extend them to a quantum particle.
In the self-dual Wu-Yang case, the $\,\SO(4)/\SO(3,1)\,$ dynamical
symmetry allows, in particular, to derive the bound-state spectrum 
 and 
the Scattering-matrix group theoretically, using the algebraic relations (\ref{JKcommrel}) and (\ref{casimirs}) \cite{Feher:1984ik,CFH1,FHsusy,FHO}.

\newpage

\subsection{The Berry phase - general theory}\label{BerryPhase1}

The Berry phase arises for systems which can be conveniently described in terms of two sets of degrees of freedom \cite{Berry}. The one is ``\textit{fast}'' moving with large differences between excitation levels, and the other is ``\textit{slow}'' with small associated energy differences. This decomposition is extensively used in molecular physics through the adiabatic or Born-Oppenheimer approximation. In a molecule, for instance, the electronic motion is described by the ``\textit{fast}'' variables $\,\vr\,$, and the nuclear motion by the ``\textit{slow}'' degrees of freedom $\,\vR\,$.

We first deal with the ``\textit{fast}'' degrees of freedom, keeping the ``\textit{slow}'' as approximately fixed. In that event, we simply solve the stationary Schr\"odinger equation for the ``\textit{fast}'' variables, with the ``\textit{slow}'' variables appearing parametrically,
\beq
\H(\vR)\Psi_{m}(\vr,\vR)=\E_{m}(\vR)\Psi_{m}(\vr,\vR)\,.\nn
\eeq
Next, we complete the analysis by allowing slow variations in time for the previously fixed variables. The (adiabatic) assumption is that the slowly varying degrees of freedom $\,\vR\,$ do not change quickly enough to induce transitions from one $\,\E_{n}\,$ level to another. Thus, the system starting in an initial eigenstate remains in this state in response to the slow change of the variables $\,\vR\,$ appearing parametrically.

As a consequence of this effective dynamics, an external vector potential called the \underline{Berry connection} is induced. It has been argued that the ``feed-back'' coming from the Berry phase modifies the (semi)classical dynamics \cite{Niu,Xiao}. The associated magnetic-like field is the \underline{Berry curvature}, and the line integral of the connection is the ``celebrated'' \underline{Berry phase} \cite{Berry,Simon}. See also \cite{Aitchison}.

For better understanding, let us study deeper the way to obtain the Berry gauge potentials. To this end, we separate the following in Abelian and non-Abelian cases.  

\hfil\break\underbar
{The Abelian gauge potential : non-degenerate states}\hfil\break

Following Berry's original paper \cite{Berry}, let us point that an $\,U(1)\,$ gauge field may appear when a single non-degenerate level is subject  to adiabatically varying external parameter. To this end, we consider a physical system described by a Hamiltonian which depends on time through the vector $\,\vR(t)\,$, 
\beq
\H=\H(\vR)\,,\quad\vR=\vR(t)\,.
\eeq
Here $\,\vR(t)$ denotes a set of $\,m\,$ classical parameters,
\beq
\vR(t)=\big(R_{1}(t),R_{2}(t),\cdots\,,R_{m}(t)\big)\,,\nn
\eeq 
slowly varying along a closed path $\,\C\,$ in the parameter space, since the system is assumed to evolve adiabatically. We introduce an instantaneous orthonormal basis constructed with the eigenstates of $\,\H(\vR(t))\,$ at each value of the parameter $\,\vR\,$. The  eigenvalue equation reads
\beq
\H(\vR(t))\vert m\,,\vR(t)\rangle=\E_{m}(\vR(t))\vert m\,,\vR(t)\rangle\,.\label{EigenF}
\eeq
The basis eigenfunctions $\,\vert m\,,\vR(t)\rangle\,$ are not completely determined by (\ref{EigenF}). Indeed, without loss of generality, while normalizing the eigenfunctions,
\beq
\langle m\,,\vR(t)\vert m\,,\vR(t)\rangle=1\,,\nn
\eeq
this implies that eigenfunctions are unique up to multiplication by a phase factor.
Moreover, for a slowly varying Hamiltonian, the quantum adiabatic theorem states that a system initially prepared to be in one of its eigenstate $\,\vert m\,,\vR(0)\rangle\,$, at $\,t=0\,$, remains in this instantaneous eigenstate along the cyclic process $\C$. Consequently, the quantum state at time $t\,$ can be written as,
\beq
\vert \Psi_{m}(t)\rangle=\exp\big(ia_{m}(t)\big)\vert m\,,\vR(t)\rangle\,,\label{QEigenState}
\eeq
where the exponential term in (\ref{QEigenState}) is the only degree of freedom we can have in the quantum state. Substituting the expression (\ref{QEigenState}) into the Schr\"odinger equation,
\beq
i\hbar\frac{\p}{\p t}\vert\Psi_{m}(t)\rangle=\H(\vR(t))\vert\Psi_{m}(t)\rangle\,,
\eeq
and projecting both sides of the equation onto $\,\langle n\,,\vR(t)\vert\,$, yields the equation for the phase,
\beq
a_{n}=i\oint_{\C}\langle m\,,\vR\,\vert\,\vnabla_{\vR}\vert\,m\,,\vR\rangle\cdot d\vR-\frac{1}{\hbar}\int_{0}^{T}\E_{n}(\vR(t'))dt'\,.
\eeq
Thus, in addition to the usual dynamical phase,
\beq
-\frac{1}{\hbar}\int_{0}^{T}\E_{n}(\vR(t'))dt'\,,
\eeq
the quantum state acquires an additional geometric phase during the evolution through a closed path in the external parameter space \cite{Berry,Simon},
\beq
\gamma_{n}(\C)=i\oint_{\C}\langle n\,,\vR\,\vert\,\vnabla_{\vR}\,\vert\, n\,,\vR\rangle\cdot d\vR=\oint_{\C}\vA_n(\vR)\cdot d\vR\,.
\eeq
The path integral in the parameter space, $\,\gamma_{n}(\vR)\,$, represents the ``celebrated'' Berry phase and the integrand,
\beq
\vA_n(\vR)=i\langle n\,,\vR\,\vert\,\vnabla_{\vR}\,\vert\, n\,,\vR\rangle\,,
\eeq
is a vector-valued function called the \underline{Berry connection} or the Berry vector potential. 

Let us note that the Berry connection transforms as a gauge vector field. Indeed, the gauge transformation
\beq
\vert\,n\,,\vR\rangle\quad\longrightarrow\quad\exp\big(i\xi(\vR)\big)\vert\,n\,,\vR\rangle\,,
\eeq
with $\,\xi(\vR)\,$ being an arbitrary smooth function, acts on the Berry connection as
\beq
\vA_n(\vR)\quad\longrightarrow\quad\vA_n(\vR)-\vnabla_{\vR}\xi(\vR)\,.
\eeq
Consequently the Berry vector potential $\,\vA_n(\vR)\,$ transforms as an $\,U (1)\,$ gauge potential.

In analogy to the electrodynamics, the gauge field tensor derived from the Berry vector potential reads
 \beq
F_{ij}^{n}=\frac{\p}{\p R^{i}}A_{j}^{n}-\frac{\p}{\p R^{j}}A_{i}^{n}\,,
\eeq
and is known as the \underline{Berry curvature}. Using Stokes's theorem, we can express the Berry phase as an integral of the Berry curvature throughout the surface $\,\S\,$ enclosed by the path $\,\C\,$,
\beq
\gamma^{n}=\frac{1}{2}\int_{\S}dR^{i}\wedge dR^{j}\,F^{n}_{ij}\,.
\eeq
It is worth noting that the Berry curvature can be viewed as a $\,U(1)\,$ gauge-invariant magnetic field in the parameter space. It is therefore observable.

\hfil\break\underbar
{The non-Abelian gauge potential : degenerate states}\hfil\break

The Berry phase admits a non-Abelian generalization when the energy levels of the Hamiltonian are degenerate \cite{WZ}. 

Thus, we now consider a quantum system described by an Hamiltonian $\,\H(\vR(t))$ with $k$-fold degenerate ground states for all values of the external parameter $\,\vR(t)\,$. For simplicity, we fix $\,k=2\,$ such that the energy levels are two-fold degenerate and the Hamiltonian has two independent eigenstates, $\,\vert n^{a},\,\vR(t)\rangle\,,\;a=1,2\,$. The eigenvalue equation now  reads as
\beq
\H(\vR(t))\vert n^{a},\,\vR(t)\rangle=\E_{n}(\vR(t))\vert n^{a},\,\vR(t)\rangle\,,
\eeq
where, without loss of generality, the eigenstates are chosen such that
\beq
\langle n_{a},\,\vR(t)\,\vert\,n^{b},\,\vR(t)\rangle=\d^{b}_{a}\,,\quad a,b=1,2\,.
\eeq
Assuming that the system starts in one of its eigenstate $\,\vert n^{a}\,,\vR(0)\rangle\,$, the adiabatic approximation implies that the system stays in its initial instantaneous eigenstate, after a cyclic tour through the space of parameters. The eigenfunctions of the system are
\beq
\vert \Psi_{m}^{a}(t)\rangle=\vert m^{b}\,,\vR(t)\rangle\,U^{a}_{b}(\vR)\,,\label{QEigenStateNA}
\eeq
where $\,U^{a}_{b}(\vR)\,\in SU(2)\,$ is an unitary matrix. Let us now substitute the expression (\ref{QEigenStateNA}) into the time-dependent Schr\"odinger equation,
\beq
i\hbar\frac{\p}{\p t}\vert\Psi_{m}^{a}(t)\rangle=\H(\vR(t))\vert\Psi_{m}^{a}(t)\rangle\,,
\eeq
and multiply it from the left by $\,\langle n^{a}\,,\vR(t)\vert\,$, one finds
\beq
\frac{\p\vR}{\p t}\,\vA^{\,b}_{c}(\vR)\,U^{a}_{b}(\vR)+i\frac{\p U^{a}_{c}(\vR)}{\p t}-\frac{1}{\hbar}\E_{n}\,U^{a}_{c}(\vR)=0\,,\quad a,b,c=1,2\,,\label{EQNA}
\eeq
where we introduced the notation,
\beq
\vA^{\,b}_{c}(\vR)=i\langle n_{c}\,,\vR(t)\vert\vnabla_{\vR}\vert n^{b}\,,\vR(t)\rangle\,.\label{NAGf}
\eeq
with $\,a,b,c=1,2\,$ denoting matrix indices. Hence, the vector potential $\,\vA(\vR)\,$ is a $\,\big(2\times2\big)\,$ anti-Hermitian matrix lying in the $\,su(2)\,$ Lie algebra. Indeed, under the non-Abelian gauge transformation,
\beq
\vert m'^{a}\,,\vR(t)\rangle=\vert m^{b}\,,\vR(t)\rangle\,U^{a}_{b}(\vR)\,,
\eeq 
the field (\ref{NAGf}) transforms as
\beq
\vA^{\,b}_{c}(\vR)\quad\longrightarrow\quad U^{-1}\left(\vA^{\,b}_{c}(\vR)-\frac{\p}{\p \vR}\right)U\,,
\eeq
and therefore defines a $\,su(2)$ valued Berry vector potential.  The associated non-Abelian Berry curvature then reads as
\beq
F_{ij}^{a}=\frac{\p}{\p R^{i}}A_{j}^{a}-\frac{\p}{\p R^{j}}A_{i}^{a}+i\left[A_i^b,\,A_j^c\right]\,.
\eeq
Writting the solution of (\ref{EQNA}) in terms of the path-ordered integrals,
\beq
U=P\exp\left(\oint\vA^{\,b}_{c}(\vR)\cdot d\vR\right)\times\exp\left(-\frac{1}{\hbar}\int_{0}^{T}\!dt'\E_{n}(\vR(t'))\right)\,.
\eeq 
It is worth remarking that the system undergoes a $\,SU(2)\,$ rotation which depends on the path taken,
\beq
\gamma_{n}=P\exp\left(\oint\vA^{\,b}_{c}(\vR)\cdot d\vR\right)\,.
\eeq
The latter defines the non-Abelian generalization of the Berry phase factor known as the Wilson loop.

Compared to the Berry phase which is always associated with a closed path, the Berry curvature is truly a local quantity. It provides a local description of the geometric properties of the parameter space while the Berry phase can be identified with the holonomy of the fiber bundle \cite{Simon}.

Also, the Berry curvature also plays the role of a (non-Abelian) magnetic-like field, which affects the particle dynamics in his neighborhood. A relevant example is provided with the ``Berry'' non-Abelian monopole-like fields arising in diatomic molecule systems \cite{MSW}, see the next section.

\newpage

\subsection{Monopole-like fields in the diatomic molecule}\label{DiatomSection}

As first investigated by Moody, Shapere and Wilczek \cite{MSW}, a ``truly'' non-Abelian gauge fields mimicking monopole-like fields can arise in a diatomic molecule system. These 
come from the non-Abelian generalization \cite{WZ} of the Berry gauge potentials. In this case, we consider sets of levels, $\,k\,$-fold degenerate, subjected to adiabatically varying external parameters. For $\,k=1\,$, a single level, we recover the $\,U(1)\,$ gauge fields discussed by Berry and Simon \cite{Berry,Simon}. For $\,k\geq2\,$, the effective gauge fields take a ``truly'' non-Abelian form.

The latter can be extended to systems where the slow dynamical variables are no longer external but are themselves quantized. This is the case, in particular, for the diatomic molecule where the quantized parameters define nuclear coordinates \cite{MSW}.

To see this, let us consider a diatom with two atomic nuclei and one or more gravitating electrons. The study of this system amounts to investigating a many-body problem which reduces, in the simplest case, to a three-body problem. Neglecting the spin degree of freedom and the relativistic effects, the Hamiltonian employed in the diatomic molecule system reads [in units $\,\hbar=1\,$],
\beq
\begin{array}{lr}
\displaystyle
\H(\vX_i,\,\vx_k) =\left\lbrace-\frac{1}{2m_{k}}\sum_{k=1}^{n}\vnabla^2_{x_k}-\sum_{i=1}^2\frac{1}{2M_i}\vnabla^2_{X_i}\right\rbrace\\[10pt]
\displaystyle
\qquad\qquad\qquad
-\sum_{i=1}^2\,\sum_{k=1}^{n}\frac{Z_i\,e}{\vert \vX_i-\vx_k\vert}+\frac{Z_1Z_2}{\vert \vX_1-\vX_2\vert}+\sum_{j=1}^{n}\,\sum_{k>j}^{n}\frac{e^2}{\vert\vx_j-\vx_k\vert}\,.
\end{array}\label{DiatomHam}
\eeq
Here the atomic number $\,Z_j\,$ corresponds to the electric charge of the \textit{jth} nucleus and the positions $\,\vX_i\,$ and $\,\vx_k\,$ denote the nuclei and the electrons coordinates, respectively. The bracketed terms in the Hamiltonian are the kinetic energy of the electron of mass $\,m\,$ plus the kinetic energy of the nuclei of mass $\,M_{i}\,$, with $\,\vnabla_{x_i}\,$ and $\,\vnabla_{X_k}\,$ referring to the Laplacians of the \textit{i}th electron and of the \textit{k}th nucleus, respectively. The two following terms in (\ref{DiatomHam}) define the classical Coulomb electron-nuclei interaction and the nucleus-nucleus interaction, respectively. The remaining term represents the electron-electron interactions. 

From now on we consider the simple configuration of the molecular ion $\,H_2^+\,$ which possesses only one gravitating electron and two identical hydrogen nuclei. Then, the electric charge of the nuclei are the same, $\,Z_{1}=Z_{2}=Z\,$, and electron-electron interactions  vanish since only one electron is considered in the present context. The total non-relativistic wave function $\,\Psi(\vx,\,\vX_i)\,$ of this diatomic molecule system is a solution of the stationary Schr\"odinger equation, 
\beq
\H\,\Psi(\vx,\,\vX_i)=\E\,\Psi(\vx,\,\vX_i)\,.
\eeq

The description of the diatomic molecule properties is commonly made using the Born-Oppenheimer approximation. Indeed, the Born-Oppenheimer or adiabatic approximation is applied to separate, in an appropriate way, the electronic motion  and the slower nuclei's degrees of freedom that couple to it, since $\,M_i\gg m\,$. To investigate electronic motions, we first assume that the nuclei positions $\,\vX_{1}\,$ and $\,\vX_{2}\,$ are fixed and correspond to infinite nuclear masses $\,M_{1}=\,M_{2}=\infty\,$. Thus for a fixed nuclear configuration, we obtain the electronic Hamiltonian, $\,\H_{el}\,$, carrying a parametrical dependence on the nuclear relative coordinate $\,\vX_{12}=\vX_1-\vX_2\,$,
\beq\begin{array}{cc}
\displaystyle
\H_{el}\big(\vx,\vX_{12}\big)=-\frac{1}{2m}\vnabla^2_{\vx}+V\big(\vx,\vX_{12}\big)\,,\\[8pt]
\displaystyle
V\big(\vx,\vX_{12}\big)=-\frac{Ze}{\vert \vX_1-\vx\vert}-\frac{Ze}{\vert \vX_2-\vx\vert}+\frac{Z^2}{\vert\vX_{12}\vert}\,.
\end{array}\label{ElHam}
\eeq
Since $\,\vX_{12}\,$ is just a parameter, then the last term in the previous potential is a constant and shifts the eigenvalues only by some constant amount. In the context of the Born-Oppenheimer approximation, we consider also
\beq
\Psi(\vx,\,\vX_i)\approx\Psi(\vx,\,\vX_{12})\,,
\eeq
where the molecular wave function, $\,\Psi(\vx,\,\vX_{12})\,$, can be expanded into a combination of the electronic wave function $\,\varphi_{m}(\vx)\,$ and the nuclear wave function $\,\chi_{m}(\vX_{12})\,$,
\beq
\Psi(\vx,\,\vX_{12}) = \sum_m\varphi_{m}(\vx,\vX_{12})\,\chi_{m}(\vX_{12})\,.
\eeq
 Note that the electronic eigenfunction depends implicitly on the nuclear relative coordinate, $\,\vX_{12}\,$, and the summation index $\,m\,$ denote the eventual energy's degeneracy of the electronic eigenstate. Hence, the electronic eigenfunction obeys to the electronic stationary Schr\"odinger equation,
\beq
\H_{el}\,\varphi_m (\vx ) = \E_{el,m}\,\varphi_m(\vx)\,,\label{SchrodEl}
\eeq
and form a complete set.
While when investigating the nuclear motions, we have to consider the electron as remaining in the same quantum eigenstate so that the nuclear wave function is a solution of the Schr\"odinger equation with an effective potential generated by the electron,  
\beq
\left(-\frac{1}{2M}\sum_{i=1}^2\vnabla^2_{X_i}+\H_{el}+\frac{Z^2}{\vert\vX_{12}\vert}\right)\chi_k (\vX_{12}) = E\,\chi_k(\vX_{12})\,.\label{SchrodNuclear}
\eeq

\begin{figure}[!h]
\begin{center}
\includegraphics[scale=0.7]{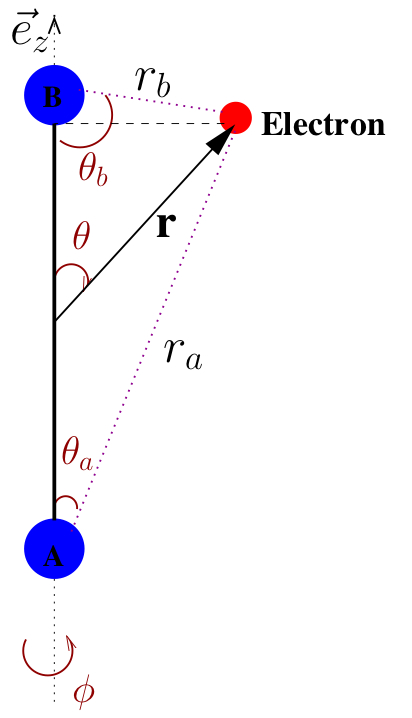}
\caption[]{Molecular ion $H_2^{+}\,$ with the set $(r,\theta,\phi)$ representing spherical coordinates and $\{\theta_{a},\,\theta_{b},\,R_{1},\,R_{2},\,r_{a},\,r_{b}\}$ providing us with elliptic coordinate system on the plan.}
\label{molH2}
\end{center}
\end{figure}

Let us first investigate the eigenvalues equation (\ref{SchrodEl}) for the molecular ion $\,H_{2}^{+}\,$ with the nuclei located on the orthogonal $\,z\,$-axis [see Figure \ref{molH2}]. We use the spherical coordinates $\,\big(r,\theta,\phi\big)\,$ [see Figure \ref{molH2}] to rewrite the electronic Hamiltonian (\ref{ElHam}) in this coordinate system,
\beq
\begin{array}{cc}
\displaystyle
\H_{el}(r,\theta,\phi)=-\frac{1}{2m}\frac{1}{r}\frac{\partial^2}{\partial r^2}r+\frac{L^2}{2m\,r^2}+V\left(r,\,\theta,\,\phi\right)\,,\\[12pt]
\displaystyle
V\big(r,\theta,\phi\big)\equiv V\big(r,\theta\big)=-\frac{Ze}{\sqrt{r^2+R_2^2+2r\,R_2\cos \theta}}-\frac{Ze}{\sqrt{r^2+R_1^2-2r\,R_1\cos \theta}}\,.
\end{array}\nn
\eeq

\noindent
The Casimir $L^2$ and the projection of the electronic orbital angular momentum $L_{z}$ read
\beq
\displaystyle
L^2=-\frac{\partial^2}{\partial \theta^2}-\frac{1}{\tan\theta}\frac{\partial}{\partial\theta}-\frac{1}{\sin^2\theta}\frac{\partial^2}{\partial\phi^2}\,,\qquad L_{z} = -i\frac{\partial}{\partial\phi}\,.
\eeq
Note that the potential, $\,V(r,\,\theta)\,$, which does not depend on the azimuthal angle $\,\phi\,\,$, is rotationally symmetric around the axis of the nuclei,
\beq
\displaystyle
\left[L_{z},\,V\left(r,\,\theta  \right)  \right]=0\,.\label{ZComm1}
\eeq
Since the component of the angular momentum $L_{z}\,$ also satisfies,
\beq
\left[L_{z},\,L^2\right]=0\,,\label{ZComm2}
\eeq
then, (\ref{ZComm1}) and (\ref{ZComm2}) yields the conservation of $\,L_{z}\,$ along the electronic motion,
\beq
\left[L_{z},\,\H_{el}\right]=0\,.
\eeq
The eigenvalues of the quantized quantity $\,L_{z}\,$ are given by the eigenvalues equation,
\beq
L_z\,\varphi_{m}(r,\,\theta,\,\phi)= m\,\varphi_{m}(r,\,\theta,\,\phi)\,.\label{eqvp1}
\eeq
The diatomic molecule therefore possesses a privileged direction carried by the axis of the nuclei [the $\,z\,$-axis]. Thus, $\,L_{z}\,$, generates an $\,SO(2)\,$ symmetry group. Moreover the configuration of the nuclei is also invariant under spatial inversion. Consequently the electronic Hamiltonian, $\,\H_{el}\,$, admits the same symmetry group $\,\G\,$ as the nuclei's configuration,
\beq
\G=SO(2)\times\big(\hbox{Parity}\big)\,.
\eeq
Taking into account the $\,SO(2)\,$ symmetry of the diatomic molecule, we can separate the electronic wave functions, $\,\varphi_m\,$, under the form of the product, 
\beq
\varphi_m\big(r,\,\theta,\,\phi\big)= g_m\big(\phi\big)f\big(r,\,\theta\big)\,.\label{sepvar}
\eeq
Injecting the expanded form (\ref{sepvar}) into the eigenvalues equation (\ref{eqvp1}), it is straightforward to obtain the normalized eigenfunction $g_{m}\,$,
\beq
g_{\pm m}\left(\phi \right)=\frac{1}{\sqrt{2\pi}}\,\exp\left(\pm i\,m\phi\right)\,,\quad m\,\in\,\IZ\,.\label{MWE}
\eeq
Now, our task is to investigate the function, $f\big(r,\,\theta\big)$, appearing as a part of the electronic wave function $\,\varphi_{m}\big(r,\,\theta,\,\phi\big)\,$ in (\ref{sepvar}). To this, we switch to elliptic coordinates, $(r,\,\theta)\longrightarrow(\xi,\,\eta)$, see Figure \ref{molH2},
\beq
\left\lbrace
\begin{array}{ll}
\displaystyle
\xi=\frac{r_a+r_b}{R}\,,\;\,\eta=\frac{r_a-r_b}{R}\,,\\[8pt]
R=R_{2}+R_{1}\,,\;\,\xi\,\in\,\left[1,\,\infty \right[\,,\;\,\eta \,\in\,\left[-1,\,1\right]\,.
\end{array}
\right.
\label{EllipCoord}
\eeq
We can now express the potential and the Laplacian operator in elliptic coordinates,
\beq
\begin{array}{lll}
\displaystyle
V\big(\xi,\eta\big)=-\frac{4eZ}{R}\frac{\xi}{\xi^2-\eta^2}\,,\\[10pt]
\displaystyle
\vnabla^2=\frac{4}{R^2\left(\xi^2-\eta^2\right)}\left\lbrace\left(\xi^2-1\right)\frac{\partial^2}{\partial\xi^2}-\left(\eta^2-1\right)\frac{\partial^2}{\partial\eta^2}\right.
\\[8pt]
\qquad\displaystyle\left.
+2\xi\,\frac{\partial}{\partial\xi}-2\eta\,\frac{\partial}{\partial\eta}+\left(\frac{1}{\xi^2-1}-\frac{1}{\eta^2-1}\right)\frac{\partial^2}{\partial\phi^2}\right\rbrace,
\end{array}
\eeq
so that the electronic Schr\"odinger equation (\ref{SchrodEl}) which takes, in elliptic coordinates, the form
\beq
\H_{el}\,g_{\pm\,m}\big(\phi\big)f\big(\xi,\,\eta\big)=\E_{el}\,g_{\pm\,m}\big(\phi\big)f\big(\xi,\,\eta\big)\,,
\eeq
separates into
\beq
\begin{array}{ll}
\left\lbrace\displaystyle
-\frac{2}{mR^2}\left(\big(\xi^2-1\big)\frac{\p^2}{\p\xi^2}+2\xi\frac{\p}{\p\xi}-\frac{m^2}{\xi^2-1}\right)-\xi^2\E_{el}-\frac{4eZ}{R}\xi\right.\\[12pt]\left.\displaystyle 
+\frac{2}{mR^2}\left(\big(\eta^2-1\big)\frac{\p^2}{\p\eta^2}+2\eta\frac{\p}{\p\eta}-\frac{m^2}{\eta^2-1}\right)+\eta^2\E_{el}
\right\rbrace g_{\pm m}\big(\phi\big)f(\xi,\,\eta)=0\,.
\end{array}
\label{HwithPhi}
\eeq
Let us note that the dependence on the azimuthal angle has disappeared from the bracketed terms in (\ref{HwithPhi}) and is replaced by the parameter $\,m^2\,$. Consequently, the electronic energy spectrum depend on $\,m^2\,$,
\beq
\displaystyle
\H_{el}\,\varphi_{\pm\,m} = \H_{el}\,\varphi_{\mp\,m}\quad\Longleftrightarrow\quad\E_{el,\pm\,m} = \E_{el,\mp\,m}\,,\label{Degeneracy}
\eeq
so that, for $\,m\neq0\,$, each electronic level is doubly degenerated.
Moreover, it is now clear that the variables $\,\xi\,$ and $\,\eta\,$ separate in the eigenfunctions $\,f(\xi,\,\eta)\,$ as 
\beq
f\left(\xi,\,\eta\right)= f_0\left(\xi\right)f_1\left(\eta\right)\,,
\eeq
where $\,f_0\,$ and $\,f_1\,$ are solutions of the following spheroidal wave equations \footnote{Spheroidal wave equations are generalization of Mathieu differential equations.},
\beq
\left\lbrace
\begin{array}{ll}
\displaystyle
\left(\left(\xi^2 -1 \right)\frac{\partial^2}{\partial \xi^2}+2\xi\,\frac{\partial}{\partial\xi}+\left(\a+\gamma\,\xi-p^2\,\xi^2-\frac{m^2}{\xi^2-1} \right) \right)f_0(\xi)=0
\\[12pt]
\displaystyle
\left(\left(\eta^2 -1 \right)\frac{\partial^2}{\partial \eta^2}+2\eta\,\frac{\partial}{\partial\eta}+\left(-\a+p^2\,\eta^2-\frac{m^2}{\eta^2-1} \right)\right)f_1(\eta)=0\,.
\end{array}\label{SpheroidalWE}
\right.
\eeq
Here $\,m^2\,$ and $\,\a\,$ are the separation constants of the differential equation, with 
\beq
\displaystyle
\gamma=2Rm\,e^2\,,\quad
\displaystyle
p^2=-\frac{R^2}{2}m\,\E_{el}\,.
\eeq
The bound states are solutions of the Schr\"odinger equation (\ref{SchrodEl}) associated with quantized negative energies, $\,\left\lbrace E_{el,m}\right\rbrace _{m\in\mathds{N}}\,$.

Collecting our results (\ref{MWE}) and (\ref{SpheroidalWE}) provides us with the complete electronic wave functions, 
\beq
\displaystyle
\varphi_{\pm m}\left(\xi,\eta,\phi\right)=\langle\left(\xi,\eta,\phi\right)\vert \pm m\rangle=\frac{1}{\sqrt{2\pi}}\,f_0\left(\xi\right)f_1\left(\eta\right)\exp\left(\pm i m \phi \right),\quad m\,\in\,\IZ\,. 
\eeq

Let us recall that  the integer $\,m\,$ which corresponds to the eigenvalue of the component of the angular momentum along the axis of symmetry $\,z\,$ is a ``good'' quantum number,
\beq
\displaystyle
L_{z}\,\varphi_m = \pm \Lambda_0\,\varphi_m\,,\quad\mbox{with}\quad\Lambda_0=\vert\,m\,\vert\,.
\eeq

When introducing the electronic spin degree of freedom, $\,S_{z}\,$, a non-vanishing additional term can be included, at each nuclear configuration, in the electronic Hamiltonian of the diatom $\,H_2^+\,$, namely
\beq
\displaystyle
\H_{so}=\mu L_{z}\cdot S_{z}\,,\quad\mu\,\in\,\IR\,,
\eeq
corresponding to the spin-orbit effects. In that event, we can show that $\,S_{z}\,$ and $\,L_{z}\,$ are separately conserved,
\beq
\begin{array}{ll}
\displaystyle
\left[S_{z},\,\H_{el}\right]= \left[S_{z},\, L_{z}\cdot S_{z}\right]=0\,\\[10pt]\displaystyle
\left[L_{z},\,\H_{el}\right]= \left[L_{z},\,L_{z}\cdot S_{z}\right]=0\,.
\end{array}
\eeq
The projection of the total angular momentum, $\,J_{z}=L_{z}+S_{z}\,$, onto the $\,z\,$-axis is therefore quantized,
\beq
\displaystyle
J_{z}\,\varphi_{\pm k}\left(\xi,\,\eta,\,\phi \right)=k\,\varphi_{\pm k}\left(\xi,\,\eta,\,\phi \right)\,,\quad k=\pm\Lambda\,,
\eeq
so that for fixed quantum number $\,\Lambda_0\,$, the eigenvalue, $\,\Lambda\,$, associated with $\,J_{z}\,$ takes the two half-integer values,
\beq
\displaystyle
\Lambda=\Lambda_0-\frac{1}{2},\quad\Lambda_0+\frac{1}{2}\,.\label{Degeneracy2}
\eeq
Consequently, the introduction of the electron spin degree of freedom does not modify the double degeneracy of the electron system. The novelty here is that even the ground level $\,\Lambda_{0}=0\,$ remains doubly degenerated. Thus, the electronic wave functions are now characterized by the quantum number $\,\Lambda\,$,
\beq
\displaystyle
\varphi_{\pm k}\left(\xi,\eta,\phi \right)=\langle\left(\xi,\eta,\phi\right)\vert \pm k\rangle=\frac{1}{\sqrt{2\pi}}f_0\left(\xi\right)f_1\left(\eta\right)\exp\left(\pm i\,k\phi \right),\quad k=\pm\Lambda\,.
\eeq
Following the Born-Oppenheimer approximation, after describing the electronic wave functions labeled by the index $\,k\,$ with the energy eigenvalues $\,\E_{el,k}\,$ which are parametric function of the relative internuclear coordinate, we are now interested on the nuclear motions described by the Schr\"odinger equation (\ref{SchrodNuclear}). Let us sandwich (\ref{SchrodNuclear}) between electronic eigenstates,
\beq
\begin{array}{ll}
\displaystyle
-\int d\vx\varphi_{n}^{\star}(\vx,\vX_{12})\big(\sum_{i=1}^2\frac{1}{2M}\vnabla^2_{X_i}-\H_{el}-\frac{Z^2}{\vert\vX_{12}\vert}\big)\sum_{m}\varphi_{m}(\vx,\vX_{12})\chi_{m}(\vX_{12})\\[12pt]
\displaystyle
=E\int d\vx\varphi_{n}^{\star}(\vx,\vX_{12})\sum_{m}\varphi_{m}(\vx,\vX_{12})\chi_{m}(\vX_{12})\,.
\end{array}
\eeq
Thus, this provides us with
\beq
\begin{array}{ll}
\displaystyle
-\sum_{i=1}^2\frac{1}{2M}\int d\vx\varphi_{n}^{\star}(\vx,\vX_{12})\vnabla^2_{X_i}\sum_{m}\varphi_{m}(\vx,\vX_{12})\chi_{m}(\vX_{12})\\[8pt]
\displaystyle
+\E_{el}\chi_{n}(\vX_{12})+\frac{Z^2}{\vert\vX_{12}\vert}\chi_{n}(\vX_{12})=E\chi_{n}(\vX_{12})\,,
\end{array}
\eeq
where the first term can be expanded using the Leibniz rule on differentiation so that we obtain for $\,\vert\varphi_{m}\rangle=\vert m\rangle\,$,
\beq
\displaystyle
-\frac{1}{2M}\big(
\sum_{m}\langle n\vert\vnabla^{2}_{\vX_{12}}\vert m\rangle\chi_{m}+2\sum_{m}\langle n\vert\vnabla_{\vX_{12}}\vert m\rangle\vnabla\chi_{m}+\vnabla^{2}_{\vX_{12}}\chi_{n}\big)
+\big(\E_{el}+\frac{Z^2}{\vert\vX_{12}\vert}\big)\chi_{n}=E\chi_{n}\,.
\nn
\displaystyle
\eeq
Hence, we derive the effective Hamiltonian describing the nuclear motions,
\beq
\displaystyle
H_{nm} =-\frac{1}{2M}\sum_k\big(\vnabla_{\vX_{12}} + \langle\, n\,\vert \,\vnabla_{\vX_{12}}\,\vert\,k\,\rangle\big)\big(\vnabla_{\vX_{12}} + \langle \,k\,\vert \,\vnabla_{\vX_{12}}\,\vert m \rangle\big)\\[8pt]
+\big(\frac{Z^2}{\vert\vX_{12}\vert}+\E_{el}\big)\delta_{nm}\,.
\nn
\eeq
In the adiabatic approximation, where the nuclei move slowly when compare to the electronic motion, the electron has to be considered to remain in the same $\,2\,$-fold degenerate $\,n\textit{th}\,$ level. Consequently, the off-diagonal transition terms are neglected, implying the relevant effective nuclear Hamiltonian,
\beq
\begin{array}{ll}
\displaystyle
H=-\frac{1}{2M}\left(\vnabla_{\vX_{12}}-i\vA(\vX_{12})\right)^2+V(\vX_{12})\,,\\[8pt]
\displaystyle
\mbox{with}\quad\vA=i\langle\, n\,\vert \,\vnabla_{\vX_{12}}\,\vert \,n\,\rangle\quad\mbox{and}\quad
V(\vX_{12})=\frac{Z^2}{\vert\vX_{12}\vert}+\E_{el}\,.
\end{array}
\eeq
Here $\,V\,$ acts as an effective scalar potential for nuclear motion and the induced gauge potential $\,\vA\,$ is a $\,(2\times2)\,$ matrix, since the state $\,\vert \,n\,\rangle\,$ belongs to a $\,2\,$-fold degenerate level, see (\ref{Degeneracy}) and (\ref{Degeneracy2}). Hence, $\,\vA\,$ transforms as a $\,U(2)\,$ gauge potential.

For the nuclear axis in the initial direction given by the polar and azimuthal angles $\,\theta=\phi=0\,$,
\beq
\vert\, n(\vec{e}_z)\,\rangle = \vert\, n(0,0)\,\rangle\,,
\eeq
we can generate a set of eigenstates adapted to nuclei pointing toward $\,(\theta,\phi)\,$ by rotating the initial eigenstate. Then, the Wigner theorem provides us with the two possible parametrizations,
\beq
\begin{array}{ll}
\displaystyle
\vert n(\theta,\phi)\rangle = \exp(iJ_3\phi)\exp(iJ_1\theta)\exp(-iJ_3\phi)\vert n(0,0)\rangle,\quad\hbox{for}\quad\theta\neq\pi\\[8pt]
\displaystyle
\widetilde{\vert n(\theta,\phi)\rangle} = \exp(iJ_3\phi)\exp(iJ_1\theta)\exp(iJ_3\phi)\vert n(0,0)\rangle,\quad\hbox{for}\quad\theta\neq 0\,.
\end{array}
\eeq
Note that the two previous parametrizations are linked by
\beq
\displaystyle
\widetilde{\vert\,n\,\rangle}=\exp(2in\phi)\vert\,n\,\rangle\,.
\eeq
Thus, the $\,U(2)\,$ gauge potentials, which are defined on the space spanned by the electronic eigenstate, depend on the geometry of the $\,2\,$-fold degenerate eigenstate space so that
\beq
\left\lbrace
\begin{array}{lll}
\displaystyle
A_r=i\langle\,n(r,\theta,\phi)\,\vert\,\partial_r\,\vert\,n(r,\theta,\phi)\rangle\\[8pt]
\displaystyle
A_{\theta}=i\langle\,n(r,\theta,\phi)\,\vert\,\partial_{\theta}\,\vert\,n(r,\theta,\phi)\rangle\\[8pt]
\displaystyle
A_{\phi}=i\langle\,n(r,\theta,\phi)\,\vert\,\partial_{\phi}\,\vert\,n(r,\theta,\phi)\rangle\,.
\end{array}
\right.
\eeq
Performing the calculation in the case of the $\,\theta\,\neq\,\pi\,$ parametrization \footnote{The procedure is exactly the same for the parametrization with $\,\theta\neq0\,$.} leads to
\beq
\left\lbrace
\begin{array}{lll}
\displaystyle
A_r= 0\,,\\[8pt]
\displaystyle
A_{\theta} = \langle\, n\left(0,0\right)\, \vert \,-\cos\phi\,J_1 +\sin\phi\,J_2\,\vert\, n\left(0,0\right)\, \rangle\,,\\[8pt]\displaystyle
A_{\phi} = \langle \,n\left(0,0\right)\,\vert\,\left( 1-\cos\theta\right)J_3+\sin\theta\left(\sin\phi\,J_1 +\cos\phi\,J_2\right)\,\vert\,n\left(0,0\right)\,\rangle\,.
\end{array}
\right.
\eeq
We posit,
\beq
J_{1}=a\sigma_{1},\quad J_{2}=b\sigma_{2},\quad J_{3}=c\sigma_{3}\,,
\eeq
and we obtain by direct calculation the shape of the non-vanishing gauge potentials induced by nuclear rotations,
\beq
\left\lbrace
\begin{array}{ll}
\displaystyle
A_{\theta}=-a\cos\phi\,\sigma_1+b\sin\phi\,\sigma_2\,,\\[8pt]\displaystyle
A_{\phi}=\sin\theta\left(a\sin\phi\,\sigma_1+b\cos\phi\,\sigma_2\right)+c\left(1-\cos\theta\right)\sigma_{3}\,.
\end{array}\label{potjbon}
\right.
\eeq
The corresponding field strength, $\,F_{\theta\phi}\,$, reads
\beq
F_{\theta\phi} = \a\sin\theta\,\sigma_3 +\left(\cos\theta -1\right)\left(\b\cos\phi\,\sigma_2 +\gamma\sin\phi\,\sigma_1\right)\,,\label{GoodFieldStrength}
\eeq
where $\a$, $\,\b\,$ and $\gamma$ satisfy the relations,
\beq
\a=c-2ab\,,\quad
\b=b-2ac\,,\quad
\gamma=a - 2bc\,.
\eeq

Let us now inquire about the real nature of the $\,U(2)\,$ gauge potentials (\ref{potjbon}) induced by nuclear motions. Are these imbedded Abelian gauge fields into $\,U(2)\,$? or not? To respond to this question, let us recall that a field strength,
\beq
\F'_{\theta\phi}=m\sigma_1+n\sigma_2+p\sigma_3\,,\label{chpgen1}
\eeq
can always be gauge-transformed so that it points in one single direction $\,\sigma_1\,$, $\,\sigma_2\,$ or $\,\sigma_3\,$ say. In the present context, we search for gauge transformations which rotate the field strength (\ref{chpgen1}) in the ``Abelian'' direction $\,\sigma_3\,$,
\beq
\F_{\theta,\phi} = f\sigma_3\,,\quad f\neq 0\,.
\eeq
In the limiting case where $\,f = 0\,$, i.e. in the null field strength configuration, the gauge potentials are pure gauge. Then, the gauge potentials must be gauge equivalent to that of the vacuum.

\noindent
We are looking for matrices $\,U\,$ taking values in $U(2)$,
\beq
U=\left(\begin{array}{cc} A  & B\\[4pt]C  & D \end{array}\right)\,,\eeq
so that
$\,
\F_{\theta\phi} = U\F'_{\theta\phi}\,U^{-1}\label{trsfj11}
\,$. Consequently, we derive from (\ref{trsfj11}) the series of constraints to be solved
\beq
(S):\,\left\lbrace\begin{array}{llll}
\left(p-f\right)A+\left(m +in \right)B=0\\[8pt]
\left(m - in\right)A-\left(f+p\right) B=0\\[8pt]
\left(m-in\right)C+\left(f -p \right)D=0\\[8pt]
\left(f+p\right)C+\left(m+in\right) D=0\,.
\end{array}\right.
\eeq
The constraints $\,(S)\,$ can be solved provided its determinant vanished,
\beq
\det(S)= 0\quad\Longleftrightarrow\quad
f^2 = m^2+n^2+p^2\label{condnorm}\,.
\eeq
We then obtain an equivalence between the length of $\,\F_{\theta\phi}\,$ and $\,\F'_{\theta\phi}\,$ which express as conservation of the length of the field strength under a gauge transformation .
Solving the constraints $\,(S)\,$, for $\,f\neq 0\,$, yields
\beq
\displaystyle
U=\sqrt{
\frac{f+p}{2f}}\left(\begin{array}{cc} \exp\left(i\mu\right)& \displaystyle
\frac{m-in}{f+p}\exp\big( i\mu\big)\\[6pt]
\displaystyle
-\frac{m+in}{f+p}\exp\big( i\nu\big)  & \exp\left(i\nu\right)
\end{array}\right)\nn
\eeq
with arbitrary real constants
\beq
\mu=arg\big(A\big)\,,\quad\nu=arg\big(D\big)\,.
\eeq
Applying the previous gauge transformation to (\ref{GoodFieldStrength}) and (\ref{potjbon}) we must take
\beq
\left\lbrace
\begin{array}{lll}
m=\gamma\sin\phi\big(\cos\theta-1\big)\\[8pt]
n=\b\cos\phi\big(\cos\theta-1\big)\\[8pt]
p=\a\sin\theta\,,
\end{array}\right.
\eeq
so that the length of the field strength reads
\beq
f^{2}\big(\theta,\phi\big)=\a^2\sin^2\theta+\big(\cos\theta-1\big)^2\big(\gamma^2\sin^2\phi+\b^2\cos^2\phi\big)\,.
\eeq
Without loss of generality, we can choose \cite{MSW,Rho},
\beq
\a=\frac{1}{2}\big(1-\k^{2}\big)\,,\quad \b=\gamma=0\, \footnote{ This choice implies that $\,a=b=\langle +\vert J_1\vert -\rangle=\pm\frac{1}{2}\vert\k\vert\,$ and $\,c=\frac{1}{2}\,$.} \quad\mbox{with}\quad\k\in\IR\, \footnote{ Since the electronic eigenstates are not eigenfunctions of angular momentum, but only of $\,J_{3}\,$, $\,\k\,$ can take any real value.}\,,
\eeq
so that applying the gauge transformation on the Berry potentials (\ref{potjbon}),
\beq
\left\lbrace
\begin{array}{ll}
\widetilde{A}_{\theta}=U\big({A}_{\theta}+i\p_{\theta}\big)U^{-1}\\[8pt]
\widetilde{A}_{\phi}=U\big({A}_{\phi}+i\p_{\phi}\big)U^{-1}\,,
\end{array}
\right.
\eeq
provides us with the gauge-equivalent potentials,
\beq
\displaystyle
\widetilde{A}_{\theta}=\mp\frac{\vert\k\vert}{2}\left(\begin{array}{cc} 0 & e^{i\big(\mu-\nu+\phi\big)}\\[6pt]
e^{i\big(\nu-\mu-\phi\big)}  & 0
\end{array}
\right)\,,
\label{potjbon2}
\eeq
and
\beq
\widetilde{A}_{\phi}=\left(\begin{array}{cc} \displaystyle\frac{1}{2}\left( 1-\cos\theta \right) & \displaystyle\mp\frac{i}{2}\vert\k\vert\sin\theta\,e^{i\left(\mu-\nu+\phi\right)}\\[6pt]
\displaystyle\pm\frac{i}{2}\vert\k\vert\sin\theta\,e^{i\left(\nu-\mu-\phi\right)}  & \displaystyle-\frac{1}{2}\left(1- \cos\theta\right) \end{array}\right)\,.\label{potjbon3}
\eeq
It is now clear that the Berry gauge potentials (\ref{potjbon}) or (\ref{potjbon2}) and (\ref{potjbon3}) become \textit{``Abelianized'' gauge potential} for $\,\k=0\,$. In that event, they represent a Dirac monopole field of unit charge imbedded into $\,U(2)\,$,
\beq
\widetilde{A}_{\theta}=0\,,\qquad\widetilde{A}_{\phi}=\frac{1}{2}\left(1-\cos\theta\right)\sigma_{3}\,.
\eeq
For $\,\vert\k\vert\neq0\,$, we obtain the \textit{truly non-Abelian case}, where the off-diagonal terms can not be eliminated \footnote{Here we fixed $\,\mu=\nu\,$.},
\beq
\left\lbrace
\begin{array}{ll}
\displaystyle
\widetilde{A}_{\theta}=\mp\frac{\vert\k\vert}{2}\big(\cos\phi\,\sigma_{1}-\sin\phi\,\sigma_{2}\big)\,,\\[8pt]\displaystyle
\widetilde{A}_{\phi} =\pm\frac{\vert\k\vert}{2}\sin\theta\big(\sin\phi\,\sigma_{1}+\cos\phi\,\sigma_{2}\big)+\frac{1}{2}\big(1-\cos\theta\big)\sigma_{3}\,.
\end{array}\right.\label{WilczekPot}
\eeq
The corresponding field strength is 
\beq
\widetilde{F}_{\theta\phi}=\frac{1}{2}\big(1-\k^{2}\big)\sin\theta\sigma_{3}\,.\label{WilczekField}
\eeq
The field strength (\ref{WilczekField}) superficially resembles to that of a monopole field but the interpretation is quite different. Indeed, $\,\vert\k\vert\neq0\,$ is not quantized here and the gauge fields induced by nuclear motions of the diatomic molecule are \textit{truly non-Abelian} \cite{MSW}. See also \cite{Zyg}.

Note that when $\,\k=\pm1\,$, the field strength vanishes and (\ref{WilczekPot}) is a gauge transform of the vacuum.

Our next step is to present the monopole-like field (\ref{WilczekPot}) in a more convenient ``hedgehog'' form. This can be achieved, by applying a suitable gauge transformations \cite{Jdiat} to the diatomic molecule gauge potential (\ref{WilczekPot}). Finally, the Berry gauge potential mimics the structure of a non-Abelian monopole \cite{tHooft,Polyakov},
\beq
\widetilde{A}_i^{\;a}=(1-\kappa)\epsilon_{iaj}\,
\frac{x^j}{r^2}\,,
\quad 
\widetilde{F}^{\;a}_{ij}=(1-\kappa^2)\epsilon_{ijk}
\frac{x^kx^a}{r^4}\,.
\label{diatfields}
\eeq
Note \textit{the presence of the unquantized constant factor $\,(1-\kappa^2)\,$ in the above magnetic field}.\\

\noindent
\underline{Classical dynamics and conserved quantities}

Now we turn to investigating the symmetries of an isospin-carrying particle, with unit charge, evolving in the monopole-like field of the diatom (\ref{diatfields}) plus a scalar potential. The Hamiltonian describing the dynamics of this particle is expressed as
\beq
\H=\frac{\vpi^{2}}{2}-\big({\gyro}/{4}\big)\epsilon_{ijk}\widetilde{F}^{\;a}_{ij}\,S^{k}+V(\vx,\,\I^{a})\,,\quad\pi_{i}=p_{i}-\widetilde{A}_{i}^{a}\,\I^{a}\,,\label{HDiatom}
\eeq 
where the spin-rotation coupling disappears when we study particle carrying null gyromagnetic ratio, $\gyro=0\,$. The resulting Hamiltonian has the same form of that of a scalar particle \footnote{i.e. particle without spin.} evolving in the same magnetic field. We define the covariant Poisson-brackets as 
\beq
\big\{M,N\big\}=D_jM\frac{\p N}{\p \pi_j}-\frac{\p M}{\p \pi_j}D_jN
+\I^a\widetilde{F}^{\;a}_{jk}\frac{\p M}{\p \pi_j}\frac{\p N}{\p \pi_k}
-\epsilon_{abc}\frac{\p M}{\p \I^a}\frac{\p N}{\p \I^b}\I^c\,,
\label{PBracketbis}
\eeq
where $D_j$ is the covariant derivative,
\beq
D_jf=\p_jf-\epsilon_{abc}\I^a\widetilde{A}_{j}^{b}\,\frac{\p f}{\p\I^c}\,.
\label{covder}
\eeq
The commutator of the covariant derivatives is recorded as
\begin{equation}
[D_i,D_j]=-\epsilon_{abc}\I^a\widetilde{F}_{ij}^{b}\,\frac{\p}{\p \I^c}\ . 
\label{covdercomm}
\end{equation}
The non-vanishing brackets are
\beq
\{x^{i},\pi_{j}\}=\d^{i}_{j}\,,\quad\{\pi_{i},\pi_{j}\}=\I^a\widetilde{F}_{ij}^{a}\,,\quad\{\I^{a},\I^{b}\}=-\epsilon_{abc}\I^{c}\,,
\eeq
and the equations of motion governing an isospin-carrying particle in the static non-Abelian gauge field (\ref{diatfields}) read
\beq\left\lbrace\begin{array}{ll}\displaystyle
\ddot{x}_{i}-\I^a\widetilde{F}_{ij}^{a}\,\dot{x}^{j}+D_{i}V=0\,,\\[10pt]\displaystyle
\dot{\I}^{a}+\epsilon_{abc}\,\I^{b}\left(\widetilde{A}_{j}^{c}\,\dot{x}^{j}-\frac{\p V}{\p \I^{c}}\right)=0\,.
\end{array}\right.\label{DiatEqM}
\eeq
The first equation in (\ref{DiatEqM}) describes the 3D real motion implying a generalized Lorentz force plus an interaction with the scalar potential; while the second equation is the Kerner-Wong equation augmented with a scalar field interaction. The latter describes, as expected, the isospin classical motion.

Let us now recall the van Holten procedure yielding the conserved quantities. The constants of the motion are expanded in powers of the momenta,
\beq
\Q\big(\vx,\vec{\I},\vpi\big)= C(\vx,\vec{\I})+C_i(\vx,\vec{\I})\pi_i+\frac{1}{2!}C_{ij}(\vx,\vec{\I})\pi_i\pi_j+
\cdots\,,\label{constexp1}
\eeq
and we require $\Q$ to Poisson-commute with the Hamiltonian,
\beq\{\Q,\;\H=\frac{\vpi^{2}}{2}+V(\vx,\,\I^{a})\}=0\,.\label{Hypothesis}\eeq
We therefore get the set of constraints which have to be solved, 
\beq\begin{array}{llll}
\displaystyle{C_iD_iV+\epsilon_{abc}\I^a\frac{\p C}{\p\I^b}\frac{\p V}{\p\I^c}}=0,& o(0)
\\[10pt]
\displaystyle{D_iC=\I^a\widetilde{F}^{a}_{ij}C_j+C_{ij}D_jV+\epsilon_{abc}\I^a\frac{\p C_i}{\p\I^b}\frac{\p V}{\p\I^c}},&
o(1)
\\[10pt]
\displaystyle
D_iC_j+D_jC_i=\I^a(\widetilde{F}^{a}_{ik}C_{kj}+\widetilde{F}^{a}_{jk}C_{ki})+C_{ijk}D_kV+\epsilon_{abc}\I^a\frac{\p C_{ij}}{\p\I^b}\frac{\p V}{\p\I^c},& o(2)
\\[10pt]
\displaystyle
D_iC_{jk}+D_jC_{ki}+D_kC_{ij}=\I^a(\widetilde{F}^{a}_{il}C_{ljk}+\widetilde{F}^{a}_{jl}C_{lki}+\widetilde{F}^{a}_{kl}C_{lij})\\[8pt]\displaystyle{\qquad\qquad\qquad\qquad\qquad\qquad+C_{ijkl}D_lV+\epsilon_{abc}\I^a\frac{\p C_{ijk}}{\p\I^b}\frac{\p V}{\p\I^c}},& o(3)
\\
\vdots\qquad\qquad\qquad\qquad\qquad\vdots&\vdots
\end{array}\label{ConstraintsDiatom}
\eeq

Turning to the zeroth-order conserved charge, we note  that, for $\kappa\neq0$, the used-to-be electric charge,
\beq
\displaystyle Q=\frac{\vx\cdot\vec{\I}}{r}\,,
\eeq
is \emph{not more covariantly conserved} in general,
\beq
\big\{Q,\H\big\}=\vpi\cdot \vD Q, 
\quad
D_jQ=\frac{\kappa}{r}\left(\I^j-Q\frac{x_j}{r}\right).
\label{noecharge}
\eeq
 An exception occurs when the isospin is aligned into the radial direction, as seen from (\ref{noecharge}). A detailed calculation shows that the equation $\,D_jQ=0\,$ can only be solved, for imbedded Abelian monopole field, when $\,\kappa=0,\pm1\,$.
 
Nor is $Q^2$  conserved,
\beq
\big\{Q^2,\H\big\}=2\kappa Q(\vpi\cdot\vD Q)\,.
\eeq
Note for further reference that,
unlike $Q^2$, the length of the isospin, $\I^2$, \emph{is} conserved, $$\{\H,\I^2\}=0\,.$$

The monopole-like gauge field (\ref{diatfields}) is rotationally symmetric
and an isospin-carrying particle moving in it admits a conserved angular momentum \cite{MSW,Jdiat}. Its form is, however, somewhat unconventional, and we re-derive it, therefore, in detail \cite{H-NGI}.

{\bf{1)}} We start our investigation with conserved quantities which are \underline{linear} in the covariant momentum. We have therefore
\beq
C_{ij}=C_{ijk}=\cdots=0\,,
\eeq
so that the series of constraints (\ref{ConstraintsDiatom}) reduce to
\beqa
\left\lbrace
\begin{array}{lll} 
\displaystyle
C_iD_iV+\epsilon_{abc}\I^a\frac{\p C}{\p\I^b}\frac{\p V}{\p\I^c}=0\,,& o(0)\\[8pt]
\displaystyle
D_iC=\I^a\widetilde{F}^{a}_{ij}C_j+\epsilon_{abc}\I^a\frac{\p C_i}{\p\I^b}\frac{\p V}{\p\I^c}\,,&
o(1)\\[8pt]
D_iC_j+D_jC_i=0\,.&
o(2)
\end{array}\label{C2}
\right.
\eeqa
We use the Killing vector generating spatial rotations,
\beq
\vC=\vn\times\vx\,.\label{rotKillingDiatom}
\eeq
Choosing $V=V(r)$, we see that, again due to the non-conservation
of $\,Q\,$, $\,D_jV\neq0\,$ in general. The zeroth-order condition $\,\vC\cdot\vD V=0\,$ in (\ref{C2}) is, nevertheless, satisfied when $V$ is a radial function independent of $\,\vec{\I}\,$, since then $\,\vD V=\vec{\nabla}V\,$, which is perpendicular to infinitesimal rotations, $\vC$. 

Evaluating the right hand side of the first-order constraint of (\ref{C2}) with 
$\,\widetilde{F}_{jk}^a\,$ as given in (\ref{diatfields}), the equation to be solved becomes
\begin{equation}
D_iC=(1-\kappa^2)\frac{Q}{r}\left((\vn\cdot\hx)\frac{x_i}{r}-n_i\right).
\label{diatC}
\end{equation}
In the Wu-Yang case, $\kappa=0$,  this equation was solved by
$\,\displaystyle C=-\vn\,\cdot\,Q{\hx}\,$. But for $\,\k\neq0\,$, the electric charge, $\,Q\,$, is not conserved, and using (\ref{noecharge}), (\ref{diatC}), as well as the relations
\beq
\left\lbrace
\begin{array}{ll}
D_i\big(\vec{\I}\cdot\vn\big)&=
\displaystyle(1-\kappa)\left(\frac{Q}{r}\,n_{i}-{\frac{\vn\cdot\hat{r}}{r}\I_{i}}
\right)\ ,
\\[8pt]
D_i\left(Q\,\vn\cdot\displaystyle{\hx}\right)&=
\displaystyle\frac{Q}{r}\left(n_i-(1+\kappa)(\vec{n}\cdot\hat{r})\frac{x_i}{r}\right)+\frac{\k}{r}(\vec{n}\cdot\hat{r})\I_{i}
\\[8pt]
\I^aF_{ij}^a&=(1-\kappa^2)\,Q\,\displaystyle\frac{\epsilon_{ijk}x_k}{r^3}\ ,
\end{array}
\label{handyform}
\right.
\eeq
 we find,
$$
-(1-\k)D_i\left(Q\,\vn\cdot\hx\right)
=\kappa D_i\big(\vec{\I}\cdot\vn\big)+D_{i}C.
$$
This allows us to infer that
\begin{equation}
C=-\Big((1-\kappa)\,Q\,{\hx}+\kappa\vec{\I}\Big)\cdot\vn\,.
\label{JC}
\end{equation}
The conserved angular momentum is, therefore,
\beq
\left\lbrace
\begin{array}{ll}
\vJ=\vx\times\vpi-\vW\,,
\\[6pt]\displaystyle
\vW=(1-\kappa)\,Q\,\hx+\kappa\vec{\I}\ =
Q\,\hx+\kappa\left(\hx\times\vec{\I}\right)\times\hx\,,
\end{array}
\right.
\label{diatangmom}
\eeq
consistently with the  results in \cite{Jdiat,Rho}. Moody, Shapere and Wilczek \cite{MSW} found the correct expression, (\ref{diatangmom}), for $\kappa=0$
but, as they say it, ``they are not aware of a canonical derivation
when $\kappa\neq0$''. 
Our construction here is an alternative to
that of  Jackiw \cite{Jdiat}, who obtained 
it using the method of Reference \cite{JackiwManton}.
In his approach, based on the study of symmetric gauge fields \cite{ForgacsManton},
each infinitesimal rotation, (\ref{rotKillingDiatom}),
is a symmetry of the monopole in the
sense that it changes the potential by a surface
term. 

\noindent
It is worth noting that comparison with the Wu-Yang case yields the ``replacement rule'',
\begin{equation}
Q\,\hx\to\vW.
\label{rule}
\end{equation}
For $\kappa=0$ we recover the Wu-Yang expression (\ref{WYAngMom}).
Eliminating $\vpi$ in favor of $\,\vp-\vA=\vpi\,$ allows us to rewrite the 
total angular momentum as
\begin{equation}
\vJ=\vx\times\vp-\vec{\I}\, ,
\end{equation}
making manifest the celebrated
``spin from isospin term'' \cite{Jackiw:1976xx}.

Alternatively, a direct calculation, using the same formulae
(\ref{noecharge})-(\ref{handyform}), allows us to
confirm that $\vJ$ 
commutes with the Hamiltonian, $\{J_i,\H\}=0$. 

Multiplying (\ref{diatangmom}) by $\,\displaystyle\hx\,$ yields, once again, the  relation  (\ref{cone}) i.e.,
\beq
\vJ\cdot\hx=-Q\,,
\eeq
the same as in the Wu-Yang case. This is, however, less useful as before, since $\,Q\,$ is not a constant of the motion so that the angle between  $\vJ$ and the
radius vector, $\vx(t)$, is not more a constant. The components of the angular momentum (\ref{diatangmom}) close, nevertheless, to $\,\so(3)\,$,
\beq
\big\{J_i,J_j\big\}=\epsilon_{ijk}J_k\,.
\eeq

In addition of $\,\vJ\,$, it is worth mentioning that the Casimir
\beq
J^{2}=\big(\vx\times\vpi\big)^2+\big(1-\k\big)^2Q^2-\k^2\vec{\I}^2-2\k\vJ\cdot\vec{\I}\,
\eeq
is obviously conserved since the angular momentum of the diatom is conserved (\ref{diatangmom}).

{\bf{2)}} Returning to the van Holten algorithm, \underline{quadratic} conserved quantities are sought by taking
\beq
C_{ijk}=C_{ijkl}\cdots=0\,.
\eeq
Consequently (\ref{ConstraintsDiatom}) reduces to 
\beqa
\left\lbrace
\begin{array}{llll}
\displaystyle{C_iD_iV+\epsilon_{abc}\I^a\frac{\p C}{\p\I^b}\frac{\p V}{\p\I^c}}=0,& o(0)
\\[8pt]
\displaystyle{D_iC=\I^a\widetilde{F}^{a}_{ij}C_j+C_{ij}D_jV+\epsilon_{abc}\I^a\frac{\p C_i}{\p\I^b}\frac{\p V}{\p\I^c}},&
o(1)
\\[8pt]
\displaystyle
D_iC_j+D_jC_i=\I^a(\widetilde{F}^{a}_{ik}C_{kj}+\widetilde{F}^{a}_{jk}C_{ki})+\epsilon_{abc}\I^a\frac{\p C_{ij}}{\p\I^b}\frac{\p V}{\p\I^c},& o(2)
\\[8pt]
\displaystyle
D_iC_{jk}+D_jC_{ki}+D_kC_{ij}=0\,.& o(3)
\end{array}\label{CDiatom}\right.
\eeqa
We consider the rank-$\,2\,$ Killing tensor,
\beq
C_{ij}=2\d_{ij}x^2-2x_{i}x_{j}\,,\label{KillingDiatom2}
\eeq
which satisfies the third-order constraint of (\ref{CDiatom}). Injecting (\ref{KillingDiatom2}) into the second-order constraint yields,
\beq
D_iC_j+D_jC_i=0\,,
\eeq
which can be solved by taking $\,C_{i}=0\,$. For radial potentials independent of $\vec{\I}$, it is straightforward to satisfy the first- and the zeroth-order constraints of (\ref{CDiatom}) with $\,C=0\,$. Thus, we obtain the conserved Casimir,
\beq
L^2=\big(\vx\times\vpi\big)^{2}=x^2\vpi^2-\big(\vx\cdot\vpi\big)^2\,,\label{Casimir1}
\eeq
which is the square of the non-conserved orbital angular momentum, $\vL=\vx\times\vpi\,$.

Since $\,J^2\,$ and $\,L^2\,$ are both conserved, it is now straightforward to identify the charge,
\beq
\Gamma=J^2-L^2=\big(1-\k\big)^2Q^2-\k^2\vec{\I}^2-2\k\vJ\cdot\vec{\I}\,,\label{Gamma}
\eeq
which is conserved along the motion in the monopole-like field of diatomic molecule. \textit{It is worth noting that the charge $\,\Gamma\,$ corresponds, in the Abelian limit with $\,\k=0\,$, to the square of the electric charge}. As the constants of the motion $\,\vJ\,$, $\,J^2\,$ and $\,L^2\,$, the charge $\,\Gamma\,$ is conserved for any radially symmetric potential, $\,V(r)\,$.

Note that $\,\Gamma\,$ can also be obtained by using the Killing vector,
\beq
\vC=2\k\big(\vx\times\vec{\I}\big)\,,
\eeq
into the van Holten algorithm (\ref{C2}).

Let us now decompose the covariant momentum, into radial and transverse components, with the vector identity,
\beq
(\vpi)^2=(\vpi\cdot{\hx})^2+(\vpi\times{\hx})^2
=\pi_r^2+\frac{L^2}{r^2}\,.
\eeq 
This hence allows us to express the diatomic molecule Hamiltonian (\ref{HDiatom}) as 
\beq
\H=\displaystyle\frac{1}{2}(\vpi\cdot{\hx})^2+ \displaystyle\frac{J^2}{2r^2}-\left\{\displaystyle\frac{(1-\kappa)^2Q^2-\kappa^2\I^2-2\kappa\vJ\cdot\vec{\I}}{2r^2}\right\}
+V(r)\,.
\label{diatraddecJackiw}
\eeq
Suggesting that the charge takes the fixed value $\,Q^2=\I^2=1/4\,$, Jackiw found a similar decomposition as (\ref{diatraddecJackiw}) \cite{Jdiat}, but this is, however, only legitimate when $\,\k=0\,$, since  $\,Q^2\,$ is not conserved for $\,\k\neq0\,$.

For $\,\k\neq0\,$, the ``good'' approach is to recognize the fixed charge $\,\Gamma\,$, which yields the nice decomposition,
\beq
\H=\displaystyle\frac{1}{2}(\vpi\cdot{\hx})^2+\frac{J^2}{2r^2}-\frac{\Gamma}{2r^2}
+V(r)\,.
\label{diatraddec}
\eeq

Let us underline that the effective field of a diatomic molecule provides us with an
interesting generalization of the Wu-Yang monopole. For $\kappa\neq0,\pm1$, it is truly non-Abelian, i.e., \emph{not reducible} to one on an $U(1)$ bundle. No covariantly constant direction field, and, therefore, \emph{no conserved electric charge} does exist in this case.

The field is nevertheless radially symmetric, but the conserved angular momentum (\ref{diatangmom}) has  a non-conventional form. 

In bundle terms, the action of a symmetry generator can be lifted to the bundle so that it preserves
the connection form which represents the potential. But the group structure may not be conserved;
this requires another, consistency condition \cite{JackiwManton}, which may or may not be satisfied. In the diatomic case, it is not satisfied when  $\kappa\neq0,\,\pm1$.

Is it possible to redefine the ``lift'' so that the group structure be preserved~? In the Abelian case, the answer can be given in cohomological terms \cite{DHInt}. If this obstruction does not vanish, it is only a {\it central extension} that acts on the bundle.

In the truly non-Abelian case, the  consistency condition involves the covariant, rather than ordinary derivative and covariantly constant sections only exist in exceptional cases -- namely when the
bundle is reducible. Thus, only some (non-central extension) acts on the bundle.

It is worth noting that for $\kappa\neq0$ the configuration (\ref{diatfields}) does not satisfy the vacuum Yang-Mills equations. It only satisfies indeed with a suitable conserved current \cite{Jdiat},
\begin{eqnarray}
\cD_iF_{ik}=j_k,
\quad
\vec{\jmath}=\frac{\kappa(1-\kappa^2)}{r^4}\,\vx\times\vec{T}\,,
\label{current}
\end{eqnarray}
Interestingly, this current can also be produced by a hedgehog Higgs field,
\begin{eqnarray}
 j_k=\big[\cD_k\Phi,\Phi\big],
\quad
\Phi^a=\frac{\sqrt{1-\kappa^2}}{r}\,\frac{x_a}{r}\;.
\end{eqnarray}

For $\,\kappa=0\,$, it is straightforward to derive the conserved Runge-Lenz vector since this case is exactly equivalent to the Wu-Yang case, an imbedded Abelian monopole. For $\,\kappa\neq0\,,\pm1$,
we derived a new conserved charge, namely $\,\Gamma\,$, which has an unconventional form, see (\ref{Gamma}). In the limit case $\,\k=0\,$, this conserved charge reduces to $\,\Gamma=Q^{2}\,$; while for $\,\k=\pm1\,$, we obtain $\,\Gamma\,\sim\vL\cdot\vec{\I}\,$. 

Let us emphasize that the derivation of the non-Abelian field configuration (\ref{diatfields})
from molecular physics \cite{MSW} indicates  that  our analysis may not be of purely academic
interest. The situation could well be analogous to what happened before with the non-Abelian
Aharonov-Bohm experiment, first put forward and studied theoretically in \cite{NABAWY,NABAHPA}, but which became recently accessible experimentally, namely by applying laser beams to cold atoms \cite{Ohberg2,Ohberg,Dalibard}. A similar technique can be used to create non-Abelian monopole-type fields \cite{Dalibard}.

\section{Supersymmetric extension of the van Holten algorithm}\label{chap:SUSY}
{\normalsize
\textit{
We investigate the super and dynamical symmetries of a fermion in external magnetic fields using a SUSY extension of the van Holten framework, based on Grassmann-valued Killing tensors. 
}}

\subsection{Supersymmetry of the monopole}\label{SUSYMonop}


In this section, we investigate the super- and the dynamical symmetries of fermions in a $\,D\,$-dimensional monopole background. Following an interesting result of D'Hoker and Vinet  \cite{DV1}, a non-relativistic spin-$\frac{1}{2}$ charged particle with  gyromagnetic ratio $\,\gyro=2\,$ interacting with a point magnetic monopole, admits an $\,\osp(1|2)\,$ supersymmetry. This was also seen in the following papers \cite{ GvH,DJ-al,RvH,HMH,Pl,Plyu,Leiva,HmonRev}.

Later, Feh\'er \cite{Feher:1987} has shown that a $\,\gyro=2\,$ spin-particle in a monopole field does not admit a Runge-Lenz type dynamical symmetry.

Another, surprising, result of D'Hoker and Vinet \cite{DV4,DV2,DV3} says, however, that a 
non-relativistic spin-$\frac{1}{2}$ charged particle with \emph{anomalous gyromagnetic ratio} $\gyro=\,4\,$, interacting with a point magnetic monopole plus a Coulomb plus a fine-tuned inverse-square potential, does have such a dynamical symmetry. This  is to be compared with the one about the ${\rm O}(4)$ symmetry of a scalar particle in such a combined field \cite{MIC,Zwanziger}. Replacing the scalar particle by a spin $1/2$ particle
with gyromagnetic ratio $\gyro=0$, one can prove that
two anomalous systems, the one with $\gyro=4$ and the one with $\gyro=0$ are, in fact, superpartners \cite{FHsusy}. Note that
for both particular $\gyro$-values, one also 
has an additional ${\rm o}(3)$ ``spin'' symmetry.

On the other hand, it has been shown by Spector \cite{Spector} that the $\,\N=1\,$ supersymmetry only allows $g=2$ and no scalar potential. Runge-Lenz and SUSY appear, hence, inconsistent.

We study the bosonic as well as supersymmetries of the Pauli-type Hamiltonian,
\beq
\H_{\gyro}=\frac{\vPi^2}{2}-\frac{e\gyro}{2}\,\vS\cdot\vB+V(r)\,,\quad\vPi=\vp-e\,\vA\,,\label{Hamiltonian}
\eeq
which describes the motion of a fermion with spin $\,\vS\,$ and electric charge $\,e\,$, in  the combined magnetic field, $\,\vB\,$, plus a spherically symmetric scalar field $V(r)$, which also includes a Coulomb term (a ``dyon'' in what follows).
In the Hamiltonian (\ref{Hamiltonian}), $\,\vPi\,$ denotes the gauge covariant momentum and the constant parameter $\,\gyro\,$ represents the gyromagnetic ratio of the spinning particle.

Let us first describe the Hamiltonian dynamics, defined by (\ref{Hamiltonian}), of the charged spin-$\frac{1}{2}$ particle, moving in the flat manifold $\,\M^{D+d}\,$. Note that $\,\M^{D+d}\,$ is the extension of the bosonic configuration space $\,\M^{D}\,$ by a $\,d\,$-dimensional internal space carrying the fermionic degrees of freedom \cite{Cariglia}. The $(D+d)$-dimensional space $\,\M^{D+d}\,$ is described by the local coordinates $\left(\,x^{\mu},\,\psi^{a}\right)$ where $\,\mu=1,\cdots,D\,$ and $\,a=1,\cdots,d\,$. The motion of the spin-particle  is, therefore, described by the curve
\beq
\tau\rightarrow\left(\,x(\tau),\,\psi(\tau)\right)\,\in\,\M^{D+d}\,.
\eeq

We choose $\,D=d=3\,$ and we focus our attention to the spin-$\frac{1}{2}$ charged particle interacting with the static $\,U(1)\,$ monopole background,
\beq\displaystyle
\vB=\vnabla\times\vA=\frac{q}{e}\frac{\vx}{r^3}\,,
\eeq
so that the system is defined by the Hamiltonian (\ref{Hamiltonian}). We introduce the covariant hamiltonian formalism extending van Holten's framework to fermions. The basic phase-space reads $\,\left(x^{j},\Pi_{j},\psi^{a}\right)\,$, where the variables $\,\psi^{a}\,$ transform as tangent vectors and satisfy the Grassmann algebra,
\beq
\psi^{i}\psi^{j} + \psi^{j}\psi^{i}=0\,.
\eeq
The internal angular momentum of the particle 
can also be described
in terms of vector-like Grassmann variables,
\beq
S^{j}=-\frac{i}{2}\epsilon^j_{\,\,kl}\psi^{k}\,\psi^{l}\,.
\eeq
Defining the covariant Poisson-Dirac brackets for functions $\,f\,$ and $\,h\,$ of the phase-space as
\beqa
\big\{f,h\big\}&=&\p_j f\,\frac{\p h}{\p \Pi_j}-\frac{\p f}{\p \Pi_j}\,\p_j h 
+eF_{ij}\,\frac{\p f}{\p \Pi_i}\frac{\p h}{\p \Pi_j}
+i(-1)^{a^{f}}\frac{\p f}{\p \psi^a}\frac{\p h}{\p \psi_{a}}\,,\label{PBrackets}
\eeqa
where $\,a^{f}=\left(0,1\right)\,$ is the Grassmann parity of the phase-space function $\,f\,$ and the magnetic field reads $\,B_{i}=(1/2)\epsilon_{ijk}F_{jk}\,$. It is straightforward to obtain the non-vanishing fundamental brackets,
\beqa
\big\{x^{i},\,\Pi_{j}\big\}=\d^{i}_{j},\quad\big\{\Pi_{i},\,\Pi_{j}\big\}=e\,F_{ij},\quad\big\{\psi^{i},\,\psi^{j}\big\}=-i\,\d^{ij}\,,\\[7pt]
\big\{S^{i},\,G^{j}\big\}=\epsilon^{\;\,ij}_{k}\,G^{k}\quad\hbox{with}\quad G^{k}=\psi^{k},\,S^{k}\,.
\eeqa
It follows that, away from the monopole's  location,  the Jacobi identities are verified \cite{Jackiw84,Chaichian}.
Thus, the equations of motion  can  be obtained in this covariant Hamiltonian framework \footnote{The dot means derivative w.r.t. the evolution parameter, $\,\frac{d}{d\tau}\,$.},
\beqa
\dot{\vG}=
\frac{e\gyro}{2}\,\vG\times\vB\,,
\label{EqM}\\[4pt]
\dot{\vPi}=
e\,\vPi\times\vB-\vnabla{V(r)}+\frac{e\gyro}{2}\,\vnabla{\left(\vS\cdot\vB\right)}\,.\label{Lorentz}
\eeqa
Equation (\ref{EqM}) shows  that the fermionic vectors $\,\vS\,$ and $\,\vpsi\,$ are conserved when the spin and the magnetic field are uncoupled, i.e. for \emph{vanishing gyromagnetic ratio}, $\,\gyro=0\,$.  Note that, in addition to the magnetic field term, the Lorentz equation (\ref{Lorentz}) also involves a potential term augmented with a spin-field interaction term (Stern and Gerlach term).

We now proceed by deducing, in a classical framework, the supersymmetries and conservation laws of the system (\ref{Hamiltonian}), using the SUSY extension of the van Holten algorithm \cite{Ngome:2010gg} developed in section \ref{vHAl}.
What is new here is that the generators of SUSY are Grassmann-valued Killing tensors.
We expand the phase-space function, associated with one (super)symmetry, in powers of the covariant momenta,
\beq
\Q\left(\vx,\,\vPi,\,\vpsi\right)=C(\vx,\,\vpsi)+\sum_{k=1}^{n-1}\,\frac{1}{k!}\,C^{i_1\cdots i_k}(\vx,\,\vpsi)\,\Pi_{i_1}\cdots\Pi_{i_k}\,.
\label{Exp}
\eeq
Note the dependence on Grassmann variables of the tensors $\,C(\vx,\,\vpsi)\,$.
Requiring that $\Q$ Poisson-commutes with the Hamiltonian,
$\big\{\H_{\gyro},\Q\big\}=0\,,$
implies the series of constraints,  
\beq
\begin{array}{llll}\displaystyle{
C_{i}\p_i V+\frac{ie\gyro}{4}\psi^l\psi^m C_j\p_j F_{lm}-\frac{e\gyro}{2}\psi^m\frac{\p C}{\p\psi^a}F_{am}
=0},&\hbox{o(0)}
\\[8pt]\displaystyle{
\p_jC=C_{jk}\p_{k}V+eF_{jk}C_k+\frac{ie\gyro}{4}\psi^l\psi^m C_{jk}\p_k F_{lm}-\frac{e\gyro}{2}\psi^m\frac{\p C_j}{\p\psi^a}F_{am}},
&\hbox{o(1)}
\\[11pt]\displaystyle{
\p_{\left(j\right.}C_{\left.k\right)}=C_{jkm}\p_{m}V+e\left(F_{jm}C_{mk}+F_{km}C_{mj}\right)}\\[8pt]
\displaystyle
\qquad\qquad\qquad\qquad+\frac{ie\gyro}{4}\psi^l\psi^m C_{ijk}\p_i F_{lm}-\frac{e\gyro}{2}\psi^m\frac{\p C_{jk}}{\p\psi^a}F_{am},
&\hbox{o(2)}\\[10pt]
\p_{\left(j\right.}C_{\left.kl\right)}= C_{jklm}\p_{m}V+e\left(F_{jm}C_{mkl}+F_{lm}C_{mjk}+F_{km}C_{mlj}\right)
\\[8pt]
\quad\qquad\qquad\qquad\qquad+\;\displaystyle{\frac{ie\gyro}{4}\psi^m\psi^n C_{ijkl}\p_i F_{mn}-\frac{e\gyro}{2}\psi^m\frac{\p C_{jkl}}{\p\psi^a}F_{am}}\,,
&\hbox{o(3)}
\\
\vdots\qquad\qquad\qquad\qquad\qquad\vdots&\vdots
\end{array}
\label{SUSYconstraints}
\eeq
This series of constraint can be truncated at a finite order $\,\textit{n}\,$ provided \textit{the higher order constraint becomes a Killing equation}. The zeroth-order equation can be interpreted as a  \textit{consistency condition between the potential and the (super)invariant}. Apart from the zeroth-order constants of the motion, i.e., such that do not depend on the  momentum, all other order-\textit{n} (super)invariants are deduced by the systematic method (\ref{SUSYconstraints}) implying rank-\textit{n} Killing tensors. Each Killing tensor solves the higher order constraint of (\ref{SUSYconstraints}) and can therefore generate a (super)invariant.

We focus our attention on searching for conserved quantities which are linear or quadratic in the covariant momenta.
Thus, we have to determine generic Grassmann-valued Killing tensors of rank-one and rank-two.
 
$\bullet$ Let us first consider the Killing equation,
\beq
\p_jC^{k}(\vx,\,\vpsi)+\p_kC^{j}(\vx,\,\vpsi)=0\,.\label{KConstraint1}
\eeq
Following Berezin and Marinov \cite{B-M}, any tensor which takes its values in the Grassmann algebra may be represented as a finite sum of homogeneous monomials,
\beq
C^{i}(\vx,\,\vpsi)=\sum_{k\geq 0}\C^{i}_{a_{1}\cdots a_{k}}(\vx)\psi^{a_{1}}\cdots\psi^{a_{k}}\,,\label{ExternalTensor}
\eeq
where the coefficients tensors, $\C^{i}_{a_{1}\cdots a_{k}}$, are completely anti-symmetric in the fermionic indices $\,\left\lbrace a_{k}\right\rbrace\,$. The tensors (\ref{ExternalTensor}) satisfy (\ref{KConstraint1}), from which we deduce that their (tensor) coefficients  satisfy
\beq
\left(\p_j\C^{k}_{a_{1}\cdots a_{m}}(\vx)+\p_k\C^{j}_{a_{1}\cdots a_{m}}(\vx)\right)\psi^{a_{1}}\cdots\psi^{a_{m}}=0\;\;\Longrightarrow\;\;\p_{i}\p_{j}\C^{k}_{a_{1}\cdots a_{m}}(\vx)=0\,,\label{KConstraint1bis}
\eeq
providing us with the most general rank-1 Grassmann-valued Killing tensor
\beq
C^{i}(\vx,\,\vpsi)=\sum_{k\geq 0}\big(M^{ij}\,x^{j}+N^{i}\big)_{a_{1}\cdots a_{k}}\psi^{a_{1}}\cdots\psi^{a_{k}}\,,\quad M^{ij}=-M^{ji}\,,\label{Exp2}
\eeq
where $\,N^{i}\,$ and the antisymmetric $\,M^{ij}\,$ define constant tensors.

$\bullet$ Let us now construct the rank-2 Killing tensors which solve the Killing equation,
\beq
\p_jC^{kl}(\vx,\,\vpsi)+\p_lC^{jk}(\vx,\,\vpsi)+\p_kC^{lj}(\vx,\,\vpsi)=0\,.\label{KConstraint2}
\eeq
Considering the expansion in terms of Grassmann degrees of freedom \cite{B-M} of the Killing tensor $\,C^{jk}(\vx,\,\vpsi)\,$, we get the coefficients tensors $\,\C ^{ij}_{a_{1}\cdots a_{k}}\,$ which are constructed as symmetrized products \cite{GR} of Yano-type Killing tensors, $\C^{i}_{\,Y}(\vx)\,$, associated with the rank-1 Killing tensors $\,\C ^{i}(\vx)\,$ obtained by (\ref{KConstraint1bis}),
\beq
\C^{ij}_{a_{1}\cdots a_{k}}(\vx)=\frac{1}{2}\left(\C^{i}_{\,Y}\widetilde{\C}^{jY}+\widetilde{\C}^{i}_{\,Y}\C^{jY}\right)_{a_{1}\cdots a_{k}}\,.\label{Coef}
\eeq
It is worth noting that the Killing tensor defined in (\ref{Coef}) is symmetric in its bosonic indices and anti-symmetric in the fermionic indices. Thus, we obtain
\beq\begin{array}{lc}
\displaystyle
C^{ij}(\vx,\,\vpsi)=\sum_{k\geq 0}\left(
M^{(i}_{\;ln}\widetilde{M}^{j)\,n}_{\;m}x^lx^m+M^{(i}_{\;ln}\widetilde{N}^{j)\,n}x^l\right.\\[8pt]
\displaystyle
\left.\qquad\qquad\qquad\qquad\qquad+N^{(i}_{\;n}\widetilde{M}^{j)\,n}_{\;m}x^m+N^{(i}_{\;n}\widetilde{N}^{j)\,n}\right)_{a_{1}\cdots a_{k}}\psi^{a_{1}}\cdots\psi^{a_{k}}\,,
\end{array}\label{Exp3}
\eeq
where $\,M^{ij}_{\;\;k}\,$, $\,\widetilde{M}^{ij}_{\;\;k}\,$, $\,N^{j}_{\;k}\,$ and $\,\widetilde{N}^{j}_{k}\,$ are skew-symmetric constants tensors. Then one can verify with direct calculations that (\ref{Exp2}) and (\ref{Exp3}) satisfy Killing equations.\\

Having constructed the generic Killing tensors (\ref{Exp2}) and (\ref{Exp3}) generating constants of the motion, we can now describe the supersymmetries of the Pauli-like Hamiltonian (\ref{Hamiltonian}). To start, we search for momentum-independent invariants, i.e. which are not derived from a Killing tensor, $\,C^{i}= C^{ij}= \cdots = 0\,$. In that event, the system of equations (\ref{SUSYconstraints}) reduces to the two constraints,
\beqa\left\lbrace
\begin{array}{lll} \displaystyle
\gyro\psi^m\frac{\p \Q_{c}(\vx,\vpsi)}{\p\psi^a}\,F_{am} =0\,,\qquad &\hbox{o(0)}&
\\[8pt]\displaystyle
\p_i\Q_{c}(\vx,\vpsi)=0\,.&\hbox{o(1)}&
\end{array}
\right.
\eeqa
For $\,\gyro=0\,$, which means no spin-gauge field coupling, it is straightforward to see that the spin vector, in particular, and all arbitrary functions $\,f\big(\vpsi\,\big)\,$ which depend only on the Grassmann variables are conserved along the motion.

\textit{For nonvanishing gyromagnetic ratio $\gyro$, only the ``chiral'' charge}
\beq
\Q_{c}=\vpsi\cdot\vS\,
\eeq
remains  conserved. The ``chiral'' charge $\,\Q_{c}\,$ can be considered as the projection of the internal angular momentum, $ \vS$, onto the internal trajectory $\,\psi(\tau)\,$. Thus, $\,\Q_{c}\,$ can be viewed as the internal analogue of the projection of the angular momentum, in bosonic sector, onto the classical trajectory $\,x(\tau)\,$.

Let us now construct superinvariants linear in the covariant momentum. $C^{ij}=\cdots =0\,$ such that (\ref{SUSYconstraints}) becomes
\beqa\left\lbrace \begin{array}{lll} 
 \displaystyle{ C^{i}\,\p_i V+\frac{ie\gyro}{4}\,\psi^l\psi^m C^j(\vx,\vpsi)\,\p_j F_{lm}-\frac{e\gyro}{2}\psi^m\frac{\p C(\vx,\vpsi)}{\p\psi^a}\,F_{am}
=0}\,,&\hbox{o(0)}& 
\\[7pt]
\displaystyle{
\p_jC(\vx,\vpsi)=eF_{jk}C^k(\vx,\vpsi)-\frac{e\gyro}{2}\psi^m\frac{\p C^j(\vx,\vpsi)}{\p\psi^a}\,F_{am}}\,, &\hbox{o(1)}& 
\\[8pt]
\p_jC^k(\vx,\vpsi)+\p_kC^j(\vx,\vpsi)=0 \,.&\hbox{o(2)}&
\label{AngMom}
\end{array}\right.
\eeqa
Choosing the non-vanishing term $\,N^{j}_{a}=\d^{j}_{a}\,$, in the general rank-1 Killing tensor (\ref{Exp2}), provides us with the rank-1 Killing tensor generating the supersymmetry transformation, 
\beq
\,C^j(\vx,\vpsi)=\d^{j}_{a}\,\psi^{a}\,.
\label{SUSYKILLING}
\eeq
By substitution of this Grassmann-valued Killing tensor into the first-order equation of (\ref{AngMom}) we get 
\beq
\vnabla C(\vx,\vpsi)=\frac{q}{2}\left(\gyro-2\right)\frac{\vx\times\vpsi}{r^3}\,.\label{PrExp}
\eeq
Consequently, a solution 
$C(\vx,\vpsi)=0$ of (\ref{PrExp}) is only obtained for a fermion with ordinary gyromagnetic ratio 
\beq
\gyro=2\,.
\label{g2}
\eeq
Thus we obtain, for $\,V(r)=0\,$, the \textit{Grassmann-odd supercharge} generating the $\N=1$ supersymmetry of the spin-monopole field system,
\beq
\Q=\vpsi\cdot\vPi\,,
\qquad
\big\{\Q,\,\Q\big\}=-2i\H_{2}\,.
\label{SC0}
\eeq
 
For nonvanishing potential, $V(r)\neq 0\,$, the zeroth-order consistency condition of (\ref{AngMom}) is expressed as \footnote{We use the identity 
$\,S^kG^j\p_j B^k=\psi^l\psi^m\,G^j\p_j F_{lm}=0\,$.}
\beq
V'(r)\frac{\vpsi\cdot\vx}{r}=0\,.
\eeq
Consequently, adding \emph{any} spherically symmetric potential $V(r)\,$ breaks the supersymmetry  generated by the Killing tensor $C^j=\d^{j}_{a}\,\psi^{a}$~:
$\N=1$ SUSY requires an ordinary gyromagnetic factor, and no additional radial potential 
is allowed \cite{Spector}.

Another Killing tensor deduced from (\ref{Exp2}) is obtained by considering the particular case with the non-null tensor $ \,N^{j}_{\;a_1a_2}=\epsilon^{j}_{\;a_1a_2}\,
$. This leads to the rank-1 Killing tensor, 
\beq
C^j(\vx,\vpsi)=\nobreak\epsilon^{j}_{\;ab}\psi^{a}\psi^{b}\,.
\eeq
In this case, the first-order constraint of (\ref{AngMom}) is solved by$\,C(\vx,\vpsi)=0\,$, provided the gyromagnetic ratio takes the value $\,\gyro=2\,$. For vanishing potential, it is straightforward to verify the zeroth-order consistency constraint and therefore to obtain \textit{the Grassmann-even supercharge},
\beq
\Q_{1}=\vS\cdot\vPi\,,\label{evenS1}
\eeq
\textit{defining the ``helicity" of the spinning particle}. As expected, the consistency condition of superinvariance under (\ref{evenS1}) is again violated for $\,V(r)\neq 0\,$, breaking the supersymmetry of the Hamiltonian $\,\H_{2}\,$, in (\ref{SC0}).

Let us now consider the rank-1 Killing vector, 
\beq
C^{j}(\vx,\vpsi)=\big(\vS\times\vx\big)^{j}\,,
\eeq 
 obtained by putting
$M^{ij}_{\;a_1a_2}=(i/2)\epsilon^{kij}\,\epsilon_{ka_1a_2}$ into the generic rank-1 Killing tensor (\ref{Exp2}). The first-order constraint 
is satisfied with $\,C(\vx,\vpsi)=0\,$, provided the particle carries  gyromagnetic ratio $\gyro=2$. Thus, we obtain the supercharge,
\beq 
\Q_{2}=(\vx\times\vPi)\cdot\vS\,,\label{q2}
\eeq
which, just like those in (\ref{SC0}) and (\ref{evenS1}) only appears when the potential is absent, $V=0$.

We consider the SUSY given when 
$\, M^{ij}_{\;a}=\epsilon^{\;\,ij}_{a}\,
$ so that the Killing tensor (\ref{Exp2}) reduces to 
\beq
C^{j}(\vx,\vpsi)=-\epsilon^{j}_{\;ka}x^{k}\psi^{a}\,.
\eeq 
The first-order constraint of (\ref{AngMom}) is solved with  $\,\displaystyle{C(\vx,\vpsi)=\frac{q}{2}\left(\gyro-2\right)\frac{\vpsi\cdot\vx}{r}}\,$. The zeroth-order consistency condition is, in this case, identically satisfied for an arbitrary  radial potential. We have thus constructed the Grassmann-odd supercharge,
\beq
\Q_{3}=(\vx\times\vPi)\cdot\vpsi+\frac{q}{2}\left(\gyro-2\right)\frac{\vpsi\cdot\vx}{r}\,,
\eeq
which is still conserved for a particle carrying an arbitrary gyromagnetic ratio $\gyro\,$. Note, that this supercharge generalizes the one obtained in the restricted case with $\gyro=2$ \cite{DJ-al}. See also \cite{HMH}.

Now we  turn to invariants which are quadratic in the covariant momentum. For this, we solve the reduced series of constraints,
\beqa\left\lbrace
\begin{array}{llll}\displaystyle{C^{i}\p_i V+\frac{ie\gyro}{4}\,\psi^l\psi^m C^j\p_j F_{lm}-\frac{e\gyro}{2}\psi^m\frac{\p C}{\p\psi^a}F_{am}
=0},&\hbox{o(0)}
\\[8pt]\displaystyle{
\p_jC=C^{jk}\p_k V+eF_{jk}C^k+\frac{ie\gyro}{4}\psi^l\psi^m\,C^{jk}\p_k F_{lm}-\frac{e\gyro}{2}\psi^m\frac{\p C^{j}}{\p\psi^a}\,F_{am}},
&\hbox{o(1)}
\\[8pt]\displaystyle{
\p_jC^k+\p_kC^j=e\left(F_{jm}C^{mk}+F_{km}C^{mj}\right)-\frac{e\gyro}{2}\psi^m\frac{\p C^{jk}}{\p\psi^a}\,F_{am}}\,,
&\hbox{o(2)}\\[8pt]\displaystyle
\p_{j}C^{km}+\p_{m}C^{jk}+\p_{k}C^{mj}=0\,.&\hbox{o(3)}
\end{array}\right.
\label{RLV}
\eeqa
We first observe that $\,C^{ij}(\vx,\vpsi)=\d^{ij}$ is a constant Killing tensor. Solving the second- and the first-order constraints of (\ref{RLV}), we obtain
\beq
C^{j}(\vx,\vpsi)=0\quad\hbox{and}\quad\displaystyle
C(\vx,\vpsi)=V(r)-\frac{e\gyro}{2}\vS\cdot\vB\,,
\eeq
respectively. The zeroth-order consistency condition is identically satisfied so we obtain the conserved energy of the spinning particle,
\beq
\E= \frac{1}{2}\vPi^{2}-\frac{e\gyro}{2}\vS\cdot\vB+V(r)\,.
\eeq
 
Next, we introduce the nonvanishing constants tensors,
$\,
M^{ijk}\!=\!\epsilon^{ijk}\,,\;\widetilde{N}^{ij}_{\;\,a}\!=\!-\epsilon_{\;\,a}^{ij}\,$, into (\ref{Exp3}) in order to derive the rank-2 Killing tensor with the property,
\beq
C^{jk}(\vx,\vpsi)=2\,\d^{jk}(\vx\cdot\vpsi)-x^j\psi^k-x^k\psi^j\,.\label{KTSUSY}
\eeq
Injecting the Killing tensor (\ref{KTSUSY}) into (\ref{RLV}), we satisfy the second-order constraints with 
\beq\displaystyle
\vC(\vx,\vpsi)=\frac{q}{2}\left(2-\gyro\right)\frac{\vpsi\times\vx}{r}\,.
\eeq 
To deduce the integrability condition of (\ref{RLV}), we require, in the first-order constraint, the vanishing of the commutator,
\beq
\left[\p_{i},\,\p_{j}\right]C(\vx)=0\quad\Longrightarrow\quad\Delta\left(V(r)-\left(2-\gyro\right)^2\frac{q^2}{8r^2}\right)=0\,.\label{Laplace0}
\eeq
Then the Laplace equation (\ref{Laplace0}) provides us with the \textit{most general form of the potential admitting a Grassmann-odd charge quadratic in the velocity}, namely with
\beq
\displaystyle{V(r)=\left(2-\gyro\right)^2\frac{q^2}{8r^2}+\frac{\a}{r}+\b}\,.\label{PotSUSY}
\eeq 
Consequently, we solve the first-order constraint with 
\beq
C(\vx,\vpsi)=\left(\frac{\a}{r}-e\gyro\vS\cdot\vB\right)\vx\cdot\vpsi\,,\label{NewSUSY}
\eeq
so that the zeroth-order consistency constraint is identically satisfied. Collecting our results leads to the Grassmann-odd conserved charge quadratic in the velocity \cite{Ngome:2010gg}, 
\beq
\Q_{4}=\left(\vPi\times(\vx\times\vPi)\right)\cdot\vpsi+\frac{q}{2}\left(2-\gyro\right)\frac{\vx\times\vPi}{r}\cdot\vpsi +\left(\frac{\a}{r}-e\gyro\vS\cdot\vB\right)\vx\cdot\vpsi\,.
\eeq
Let us underline that the conserved charge $\,\Q_4\,$ which is \emph{not} a square root of the Hamiltonian $\,\H_g\,$ remains conserved without restriction on the gyromagnetic factor, $\gyro\,$. We can also remark that for $\,\gyro=0\,$, this charge coincides with the scalar product of the \textit{separately conserved Runge-Lenz vector \footnote{The case of spinning particle with null gyromagnetic ratio, $\,\gyro=0\,$, coincides with a spinless particle.} }\cite{MIC,Zwanziger} \textit{by the Grassmann-odd vector}:
\beq
\left. \Q_{4} \right|_{g = 0} = \vK_{s = 0}\cdot\vpsi\,.
\eeq 

The supercharges $\,\Q\,$ and $\,\Q_j\,$ with $\,j= 0,\cdots,3\,$, previously determined, form together, for ordinary gyromagnetic ratio, the classical superalgebra,
\beqa\begin{array}{lll}\displaystyle{
\big\{\Q_0,\,\Q_0\big\}=\big\{\Q_0,\,\Q_1\big\}=\big\{\Q,\,\Q_1\big\}=\big\{\Q_1,\,\Q_1\big\}=\big\{\Q_2,\,\Q_2\big\}=0}\,,\\[8pt]\displaystyle{

\big\{\Q_0,\,\Q\big\}=i\Q_{1}\,,\quad\big\{\Q_0,\,\Q_2\big\}=\big\{\Q_2,\,\Q_3\big\}=0}\,,\\[8pt]\displaystyle{

\big\{\Q_0,\,\Q_3\big\}=i\Q_{2}\,,\quad
\big\{\Q,\,\Q\big\}=-2i\H_{2}}\,,\\[8pt]\displaystyle{

\big\{\Q,\,\Q_2\big\}=\big\{\Q_1,\,\Q_3\big\}=\Q_4}\,,\\[8pt]\displaystyle{

\big\{\Q,\,\Q_3\big\}=2i\Q_1\,,\quad\big\{\Q_1,\,\Q_2\big\}=i\Q_3\Q\,,\quad\big\{\Q_3,\,\Q_3\big\}=i\left(2\Q_{2}-\Q_5\right)}\,,
\end{array}\label{SusyAlgbra}
\eeqa
where $\Q_5$ is a bosonic supercharge that we will construct below [\ref{Msquare}]. From (\ref{SusyAlgbra}) it follows that the linear combination
$\Q_Y = \Q_3 - 2 \Q_0$ has the special property that its bracket with the standard supercharge $\Q$ vanishes:
\beq
\big\{ \Q_Y, \Q \big\} = 0.
\eeq
Indeed, $\Q_Y$ is precisely the Killing-Yano supercharge constructed by De Jonghe, Macfarlane, Peeters and van Holten \cite{DJ-al}.\\

Let us now investigate the bosonic symmetries of the Pauli-like Hamiltonian (\ref{Hamiltonian}). We use the generic Killing tensors previously constructed [ cf. (\ref{Exp2}) and (\ref{Exp3})] to derive the associated bosonic constants of the motion.

 Firstly, we describe the rotationally invariance of the system by solving the reduced series of constraints (\ref{AngMom}).
For this, we consider the Killing vector provided by the replacement,
$\,
M^{ij}=-\epsilon^{ij}_{\;\;k}n^{k}\,
$ 
into  (\ref{Exp2}). Thus for any unit vector $\,\vn\,$, we obtain the generator of space rotations around $\,\vn\,$,
\beq
\vC(\vx,\vpsi)=\vn\times\vx\,.\label{KillingV} 
\eeq
Inserting the previous Killing vector in the first-order equation of (\ref{AngMom}) yields
\beq
C(\vx,\vpsi)=c(\vpsi)-q\frac{\vn\cdot\vx}{r}\,.
\eeq
Moreover the zeroth-order consistency condition of (\ref{AngMom}) requires for arbitrary radial potential,
\beq
c(\vpsi)=\vS\cdot\vn\,.
\eeq
Collecting our results provides us with the total angular momentum, which is plainly conserved for arbitrary gyromagnetic ratio,
\beq
\displaystyle
\vJ=\vL+\vS=
\vx\times\vPi-q\,\hx+\vS\,.
\label{AngMomentum}
\eeq
In addition to the typical monopole term, the conserved angular momentum also involves the spin vector, $\,\vS\,$. It
 generates an $\,\ort(3)_{rotations}\,$ bosonic symmetry algebra,
\beq
\big\{J^{i},J^{j}\big\}=\epsilon^{ijk}J^{k}\,.
\eeq
In the particular case of vanishing gyromagnetic factor $\gyro=0$, the usual monopole angular momentum $\vL\,$ and the internal spin angular momentum $\vS\,$ are separately conserved involving an
\beq
\ort(3)_{rotations}\oplus \ort(3)_{spin}
\eeq
symmetry algebra\,.\\

We turn into invariants which are quadratic in the covariant momenta. Then, we have to solve the series of constraints (\ref{RLV}). We first observe that for $\,M^{jmk}\!=\!\widetilde{M}^{jmk}\!=\!\epsilon^{jmk}\,$,
the Killing tensor (\ref{Exp3}) reduces to the rank-2 Killing-St\"ackel tensor,
\beq
C^{ij}(\vx,\vpsi)=2\d^{ij}\,\vx^{\,2}-2x^{i}x^j\,.\label{Momsquare}
\eeq
Inserting (\ref{Momsquare}) into the second- and in the first-order constraints of (\ref{RLV}), we get for any gyromagnetic factor and for any arbitrary radial potential,
\beq
\vC(\vx,\vpsi)= 0\quad\hbox{and}\quad C(\vx,\vpsi)=-\gyro q\,\frac{\vx\cdot\vS}{r}\,.
\eeq
Hence, we obtain the Casimir
\beq
\Q_{5}=\vJ^{2}-q^2+\left(\gyro-2\right)\vJ\cdot\vS-\gyro\Q_{2}\,.\label{Msquare}
\eeq
The bosonic supercharge $\,\Q_{5}$ is, as expected, \textit{the square of the total angular momentum, augmented with another, separately conserved charge} \cite{Ngome:2010gg},
\beq
\left(\gyro-2\right)\vJ\cdot\vS-\gyro\Q_{2}\,.\label{Msquarebis}
\eeq

\noindent
$\bullet$ Indeed, for $\gyro=0\,$, (\ref{Msquarebis}) directly implies that the product, $\,\vJ\cdot\vS\,$, and hence the spin vector, $\,\vS\,$, are separately  conserved.

\noindent
$\bullet$ For $\gyro=2\,$, we recover the conservation of the supercharge $\,\Q_{2}\,$ [ cf. (\ref{q2})].

\noindent
$\bullet$ For the anomalous gyromagnetic ratio  $\gyro=4\,$, we obtain that
$\,
\vJ\cdot\vS-2\Q_{2}
\,$ is a constant of the motion.\\

Now we are interested in the hidden symmetry generated by a conserved Laplace-Runge-Lenz-type vector. Then, we introduce into the algorithm (\ref{RLV}) the generator,
\begin{eqnarray}\displaystyle{ C^{ij}(\vx,\vpsi)= 2\,\d^{ij}\,\vn\cdot\vx-n^{i}x^j-n^{j}x^i}\,,
\label{RL1}
\end{eqnarray}
easily obtained by choosing the non-vanishing,
$\,\widetilde{N}^{ij}\!=\!\epsilon^{imj}n^m\;\hbox{and}\;M^{ijm}\!=\!\epsilon^{ijm}\,$, into the generic rank-2 Killing tensor (\ref{Exp3}).
Inserting (\ref{RL1}) into the second-order constraint of (\ref{RLV}), we get
\beq
\vC(\vx,\vpsi)= q\frac{\vn\times\vx}{r}+\vC(\vpsi\,)\,.
 \label{RL2}
\eeq
We solve the first-order constraint of (\ref{RLV}) by expanding $\,C(\vx,\,\vpsi)\,$ in terms of Grassmann variables \cite{B-M},
\beq
C(\vx,\,\vpsi)=C(\vx)+\sum_{k\geq 1}C_{a_{1}\cdots a_{k}}(\vx)\psi^{a_{1}}\cdots\psi^{a_{k}}\,.\label{Expansion}
\eeq
Consequently, the first- and the zeroth-order equations of (\ref{RLV}) can be classified order-by-order in Grassmann-odd variables. Thus, inserting  (\ref{RL2}) in the first-order equation, and  requiring again the vanishing of the commutator,
\beq
\left[\p_{i},\,\p_{j}\right]C(\vx)=0\;\Longrightarrow\;\Delta\left( V(r)-\frac{q^2}{2r^2}\right)=0\,,
\label{Laplace}
\eeq
we deduce the most general radial potential admitting a conserved Laplace-Runge-Lenz vector in the fermion-monopole interaction, namely
\beq
V(r)=\frac{q^2}{2r^2}+\frac{\mu}{r}+\gamma\,,\quad\mu\,,\gamma\in\IR\,.\label{Potential}
\eeq
Investigating the first term on the right-hand side of (\ref{Expansion}), we obtain
\beq
\displaystyle
C(\vx)=\mu\frac{(\vn\cdot\vx)}{r}\,.
\eeq
Introducing (\ref{RL2}) and (\ref{Potential}) into the first-order constraint of (\ref{RLV}), on one hand, provides us with
\beq
\displaystyle
\vC(\vpsi\,)=-\frac{\gyro}{2}\vn\times\vS\,,
\eeq
and on the other hand with
\beq\begin{array}{cc}
\displaystyle
\sum_{k\geq 1}C_{a_{1}\cdots a_{k}}(\vx)\psi^{a_{1}}\cdots\psi^{a_{k}}=-\frac{e\gyro}{2}\left(\vS\cdot\vB\right)\left(\vn\cdot\vx\right)-\frac{\gyro q}{2}\left(1-\frac{\gyro}{2}\right)\frac{\vn\cdot\vS}{r}+C(\vpsi)\,,\\[19pt]
\hbox{with}\quad\gyro\big(\gyro-4\big)=0\,.
\end{array}
\label{RL3}
\eeq
Let us precise that the zeroth-order consistency condition of (\ref{RLV}) is only satisfied for 
\beq
\displaystyle
C(\vpsi)=\frac{\mu}{q}\vS\cdot\vn\,.
\eeq
Collecting our results, (\ref{RL1}), (\ref{RL2}), (\ref{Potential}) and (\ref{RL3}), we obtain a conserved Runge-Lenz vector
if and only if
\beq
\gyro=0\qquad\hbox{or}\qquad\gyro=4\,;
\label{g04}
\eeq
we get namely
\beq
\vK_{\gyro}=\vPi\times\vJ+\mu\,\hx+\left(1-\frac{\gyro}{2}\right)\vS\times\vPi-\frac{e\gyro}{2}\left(\vS\cdot\vB\right)\vx-\frac{\gyro q}{2}\left(1-\frac{\gyro}{2}\right)\frac{\vS}{r}+\frac{\mu}{q}\vS\,.
\eeq

Note that the spin angular momentum which generates the extra ``spin'' symmetry for vanishing gyromagnetic ratio is no more separately conserved for $\gyro=4$. Then, an interesting question is to know if the extra ``spin'' symmetry of $\gyro=0$
is still present for the anomalous superpartner $\gyro=4\,$ in some ``hidden'' way.

Let us consider the ``spin'' transformation generated by the rank-2 Killing tensor with the property,
\beq
C^{mk}(\vx,\vpsi)=2\d^{mk}\big(\vS\cdot\vn\big)-\frac{g}{2}\big(S^{m}n^{k}+S^{k}n^{m}\big)\,.\label{KT2}
\eeq
The rank-$2$ Killing tensor (\ref{KT2}) which can be separated as $\;C^{mk}=C^{mk}_{+}\,+\,C^{mk}_{-}\;$ is obtained by putting
\beq
\begin{array}{cc}
\displaystyle
N^{jk}_{+}=\frac{g}{2}\epsilon^{\;jk}_{l}\,n^l\,,
\qquad
\widetilde{N}^{jk}_{+\,\;a}=-\frac{i}{2}\epsilon^{jk}_{\;\,m}\,\epsilon^{m}_{\;\,a_1 a_2}\,,
\\[8pt]
\displaystyle
N^{jkl}_{-}=\big(1-\frac{g}{2}\big)\epsilon^{jkl}\,,
\qquad
\widetilde{N}^{jkl}_{-\;\,a}=-\frac{i}{4}\epsilon^{jkl}\,n_{m}\,\epsilon^{m}_{\;\,a_1 a_2}\,,
\end{array}
\eeq
into the general rank-2 Killing tensor (\ref{Exp3}). Inserting (\ref{KT2}) into the second-order constraint of (\ref{RLV}) leads to
\beq
\vC(\vx,\vpsi)=-\frac{qg}{2}\frac{\big(\vS\times\vn\big)}{r}+\vC(\psi)\quad\hbox{and}\quad g(g-4)=0\,.
\eeq
We use the potential  (\ref{Potential}) to solve the first-order equation of (\ref{RLV}),
\beq
\begin{array}{ll}
\displaystyle
C(\vx,\vpsi)=\left(2V(r)-\frac{q^{2}g^{2}}{8r^{2}}-\frac{\mu g^2}{4r}\right)\vS\cdot\vn+c(\psi)\,,
\\[12pt]
\displaystyle
\vC(\psi)=\frac{\mu g}{2q}\vn\times\vS\qquad\hbox{and}\qquad\gyro\big(\gyro-4\big)=0\,.
\end{array}
\eeq
The zeroth-order consistency condition is satisfied with
\beq
\displaystyle
c(\psi)=-\frac{\gyro^{2}}{8}\frac{\mu^{2}}{q^{2}}\vS\cdot\vn\,,
\eeq
so that collecting our results provides us with the conserved ``spin" vector,
\beq\begin{array}{ll}
\displaystyle
\vo_{\gyro}=\left(\vPi^{2}+\big(2-\frac{\gyro^{2}}{4}\big)V(r)\right)\vS-\frac{\gyro}{2}\big(\vPi\cdot\vS\big)\vPi+\frac{\gyro}{2}\big(\frac{q}{r}+\frac{\mu}{q}\big)\vS\times\vPi\\[12pt]
\displaystyle
\qquad
-\frac{\gyro^{2}}{4}\big(\frac{\mu^{2}}{2q^{2}}-\gamma\big)\vS
\qquad
\displaystyle\hbox{with}\qquad\gyro\big(\gyro-4\big)=0\,.
\end{array}
\eeq
In conclusion, the additional $\,\ort(3)_{spin}\,$ ``spin'' symmetry is recovered in the same particular cases of anomalous gyromagnetic ratios $0$ and $4$ [cf. (\ref{g04})]. 

$\bullet$ For $\gyro=0$, in particular,
\beq
\vo_{0}=2\E\,\vS\,.
\eeq

$\bullet$ For $\gyro=4$, we find an expression equivalent to that of D'Hoker and Vinet \cite{DV2}, namely 
\beq
\vo_{4}=\left(\vPi^{2}-2V(r)\right)\vS-2\big(\vPi\cdot\vS\big)\,\vPi+2\big(\frac{q}{r}+\frac{\mu}{q}\big)\vS\times\vPi-4\left(\frac{\mu^{2}}{2q^{2}}-\gamma\right)\vS\,.
\eeq
Note that this extra symmetry is generated by a
\emph{Killing tensor}, rather than a Killing vector, as for ``ordinary'' angular momentum. Thus, for sufficiently low energy, the motions are bounded and the conserved vectors $\,\vJ\,$, $\,\vK_{\gyro}\,$ and $\,\vo_{\gyro}\,$ generate an
\beq
\ort(4)\oplus \ort(3)_{spin}
\eeq
bosonic symmetry algebra.\\

So far we have seen that, for a spinning particle with a single Grassmann variable, SUSY and dynamical symmetry are inconsistent, since they require different values for the $\gyro$-factor. Now, adapting the idea of D'Hoker and Vinet to our framework, we show that the
two contradictory conditions can be conciliated by doubling the odd degrees of freedom. The
systems with $\gyro=0$ and $\gyro=4$ will then become
superpartners inside a unified  $\,\N=2\,$ SUSY system \cite{FHsusy}.

We consider, hence, a charged  spin-$\frac{1}{2}$ particle moving in a flat manifold $\,\M^{D+2d}\,$, interacting with a static magnetic field $\,\vB\,$. The fermionic degrees of freedom are now carried by a $2d$-dimensional internal space \cite{Bellucci,Kochan,Gonzales:2009ye,Avery}. This is to be compared with the $d$-dimensional internal space sufficient to describe the $\,\N=1\,$ SUSY of the monopole. In terms of Grassmann-odd variables $\,\psi_{1,2}\;$, the local coordinates of the fermionic extension $\,\M^{2d}\,$ read $\left(\psi^{a}_{1},\,\psi^{b}_{2}\right)$ with $\,a,b=1,\cdots,d\,$. The system is still described by the Pauli-like Hamiltonian (\ref{Hamiltonian}). Choosing $\,d=3\,$, we consider the fermion $\xi_{\a}\,$ which is a two-component spinor, $\,\xi_{\a}=\left(\begin{array}{c} \psi_{1}\\\psi_{2}\end{array}\right)\,$, and whose conjugate is $\bar{\xi}^{\a}\,$ \cite{Salomonson}. Thus, we have a representation of the spin angular momentum,
\beq
S^{k}=\frac{1}{2}\bar{\xi}^{\a} \,\sigma^{k\;\,\b}_{\,\a}\,\xi_{\b}\quad\hbox{with}\quad\a,\b=1,2\,,
\eeq 
where the $\,\sigma^{k\;\,\b}_{\,\a}\,$ with $\,k=1,2,3\,$ define the standard Pauli matrices. Defining the covariant Poisson-Dirac brackets as
\beq
\begin{array}{ll}
\displaystyle
\big\{f,h\big\}=\p_j f \frac{\p h}{\p\Pi_j}-\frac{\p f}{\p\Pi_j}\p_j h 
+e\epsilon_{ijk}B^k\frac{\p f}{\p\Pi_i}\frac{\p h}{\p\Pi_j}\\[12pt]
\displaystyle
\qquad\qquad+\;i(-1)^{a^{f}}\left(\frac{\p f}{\p \xi_{\a}}\frac{\p h}{\p\bar{\xi}^{\a}}+\frac{\p f}{\p \bar{\xi}^{\a}}\frac{\p h}{\p\xi_{\a}}\right)\,,
\end{array}
\eeq
we deduce the non-vanishing fundamental brackets,
\beqa\begin{array}{ll}
\big\{x^{i},\Pi_{j}\big\}=\d^{i}_{j},\quad\big\{\Pi_{i},\Pi_{j}\big\}=e\,\epsilon_{ijk}B^k,\quad\big\{\xi_{\a},\bar{\xi}^{\b}\big\}=-i\d^{\;\b}_{\a},\\[8pt]\displaystyle{
\big\{S^{k},S^{l}\big\}=\epsilon^{kl}_{\;\;m}S^{m},\quad\big\{S^{k},\bar{\xi}^{\b}\big\}=-\frac{i}{2}\bar{\xi}^{\mu}\sigma^{k\;\,\b}_{\,\mu},\quad\big\{S^{k},\xi_{\b}\big\}=\frac{i}{2}\sigma^{k\;\,\nu}_{\,\b}\xi_{\nu}}\,.
\end{array}
\eeqa
We also introduce an auxiliary scalar field, $\,\Phi(r)\,$, satisfying \textit{the ``self-duality'' or
`` Bogomolny'' relation}\footnote{See \cite{FHsusy} to justify terminology.},
\beq
\big\{\Pi^{k},\Phi(r)\big\}=\pm eB^{k}\,.\label{SelfDuality}
\eeq
This auxiliary scalar field also defines a square root of the external potential of the system so that
\beq
\displaystyle
\frac{1}{2}\Phi^{2}(r)=V(r)\,.
\eeq
As an illustration we obtain the potential \footnote{The constant is $\,\displaystyle{\gamma=\frac{\mu^2}{2q^2}}\,$.} defined in (\ref{Potential}) by considering the auxiliary field
\beq
\displaystyle
\Phi(r)=\pm\left(\frac{q}{r}+\frac{\mu}{q}\right)\,. 
\eeq

In order to investigate the $\,\N=2\,$ supersymmetry of the Pauli-like Hamiltonian (\ref{Hamiltonian}), we outline the algorithm developed we use to construct supercharges linear in the gauge covariant momentum,
\beqa\left\lbrace \begin{array}{lll} 
\displaystyle{\mp e\Phi(r)\,B^{j}C^j+\frac{ie\gyro}{4}B^{k}\left(\bar{\xi}^{\mu}\sigma^{k\,\nu}_{\mu}\frac{\p C}{\p\bar{\xi}^{\nu}}-\frac{\p C}{\p\xi_{\mu}}\sigma^{k\,\nu}_{\mu}\xi_{\nu}\right)
-\frac{e\gyro}{4}\,\bar{\xi}^{\mu}\sigma^{k\,\nu}_{\mu}\xi_{\nu}\,C^{j}\p_{j}B^{k}
=0}\,,&\hbox{o(0)}& 
\\[12pt]
\displaystyle{
\p_{m}C=e\,\epsilon_{mjk}B^kC^j+i\frac{e\gyro}{4}B^k\left(\bar{\xi}^{\mu}\sigma^{k\,\nu}_{\mu}\frac{\p C^m}{\p\bar{\xi}^{\nu}}-\frac{\p C^m}{\p\xi_{\mu}}\sigma^{k\,\nu}_{\mu}\xi_{\nu}\right) } \,, &\hbox{o(1)}& 
\\[12pt]
\p_jC^k(x,\xi,\bar{\xi})+\p_kC^j(x,\xi,\bar{\xi})=0\,.&\hbox{o(2)}&
\label{2susy}
\end{array}\right.
\eeqa
Let us first consider the Killing spinor,
\beq
C_{\b}^j=\frac{1}{2}\sigma^{j\,\a}_{\b}\,\xi_{\a}\,.
\eeq
Inserting this Killing spinor into the first-order equation of (\ref{2susy}) provides us with
\beq
\p_m C_{\b}=-\frac{i}{2}e\,B_m\,\xi_{\b}\quad\hbox{and}\quad\gyro=4\,,
\eeq
which can be solved using \textit{the  self-duality relation (\ref{SelfDuality})}. Thus, we get
\beq
\displaystyle
C_{\b}(\vx,\vxi)=\pm\frac{i}{2}\Phi(r)\,\xi_{\b}\,,
\eeq
provided the anomalous gyromagnetic factor is $\,g=4\,$. The zeroth-order constraint of (\ref{2susy}) is identically satisfied, so that collecting our results provides us with the supercharge,
\beq
\Q_{\b}=\frac{1}{2}\Pi_j\,\sigma^{j\,\a}_{\b}\,\xi_{\a}\pm\frac{i}{2}\Phi(r)\xi_{\b}\,.
\label{2susy2}
\eeq
To obtain the supercharge conjugate to (\ref{2susy2}), we consider the conjugate Killing spinor,
\beq
\bar{C}^{k\,\b}=\frac{1}{2}\bar{\xi}^{\a}\,\sigma_{\a}^{k\,\b}\,.
\eeq
In the case of anomalous value of the gyromagnetic ratio $\,g=4\,$, the first-order equation of (\ref{2susy}) is solved by using the Bogomolny equation (\ref{SelfDuality}). This leads to the conjugate
\beq
\displaystyle
\bar{C}^{\b}(\vx,\vxi)=\mp\frac{i}{2}\Phi(r)\bar{\xi}^{\b}\,.
\eeq
The zeroth-order consistency constraint is still satisfied, so we obtain the odd-supercharge,
\beq
\bar{\Q}^{\b}=\frac{1}{2}\bar{\xi}^{\a}\,\sigma_{\a}^{k\,\b}\Pi_{k}\mp\frac{i}{2}\Phi(r)\,\bar{\xi}^{\b}\,.
\label{2susy3}
\eeq
The supercharges $\,\Q_{\b}\,$ and $\,\bar{\Q}^{\b}\,$ are, both, square roots of the Pauli-like Hamiltonian $\,\H_{4}\,$,
\beq
\big\{\bar{\Q}^{\b},\Q_{\b}\big\}=-i\H_{4}\,\Id\,,
\eeq
and therefore \textit{generate the $\N=2$ supersymmetry of the spin-monopole field system.} It is worth noting that defining the rescaled, 
\beq
\displaystyle
\bar{\U}^{\b}=\bar{\Q}^{\b}\frac{1}{\sqrt{\H_4}}\qquad\hbox{and}\qquad
\displaystyle
\U_{\b}=\frac{1}{\sqrt{\H_4}}\Q_{\b}\,,
\eeq
it is straightforward to get,
\beq\displaystyle{
\H_0=\bar{\U}^{\b}\,\H_4\,\U_{\b}}\,,
\eeq
which make manifest the fact that the two anomalous cases $\,\gyro=0\,$ and $\,\gyro=4\,$ can be viewed as superpartners \footnote{With The scalar $\,\bar{\xi}^{\b}\xi_{\b}=2\,$.}, see \cite{FHsusy}. Moreover, in our enlarged system, the following bosonic charges
\beqa\begin{array}{lll}\displaystyle{
\vJ=\vx\times\vPi-q\,\hx+\vS}\,,\\[8pt]\displaystyle{
\vK=\vPi\times\vJ+\mu\,\hx-\vS\times\vPi-2e\left(\vS\cdot\vB\right)\vx+2q\frac{\vS}{r}+\frac{\mu}{q}\vS}\,,\\[12pt]\displaystyle{
\vo=\bar{\Q}^{\b}\,\vsigma_{\b}^{\;\a}\,\Q_{\a}=\frac{1}{2}\left(\Phi^2(r)-\vPi^{2}\right)\vS+\big(\vPi\cdot\vS\big)\vPi\mp\Phi(r)\,\vS\times\vPi,}
\end{array}
\eeqa
remain conserved such that they form, together with the supercharges $\,\Q_{\b}\,$ and $\,\bar{\Q}^{\b}\,$, the classical symmetry superalgebra \cite{DV2,FHsusy},
\beqa\begin{array}{llll}\displaystyle{
\big\{\bar{\Q}^{\b},\Q_{\b}\big\}= -i\H_4\,\Id\,,\quad\big\{\bar{\Q}^{\b},\bar{\Q}^{\b}\big\}=\big\{\Q_{\b},\Q_{\b}\big\}=0\,,\quad\big\{\bar{\Q}^{\b},J^k\big\}=\frac{i}{4}\bar{\Q}^{\a}\sigma^{k\,\b}_{\a}}\,,\\[12pt]\displaystyle{
\big\{\Q_{\b},J^k\big\}=-\frac{i}{4}\sigma^{k\,\a}_{\b}\Q_{\a}\,,\quad\big\{\bar{\Q}^{\b},K^j\big\}=-\frac{i}{4}\,\frac{\mu}{q}\,\bar{\Q}^{\a}\sigma^{j\,\b}_{\a}\,,\quad\big\{\Q_{\b},K^j\big\}=\frac{i}{4}\,\frac{\mu}{q}\,\sigma^{j\,\a}_{\b}\Q_{\a}}\,,\\[12pt]\displaystyle{
\big\{\bar{\Q}^{\b},\Omega^k\big\}=-i\H_4\,\bar{\Q}^{\a}\sigma^{k\,\b}_{\a}\,,\quad\big\{\Q_{\b},\Omega^k\big\}=i\H_4\,\sigma^{k\,\a}_{\b}\Q_{\a}\,,\quad\big\{\Omega^i,K^j\big\}=\frac{\mu}{q}\epsilon^{ijk}\,\Omega^k}\,,\\[12pt]\displaystyle{
\big\{K^i,K^j\big\}=\epsilon^{ijk}\left[\left(\frac{\mu^2}{q^2}-2\H_4\right)J^k+2\Omega^k\right]\,,\quad\big\{\Omega^i,\Omega^j\big\}=\epsilon^{ijk}\,\H_4\,\Omega^k}\,,\\[12pt]\displaystyle{
\big\{J^i,\Lambda^j\big\}=\epsilon^{ijk}\Lambda^k\quad\hbox{with}\quad\Lambda^l=J^l,K^l,\Omega^l}\,.
\end{array}\nonumber
\eeqa

We have shown, in this section, that the Runge-Lenz-type dynamical symmetry and the additional extra ``spin" symmetry both require instead an anomalous gyromagnetic ratio,
\beq
\gyro=0\quad\hbox{or}\quad\gyro=4\,.
\eeq
These particular values of the $\gyro$-factor come from the effective coupling of the form $F_{ij}\mp\epsilon_{ijk}D_k\Phi$, which add or cancel for self-dual fields \cite{FHsusy},
\beq
F_{ij}=\epsilon_{ijk}D_k\Phi\,. 
\eeq
Moreover, the super- and the bosonic symmetry can be combined in this enlarged fermionic space and provides us with an $\,\N=2\,$ SUSY, as proposed by D'Hoker and Vinet \cite{DV2}. See also \cite{FHsusy,FHO,Feher:1989,Bloore:1992fv}.

At last, let us remark that confining the spinning particle onto a sphere of fixed radius $\,\rho\,$ implies the set of constraints \cite{DJ-al},
\beq
\vx^2=\rho^2\,,\quad\vx\cdot\vpsi=0\quad\hbox{and}\quad\vx\cdot\vPi=0\,.
\eeq  
This freezes the radial potential to a constant,
and we recover the $\,\N=1\,$ SUSY  described by the supercharges $\,\Q\,$, $\,\Q_1\,$ and $\,\Q_2\,$ for ordinary gyromagnetic factor $\,\gyro=2\,$.

\newpage
\subsection{$\,\N=2\,$ SUSY in the plane}\label{S7}

The planar system consisting of a spinning particle interacting with a static magnetic field in the plane exhibits more symmetries as its higher-dimensional counterpart. Indeed, the $\N=2\,$ supersymmetry, here, is realized without doubling the Grassmann-variable of the internal space as it was the case in three-dimensional space system, see section \ref{SUSYMonop}. Such an ``exotic'' supersymmetry, which is realized in two different ways, is only possible in two spatial dimensions \cite{Duval:1993,Duval:1995dq,Duval:2008gf}. This is one more indication of the particular status of two-dimensional physics. 

To see this, we investigate the two dimensional model given by the Pauli-like Hamiltonian,
\beq
\H = \frac{1}{2}\,\Pi^2-\frac{e\gyro}{2}\, S B + V(r)\,,
\label{a.1}
\eeq
where the magnetic field simplifies into \footnote{We dropped the irrelevant third $\,z\,$-direction.}
\beq
F_{ij} = \varepsilon_{ij}B = \p_i A_j - \p_j A_i\,,
\eeq
and the spin tensor is actually a scalar
\beq
S = - \frac{i}{2}\, \varepsilon_{ij}\psi^{i}\psi^{j}.
\label{a.2}
\eeq
The fundamental brackets remain the same as in (\ref{PBrackets}), and the spatial and the internal motions of the particle are governed by the following equations,
\beq
\begin{array}{ll}
\displaystyle
\ddot{x}_{k}=\frac{e\gyro}{2}S\p_{k}B+eB\varepsilon_{kj}\dot{x}^j+\p_{k}V\,,\\[8pt]
\displaystyle
\dot{\psi}_{i}=\frac{e\gyro}{2}B\varepsilon_{ij}\psi^j\,,\quad
\dot{S}=0\,.
\end{array}\label{EqMPlanar}
\eeq
Observe the conservation of the spin $S$ along the particle motion and let us recall that all quantities quadratic in the Grassmann variables are proportional to $S$.\\

We search for dynamical quantities which are constants of the motion, for the planar system, by solving the series of constraints:
\beqa
\begin{array}{lllll}
\displaystyle{ C_i \p_i \H + i \frac{\p \H}{\p \psi_i} \frac{\p C}{\p \psi_i} = 0}\,,&\hbox{o(0)}&\\[10pt]

\displaystyle{ \p_i C = e F_{ij} C_j + i \frac{\p \H}{\p \psi_j} \frac{\p C_i}{\p \psi_j} + C_{ij} \p_j \H }\,,&\hbox{o(1)}&\\[10pt]

\displaystyle{ \p_i C_j + \p_j C_i = e \left( F_{ik} C_{kj} - C_{ik} F_{kj} \right) + i \frac{\p \H}{\p \psi_k} \frac{\p C_{ij}}{\p \psi_k}
 + C_{ijk} \p_k \H}\,,&\hbox{o(2)}&\\[10pt]
 
 \p_i C_{jk} + \p_j C_{ki} + \p_k C_{ij} = C_{ijkl} \p_l \H + (\mbox{terms linear in $C_{lmn}$})\,.&\hbox{o(3)}&\\
\vdots\qquad\qquad\qquad\qquad\qquad\vdots&\vdots
\end{array}\label{vHeq1}
\eeqa
Using the equality
\beq
i \frac{\p\H}{\p \psi^i} = - \frac{e\gyro}{2}\, F_{ij} \psi^{j} = - \frac{e\gyro}{2}\, B \varepsilon_{ij} \psi^{j}\,,
\label{a.5}
\eeq
the zeroth-order constraint in (\ref{vHeq1}) becomes
\beq
\frac{e\gyro}{2}\, B \varepsilon_{ij} \psi_j \frac{\p C}{\p \psi_i} = C_i \left(\p_i V - \frac{e\gyro}{2}\, S\, \p_i B \right),
\label{a.4}
\eeq
complemented by the first-order equation of (\ref{vHeq1})
\beq
\p_i C = eB \left( \varepsilon_{ij} C_j + \frac{\gyro}{2}\, \varepsilon_{jk} \psi_j \frac{\p C_i}{\p \psi_k} \right)
 + C_{ij} \left(  \p_j V - \frac{e\gyro}{2}\, S  \p_j B \right).
\label{a.7}
\eeq
Similarly the second and higher-order equations take the form
\beq
\p_{\left(i\right.}C_{\left.j\right)}= eB\left(\varepsilon_{ik} C_{kj} + \varepsilon_{jk} C_{ki} + \frac{\gyro}{2}\,\varepsilon_{jk} \psi_j \frac{\p C_i}{\p \psi_k} 
 \right) + C_{ijk} \left(\p_k V - \frac{e\gyro}{2}\,S  \p_k B\right)\,.
\label{a.8}
\eeq
For radial functions $\,V(r)\,$ and $\,B(r)\,$,
\beq
\p_i V =\frac{x_i}{r}V^{\prime}\,,\qquad
\p_i B = \frac{x_i}{r}B^{\prime}\,,
\eeq
hence
\beq
\left(  \p_j V - \frac{e\gyro}{2}\, S  \p_j B \right)C_{i...j}=\frac{x_j}{r} \left(V^{\prime} - \frac{e\gyro}{2}\, S B^{\prime}\right)C_{i...j}.
\label{a.9}
\eeq

Let us now consider some specific cases. To this, we introduce the universal generalized Killing vectors,
\beq
\displaystyle
C_i = \left\lbrace\gamma_i,\; \varepsilon_{ij} x^{j},\;  \psi_i,\; \varepsilon_{ij} \psi^{j}\right\rbrace\,,
\label{a.10}
\eeq
where $\gamma_i$ denotes a constant vector. 

\vskip2mm
$\bullet$
A constant Killing vector $\gamma_i$ gives a constant of the motion only if we can find solutions for the equations
\beq
\p_i C = e B \varepsilon_{ij} \gamma_j, \hspace{2em} 
B  \varepsilon_{ji} \psi_i \frac{\p C}{\p \psi_j} = \gamma_i \left( \frac{2}{e\gyro}\, \p_i V- S \p_i B \right).
\label{a.a1}
\eeq
Now for a Grassmann-even function
\beq
C = c_0 + c_2 S\,,
\eeq
the left-hand side of the second equation in (\ref{a.a1}) vanishes, therefore we
must require $B$ and $V$ to be constant. This leads to the solution
\beq
C = - e B \varepsilon_{ij} \gamma^i x^j, \hspace{2em} V =\const, \hspace{2em} B = \const\,.
\label{a.a2}
\eeq
The corresponding constant of the motion, $\,\zeta\,$, is identified with the \textit{``magnetic translations''} \cite{Kostel},
\beq
\displaystyle
\zeta=\gamma^i P_i\qquad\mbox{with}\qquad
P_i = \Pi_i - e B \varepsilon_{ij} x^j\,.
\label{a.a3}
\eeq

\vskip2mm
$\bullet$
Next we consider the linear Killing vector $C_i = \varepsilon_{ij} x^j$, with all higher-order coefficients $C_{ij...} = 0$. 
Again for Grassmann-even $C$ the left-hand side of equation (\ref{a.4}) vanishes, and we get the condition
\beq
\varepsilon_{ij} x_i \p_j B = \varepsilon_{ij} x_i \p_j V = 0,
\label{a.a4}
\eeq
which is automatically satisfied for radial functions $B(r)$ and $V(r)$. Therefore we only have to solve the equation (\ref{a.7}): 
\beq
\p_i C = - e B  x_i = - \frac{e x_i}{r}\,\big(rB\big). 
\label{a.a5}
\eeq
We infer that $C(r)$ is a radial function, with
\beq
C^{\prime} = -e rB\,.
\eeq
Therefore $\,C\,$ is given by \textit{the magnetic flux through the disk $D_r$ centered at the origin with radius $r$}:
\beq
C = - \frac{e}{2\pi}\, \int_{D_r} B(r) d^2x \equiv - \frac{e}{2\pi}\, \Phi_B(r).
\label{a.a7}
\eeq
We then find the constant of the motion representing the angular momentum \cite{Ngome:2010gg},
\beq
L = \varepsilon_{ij} x_i \Pi_j +\frac{e}{2\pi}\,  \Phi_B(r),
\label{a.a8}
\eeq
associated with the
$\,
\ort(2)_{rotations}
\,$ symmetry group.\\

$\bullet$
There are two Grassmann-odd Killing vectors, the first one being $C_i = \psi_i$. With this Ansatz, we get for the 
scalar contribution to the constant of the motion the constraints
\beq
\frac{e\gyro}{2} B\, \varepsilon_{ij} \psi_j\, \frac{\p C}{\p \psi_i} = \psi_i \p_i V\qquad\mbox{and}\qquad
\p_i C = \frac{eB}{2} \left( 2 - \gyro \right) \varepsilon_{ij} \psi_j\,.
\label{a.b1}
\eeq
It follows that either $\,\gyro = 2\,$ and $\,\big(C,\,V\big)\,$ are constant, in which case one may take $C = V = 0$,  
or $\,\gyro \neq 2\,$ and $\,C\,$ is of the form
\beq
C = \varepsilon_{ij} K_i(r) \psi_j\,
\quad\hbox{with}\quad 
\p_i V = - \frac{e\gyro}{2}\, B K_i\,, \hspace{2em} 
\p_i K_j = \frac{(2-\gyro)eB}{2}\, \delta_{ij}.
\label{a.b3}
\eeq
This is possible only if $B$ is constant and
\beq
K_i = \frac{eB(2-\gyro)}{2}\,x_i \equiv \kappa x_i\,, \hspace{2em} 
V(r) = \frac{\gyro(\gyro-2)}{8}\, e^2 B^2 r^2 = - \frac{e\gyro \kappa}{4\pi}\, \Phi_B(r).
\label{a.b4}
\eeq
It follows that we have a conserved supercharge of the form,
\beq
Q = \psi^{i} \left( \Pi_i - \kappa \varepsilon_{ij} x^{j}\right).
\label{a.b5}
\eeq
The bracket algebra of this supercharge takes the form
\beq
i \left\{ Q, Q \right\} = 2\H + (2 - g)eBJ, \hspace{2em} J = L + S.
\label{a.b6}
\eeq
Of course, as $S$ and $L$ are separately conserved, $J$ is a constant of the motion as well. It is now easy to see that for ordinary $\,\gyro\,$-factor the supercharge $\,\Q\,$ in (\ref{a.b5}) is a square root of the Hamiltonian.

$\bullet$ Let us remark that for the anomalous gyromagnetic ratio $\,\gyro=1\,$,
we construct the conserved conformal supercharge,
\beq
\S=\vx\cdot\vpsi-t\Q\,,
\eeq
obtained by using the internal equation of the motion in (\ref{EqMPlanar}).
\vspace{1ex}

$\bullet$
Finally we consider the dual Grassmann-odd Killing vector $C_i = \varepsilon_{ij} \psi_j$. Then the 
constraints (\ref{a.4}) and (\ref{a.7}) become
\beq
\frac{eg}{2}\,B\,\frac{\p C}{\p \psi_i} = \p_i V, \hspace{2em}
\p_i C = \frac{(g-2)eB}{2}\, \psi_i, 
\label{a.c1}
\eeq
implying that
\beq
C = N_i(x) \psi_i\quad\mbox{and}\quad
\frac{eg}{2}\, B\, N_i = \p_i V, \hspace{2em}
\p_i N_j = \frac{(g-2) eB}{2}\, \delta_{ij}.
\label{a.c2}
\eeq
As before, the magnetic field $\,B\,$ must be constant and the potential is identical to (\ref{a.b4}),
\beq
N_i = - \kappa x_i, \hspace{2em} 
V = - \frac{eg\kappa}{4\pi}\, \Phi_B(r) = \frac{g(g-2)}{8}\, e^2 B^2 r^2.
\label{a.c3}
\eeq
Thus, we find the dual conserved supercharge \footnote{The cross product of two planar vectors, $\,\va\times\vb=\varepsilon_{ij}a^{i}b^{j}\,$, again defines a scalar.} \cite{HmonRev},
\beq
\Q^{\star} = \varepsilon_{ij} \psi_i \left( \Pi_j - \kappa \varepsilon_{jk} x_k \right) 
 = \psi_i \left( \varepsilon_{ij} \Pi_j + \kappa x_i \right)\,,
\label{a.c4}
\eeq
corresponding, in the case of ordinary gyromagnetic ratio, to the ``twisted'' supercharge used by Jackiw \cite{Jackiw:1984ji} to describe the Landau states in a constant magnetic field. Moreover $\,\Q^{\star}\,$ satisfies the bracket relations
\beq
i \left\{ \Q^{\star}, \Q^{\star} \right\} = 2\H + (2-g) eBJ, \hspace{2em}
i \left\{ Q, \Q^{\star} \right\} = 0.
\label{a.c5}
\eeq
Thus the harmonic potential (\ref{a.b4}) with constant magnetic field $B$ allows a classical $\N = 2$ supersymmetry 
with supercharges $\big(Q, \Q^{\star}\big)$, whilst the special conditions $g = 2$ and $V = 0$ allows for $\N = 2$
supersymmetry for any $B(r)$. 

$\bullet$ As a consequence of the conservation of the ``twisted'' supercharge, we construct for $\,\gyro=1\,$, the associate conserved conformal supercharge
\beq
\S^{\star}=\vx\times\vpsi+t\Q^{\star}\,.
\eeq
Thus, for the non-ordinary gyromagnetic ratio $\,\gyro=1\,$, the supercharges $\,\Q\,$, $\,\Q^{\star}\,$, $\,\S\,$ and $\,\S^{\star}\,$ extend the $\,o(2, 1)\,$ algebra into an $\,\osp(1,1)\,$ superalgebra and satisfy the commutation relations,
\beqa
\begin{array}{lllll}
\displaystyle
\big\{\Q,\Q\big\}=\{\Q^{\star},\Q^{\star}\big\}=-i\big(2\H+eBJ\big)\,,\quad\big\{\S^{\star},\S^{\star}\big\}=-it^2\big(2\H+eBJ\big)\,,\\[8pt]
\displaystyle
\big\{\Q^{\star},\S^{\star}\big\}=-\big\{\Q,\S\big\}=-it\big(2\H+eBJ\big)+i\vx\cdot\vPi\,,\quad
\big\{\Q,\Q^{\star}\big\}=0\,,\\[8pt]
\displaystyle
\big\{\Q^{\star},\S\big\}=\big\{\Q,\S^{\star}\big\}=-i\big(L+2S\big)\,,\quad\big\{\S,\S^{\star}\big\}=2it\big(L+2S\big)\,,\\[8pt]
\displaystyle
\big\{\S,\S\big\}=-it^2\big(2\H+eBJ\big)+2it\,\vx\cdot\vPi-ir^2\,.
\end{array}
\eeqa
The van Holten recipe is therefore relevant to study planar fermions in an arbitrary planar magnetic field, i.e. one perpendicular to the plane. As an illustration, we have shown, for ordinary gyromagnetic factor, that in addition to the usual supercharge (\ref{a.b5}) generating the supersymmetry,  the system also admits another square root of the Pauli Hamiltonian $\,\H\,$ \cite{HmonRev}. This happens due to the existence of a dual Killing tensor generating the ``twisted'' supercharge.

\section{Non-commutative models}\label{chap:NCmodels}

{\normalsize
\textit{
A non-commutative oscillator with no kinetic term but with a certain momentum-dependent potential is constructed. The classical trajectories followed by a non-commutative particle in this oscillator field lie on (arcs of) ellipses.
}}

\subsection{Non-commutative oscillator with Kepler-type dynamical symmetry}

In recent years, a remarkable non-commutative model was derived in the context of solid state physics by Chang and Niu \cite{Niu}. They stated that the semiclassical analysis of a Bloch electron in a three-dimensional crystal lattice reveals an extra ``Berry phase'' term, $\,\vTheta\,$, which can take a monopole-like form in the band structure. The study of the wave-packet dynamics of this Bloch electron, under perturbations slowly varying in space and in time, leads to the equations of the motion in the \textit{mth} band [in units $\hbar=1$],
\beq
\dot{\vk}=-e\vE-e\dot{\vx}\times\vB\big(\vx\big)\,,\quad 
\dot{\vx}=\frac{\p \E_{m}\big(\vk\big)}{\p\vk}-\dot{\vk}\times\vTheta(\vk)\,.\label{BlochDyn}
\eeq
Here $\,\E_{m}\big(\vk\big)\,$, $\,\vx\,$ and $\,\vk\,$ denote the Bloch electron's band energy, the intracell position and the quasi-momentum, respectively. Note that in the right hand side equation of (\ref{BlochDyn}), the electron velocity gains an anomalous velocity term, $\,\dot{\vk}\times\vTheta(\vk)\,$, which is the mechanical counterpart of the anomalous current.

In a magnetic field-free theory [with $\,\vB=\vec{0}\,$], the equations (\ref{BlochDyn}) can also be deduced using the symplectic closed two-form,
\beq
\Omega=dp_i\wedge dx_i+\frac{1}{2}\epsilon_{ijk}\Theta^{i}dp_j\wedge dp_k\,,\label{Symp2form}
\eeq 
where the ``extra'' term induced by the Berry phase yields the position coordinates non-commutative \cite{Niu,Xiao},
\beqa
\{x_i,x_j\}=\epsilon_{ijk}\Theta_k=\Theta_{ij},
\qquad
\{x_i,p_j\}=\delta_{ij},
\qquad 
\{p_i,p_j\}=0\,.
\eeqa
Applying the Jacobi identities to the coordinates, we get
\beq
\left\lbrace
\begin{array}{ll}
\displaystyle
0=\{p_{i},\{x_{j},x_{k}\}\}_{cyclic}=-\epsilon_{jkm}\frac{\p\Theta^{m}}{\p x_{i}}\,,\\[10pt]
\displaystyle
0=\{x_{i},\{x_{j},x_{k}\}\}_{cyclic}=\frac{\p\Theta^{ij}}{\p p_{k}}+\frac{\p\Theta^{jk}}{\p p_{i}}+\frac{\p\Theta^{ki}}{\p p_{j}}\,.\end{array}
\right.
\eeq
Then, the vector field $\,\vTheta\,$ has the property \footnote{For a more general theory which also includes magnetic fields, see, e.g.,
\cite{Niu,Xiao,NCRev}. For simplicity, the mass has been chosen unity.} to be only momentum-dependent \cite{BeMo},
\beq
\Theta_i=\Theta_i\big(\vp\big)\,,
\eeq
and also requires the consistency condition
\beq
\vnabla_{\vp}\cdot\vTheta\big(\vp\big)=0\,,
\eeq
which can be interpreted as a field Maxwell equation in the dual momentum space. Choosing, for example, the non-commutative vector aligned in the third direction,
\beq
\Theta_i=\theta\delta_{i3}\,,
\quad\theta=\const\,,\label{vertNC}
\eeq
the $3$-dimensional theory reduces to the planar mechanics based on ``exotic'' Galilean symmetry \cite{LSZ,NCRev,NCRev1,Chaichian0,Nair,Scholtz1,Scholtz2,NCRev2}. As an application of (\ref{vertNC}), some interesting results, including  perihelion point precession of the planetary orbit, can be derived \cite{RoVe} when taking into account the Kepler potential,
\beq
V\big(r\big)\propto r^{-1}\,.
\eeq
Other applications of (\ref{vertNC}) concern, for example, the Quantum Hall Effect \cite{DJTCS,NCRev,H2002}.

Such a choice only allows for axial symmetry, though. In our theory, however, we restore the full rotational symmetry by choosing instead $\,\vTheta\,$ to be a ``monopole in $\vp$-space'' \cite{BeMo},
\beq
\Theta_i=\theta\,\frac{p_i}{p^3}\,,
\qquad
\theta=\const\,,
\label{monoptheta}
\eeq
where $\,p=|\vp|\,$. Indeed, away from the origin, the dual monopole (\ref{monoptheta}) is the only spherically symmetric possibility consistent with the Jacobi identities \footnote{See the equivalent demonstration in real $\vx$-space in section \ref{DiracMonop}.}. Let us mention that the $\,\vp\,$-monopole form in (\ref{monoptheta}) has already been observed experimentally by Fang et al. in the context of anomalous Hall effect in the metallic ferromagnet \nobreak{SrRuO3} \cite{Fang:2003ir}.

As expected, (\ref{monoptheta}) corresponds to extra, ``monopole'' term in the symplectic structure (\ref{Symp2form}) which is in fact that of a mass-zero spin-$\theta$ coadjoint orbit of the Poincar\'e group. The orbit is indeed that of the $\,\ort(4,2)\,$ conformal group \cite{Penrose,Penrose1,CFH3}.

We can now study the $3D$ mechanics with non-commutativity  (\ref{monoptheta}), augmented with the Hamiltonian,
\beq
\H=
\frac{p^2}{2}+V(\vx,\vp)\,,
\eeq
where we allowed that the potential may also depend on the momentum variable, $\,\vp\,$ \footnote{Note that momentum-dependent potentials are frequently used in nuclear physics and correspond to non-local interactions.}.

The equations of motion of the system read
\beq
\dot{x}_i={p_i}{}+\frac{\p V}{\p p_i}+\theta\epsilon_{ijk}\frac{p_k}{p^3}\frac{\p V}{\p x_j},
\qquad
\dot{p}_i=-\frac{\p V}{\p x_i}\,,
\eeq
where, in the first relation, the  ``anomalous velocity terms'' is due to our assumptions (\ref{monoptheta}). 

We are particularly  interested in finding conserved quantities. This task is conveniently achieved by using van Holten's covariant framework \cite{vH}, which amounts to searching for an expansion into integer powers of the momentum,
\beq
Q=C_0(\vx)+C_i(\vx)p_i+ \frac{1}{2!}C_{ij}(\vx)p_ip_j+\frac{1}{3!}C_{ijk}(\vx)p_ip_jp_k+\dots\,.
\eeq
Requiring $\,Q\,$ to Poisson-commute with the Hamiltonian yields an infinite series of constraints. However, the expansion can be truncated at a finite order $\,n\,$, provided to satisfy the Killing equation,
$\,
D_{(i_1}C_{i_2\dots i_n)}=0\,
$, when we can set $\,C_{i_1\dots i_{n+1}\dots}=0\,$.

Let us assume that the potential has the form $\,V\big(|\vx|,|\vp|\big)\,$,
and try to find the conserved angular momentum, associated with the Killing vector $\,\vC=\vn\times\vx\,$, which represents space rotations around $\vn$. An easy calculation shows that the procedure fails to work, however, owing to the $\,\vp\,$-monopole term. We propose, therefore,  to work instead in a ``dual'' framework \cite{ZHN}, i.e. in momentum space, and search for conserved quantities expanded 
rather into powers of the position,
\beq
Q=C_0(\vp)+C_i(\vp)x_i+\frac{1}{2!}C_{ij}(\vp)x_ix_j+
\frac{1}{3!}C_{ijk}(\vp)x_ix_jx_k\dots\,.
\label{pConsQuant}
\eeq
Then, the covariant van Holten algorithm, presented in section \ref{vHAl}, is replaced by
\beq
\begin{array}{llll}
C_i\left(\displaystyle{p_i}{}+ \displaystyle\frac{\p V}{\p p_i}\right)&=&0&\hbox{o(0)}
\\[8pt]
\displaystyle\frac{1}{r}
\displaystyle\frac{\p V}{\p r}\left(
\theta\epsilon_{ijk} \displaystyle\frac{p_k}{p^3}C_i
-\displaystyle\frac{\p C}{\p p_j}\right)
+C_{ij}\left(\displaystyle{p_i}{}
+\displaystyle\frac{\p V}{\p p_i}\right)&=&0
&\hbox{o(1)}
\\[8pt]
\displaystyle\frac{1}{r}
\displaystyle\frac{\p V}{\p r}\left(
\theta\displaystyle\frac{p_m}{p^3}\big(\epsilon_{ijm}C_{ik}
+\epsilon_{ikm}C_{ij}\big)
-\big(\displaystyle\frac{\p C_k}{\p p_j}+
\displaystyle\frac{\p C_j}{\p p_k}\big)
\right)
+C_{ijk}
\left(\displaystyle{p_i}{}
+\displaystyle\frac{\p V}{\p p_i}\right)&=&0\,
&\hbox{o(2)}
\\[8pt]
\displaystyle\frac{1}{r}
\displaystyle\frac{\partial V}{\partial r}\left(
\theta\displaystyle\frac{p_m}{p^3}
\big(\epsilon_{lim}C_{ljk}+\epsilon_{ljm}C_{lki}
+\epsilon_{lkm}C_{lij}\big)
-\big(
\displaystyle\frac{\partial C_{ij}}{\partial p_k}+ \displaystyle\frac{\partial C_{jk}}{\partial p_i}+%
\displaystyle\frac{\partial C_{ki}}{\partial p_j}\big)\right)+
\nn
\\
\hskip50mm
C_{lijk}\left(\displaystyle{p_l}{}+
\displaystyle\frac{
\partial V}{\partial p_l}\right)&=&0
&\hbox{o(3)}
\\[8pt]
\vdots\qquad\qquad\qquad\qquad\qquad\vdots&\vdots
\end{array}\label{NCconstraints}
\eeq
where $r=|\vx|$. The expansion (\ref{pConsQuant}) can again be truncated at a finite order $\,n\,$, provided the higher order constraint of the previous series of constraints transforms into a dual Killing equation,
\beq
\p_{(p_{i_1}}C_{p_{i_2}\dots p_{i_n})}=0\,.
\eeq
Then, for linear conserved quantities, $\,Q=C_0(\vp)+C_i(\vp)x_i\,$, we can set $C_{ij}=C_{ijk}=\dots0$. The dual algorithm therefore reduces to
\beq
\left\lbrace
\begin{array}{lll}
\displaystyle
C_i\left(\displaystyle{p_i}{}+ \displaystyle\frac{\p V}{\p p_i}\right)=0\,,&\hbox{o(0)}
\\[8pt]
\theta\epsilon_{ijk} \displaystyle\frac{p_k}{p^3}C_i
-\displaystyle\frac{\p C}{\p p_j}=0\,,
&\hbox{o(1)}
\\[8pt]
\displaystyle\frac{\p C_k}{\p p_j}+\frac{\p C_j}{\p p_k}
=0\,.
&\hbox{o(2)}
\end{array}
\right.
\eeq
Introducing the dual Killing vector $$\,\vC=\vn\times\vp\,$$ into the previous algorithm provides us with
\beq
C=\theta\,\vn\cdot\hp\,,\qquad
\hp=\frac{\vp}{p}\,.
\eeq
Thus, we obtain the conserved angular
momentum,
\beq
\vJ=\vL-\theta\,\hp=\vx\times\vp-\theta\,\hp\,,
\label{pAngMom}
\eeq
which is what one would expect, due to the ``monopole in $\,\vp\,$-space'', whereas the non-commutative parameter, $\theta$, behaves as the ``monopole charge'' \cite{Cortes}.

The next step is to inquire about second order conserved quantities. Then, the series of constraints which has to be solve read 
\beq
\left\lbrace
\begin{array}{llll}
\displaystyle
C_i\left(\displaystyle{p_i}{}+ \displaystyle\frac{\p V}{\p p_i}\right)=0\,,&\hbox{o(0)}
\\[8pt]
\displaystyle
\frac{1}{r}\frac{\p V}{\p r}\left(
\theta\epsilon_{ijk}\frac{p_k}{p^3}C_i
-\frac{\p C}{\p p_j}\right)
+C_{ij}\left(p_i+\frac{\p V}{\p p_i}\right)=0\,,
&\hbox{o(1)}
\\[8pt]
\displaystyle
\theta\frac{p_m}{p^3}\big(\epsilon_{ijm}C_{ik}
+\epsilon_{ikm}C_{ij}\big)
-\big(\frac{\p C_k}{\p p_j}+\frac{\p C_j}{\p p_k}\big)
=0\,,
&\hbox{o(2)}\\[8pt]
\displaystyle
\frac{\partial C_{ij}}{\partial p_k}+\frac{\partial C_{jk}}{\partial p_i}+\frac{\partial C_{ki}}{\partial p_j}=0\,.
&\hbox{o(3)}
\end{array}
\right.\label{NCConstraints2}
\eeq
Remark that usually the Runge-Lenz vector is generated by the rank-$\,2\,$ Killing tensor
$\,
C_{ij}=2\delta_{ij}\vn\cdot\vx-n_ix_j-n_jx_i\,
$
where $\,\vn\,$ is some fixed unit vector \cite{vH}. 
Not surprisingly, the original procedure fails once again. The  dual procedure works, though. The dual two-tensor
\beq
C_{ij}=2\delta_{ij}\vn\cdot\vp-n_ip_j-n_jp_i\,,
\label{pRLK}
\eeq
verifies the dual Killing equation of order 3 in (\ref{NCConstraints2}). Then the order-2 equation yields
\beq
\vC=\theta\frac{\vn\times\vp}{p}\,.\label{NCR1}
\eeq
Inserting into the first-order constraint of (\ref{NCConstraints2}) and assuming $\,\p_rV\neq0\,$, the constraint is satisfied with
\beq
C=\alpha\,\vn\cdot\hp\,\label{NCR2}
\eeq
$\alpha$ being an arbitrary constant, \emph{provided} the momentum-dependent potential and the Hamiltonian take the form
\beq
V=
\frac{\vx{}^2}{2}-\frac{p^2}{2}+\frac{\theta^2}{2p^2}
+\frac{\alpha}{p}\quad\hbox{and}\quad\H=\frac{\vx^2}{2}+\frac{\theta^2}{2p^2}+
\frac{\alpha}{p}\,,
\label{Vee}
\eeq
respectively. 
Then the dual algorithm provides us with the Runge-Lenz-type vector
\beq
\vK=\vx\times\vJ-\alpha\hp
\,. 
\label{pRL}
\eeq
Its conservation can also be checked by a direct 
calculation, using the equations of the motion,
\beq
\dot{\vx}=\theta\,\frac{\vx\times{\vp}}{p^3}-\left(\frac{\theta^2}{p^4}+\frac{\alpha}{p^3}\right)\vp\,,
\qquad
\dot{\vp}=-{\vx}{}\,,
\label{expleqmot}
\eeq
where the anomalous velocity term in the first relation is transversal.

Note that the $\,(-p^2/2)\,$ term in the potential cancels the usual kinetic term, and our system  describes a \emph{non-relativistic,
non-commutative particle with no mass term in an oscillator field, plus some momentum-dependent interaction}.

Writing the Hamiltonian as
\beq
\H=\frac{\vx^2}{2}+\frac{\theta^2}{2}\left(\frac{1}{p}+
\frac{\alpha}{\theta^2}\right)^2-\frac{\alpha^2}{2\theta^2}\,
\eeq
shows, moreover, that $\,\displaystyle\H\geq -\frac{\alpha^2}{2\theta^2}\,$ with equality only attained when $\,\displaystyle p=-\frac{\theta^2}{\alpha}\,$, which plainly requires $\alpha<0$.
 
It is easy to understand the reason why our modified algorithm did work~: calling 
\beq
\left\lbrace
\begin{array}{ll}
\vp\;\longrightarrow\;\vR\qquad\mbox{``position''}\\[6pt]
-\vx\;\longrightarrow\;\vP\qquad\mbox{``momentum''}\,,
\end{array}
\right.
\eeq
the system can also be interpreted as an ``ordinary''
(i.e. massive and commutative) 
\emph{non-relativistic charged particle in the  field of a Dirac monopole of strength 
$\theta$, augmented with an inverse-square plus
a Newtonian potential}. This is 
the well-known ``McIntosh-Cisneros -- Zwanziger'' 
(MICZ) system \cite{MIC,Zwanziger},
for which the fine-tuned inverse-square potential is known to cancel
the effect of the monopole, allowing for a Kepler-type dynamical symmetry \cite{MIC,Zwanziger}. The angular momentum, (\ref{pAngMom}), and the
 Runge-Lenz vector, (\ref{pRL}), are, in particular, that  of the MICZ problem \cite{MIC,Zwanziger} in ``dual'' momentum-space.

The conserved quantities provide us with valuable information on the motion. Mimicking what is done for the MICZ case, we note that
\beq
\vJ\cdot\hp=-\theta
\eeq
implies that the vector $\vp$ moves on a cone of opening angle
$
\arccos\big(-{\theta}/{J}\big)\,.
$
On the other hand, defining the conserved vector
\beq
\vN={\alpha}\vJ-{\theta}\vK\,,
\eeq
we construct the constant, 
\beq
\vN\cdot\vp=\theta(J^2-\theta^2)=\theta L^2\,,
\eeq
so that the $\vp$-motion lies on the plane perpendicular to $\vN$. \emph{The trajectory in $p$-space belongs therefore to a conic section}. 

For the MICZ problem, this is the main result, but
 for us here our main interest lies in finding the real space trajectories, $\vx(t)$. By (\ref{expleqmot}), this amounts to find the [momentum-] \emph{``hodograph''} of the MICZ problem. Curiously, while the hodograph of the Kepler problem is well-known, it is actually a circle or a circular arc, we could not find the corresponding result in the vast literature of MICZ system.

Returning to our notations, we note that due to
\beq
\vN\cdot\vx=0\,,
\eeq
the $\vx(t)$-trajectories also belongs to an oblique plane, whose normal is $\,\vN={\alpha}\vJ-{\theta}\vK\,$. We can thus conveniently study the problem in an adapted coordinate system. One proves indeed that
\beq
\begin{array}{ll}
\displaystyle
\Big\{\hat{\imath},\hat{\jmath},\hat{k}\Big\}=
\Big\{\frac{1}{|\epsilon L|}\vec{K}\times\vec{J}
,\, \frac{1}{|\lambda \epsilon|}(2\theta \H\vec{J}+\alpha \vec{K}),\,\frac{1}{|\lambda L|}(\alpha\vec{J}-\theta\vec{K})\Big\}\\[14pt]
\mbox{with}\quad\lambda^2=\alpha^2+2\H\theta^2\,,\quad\epsilon^2=\alpha^2+2\H J^2\quad\mbox{and}\quad L^2=J^2-\theta^2\,,
\end{array}\label{basis}
\eeq
is a convenient orthonormal basis to study the $\,\vx\,$-trajectories. Here we recognize, in $\,\hat{k}\,$, $\,\vN/N\,$ in particular.

$\bullet$ Firstly, projecting onto these axis,
\beq
\left\lbrace
\begin{array}{ll}
p_z=\vp\cdot\hat{k}=\theta L/|\lambda|=\const\,,\\[8pt]
p_x=\vec{p}\cdot\hat{\imath}\,,\\[8pt]
p_y=\vec{p}\cdot\hat{\jmath}\,,
\end{array}\right.
\eeq
we find the equation
\beq
\displaystyle
\frac{\left(p_y+\displaystyle\frac{|\epsilon|\alpha}{2|\lambda|\H}\right)^2}{\lambda^2/4\H^2}-\frac{p_x^2}{L^2/2\H}=1,
\eeq
which is the equation of a hyperbola or of an ellipse in momentum space, depending on the sign of $\H$, positive or negative. For vanishing $\H$ one gets a parabola. This confirms what is known for the MICZ problem \cite{MIC,Zwanziger}, and is consistent with what we deduced geometrically.

$\bullet$ Next, projecting the $\,\vx\,$-motion onto the orthonormal basis (\ref{basis}) yields
\beq
\displaystyle
X=\vec{x}\cdot\hat{\imath}=-\frac{2|L|}{|\epsilon|}(\H-\frac \alpha{2p})\,,\qquad
Y=\vec{x}\cdot\hat{\jmath}=-\frac{|\lambda|}{|\epsilon|}\frac{\vec{x}\cdot\vec{p}}p\,,\qquad
Z=\vec{x}\cdot \hat{k}=0\,.
\eeq
An easy calculation leads to the equation 
\beq
\big(X+\frac{|\epsilon|L}{J^2}\big)^2+\frac{\alpha ^2L^2}{\lambda^2J^2}Y^2=
\frac{L^2\alpha^2}{J^4}
\eeq
which always describes an ellipse or an arc of ellipse, since
\beq
\lambda^2=\alpha^2+2\H\theta^2\geq0\,.
\eeq
The center has been shifted along the axis $\hat{\imath}$ by the quantity $
\,\big(-{|\epsilon|L}/{J^2}\big)\,
$
and the major axis is directed along $\hat{\jmath}$. Note that, unlike as in $\vp\,$-space, the $\vx$-trajectories are always bounded.

When the energy is negative, $\H<0$, which is only possible when the Newtonian potential is attractive, $\,\a<0\,$, the $\vx$-trajectories are full ellipses. The origin is inside the ellipse~:
\begin{figure}[!h]
\begin{center}
\includegraphics[scale=.56]{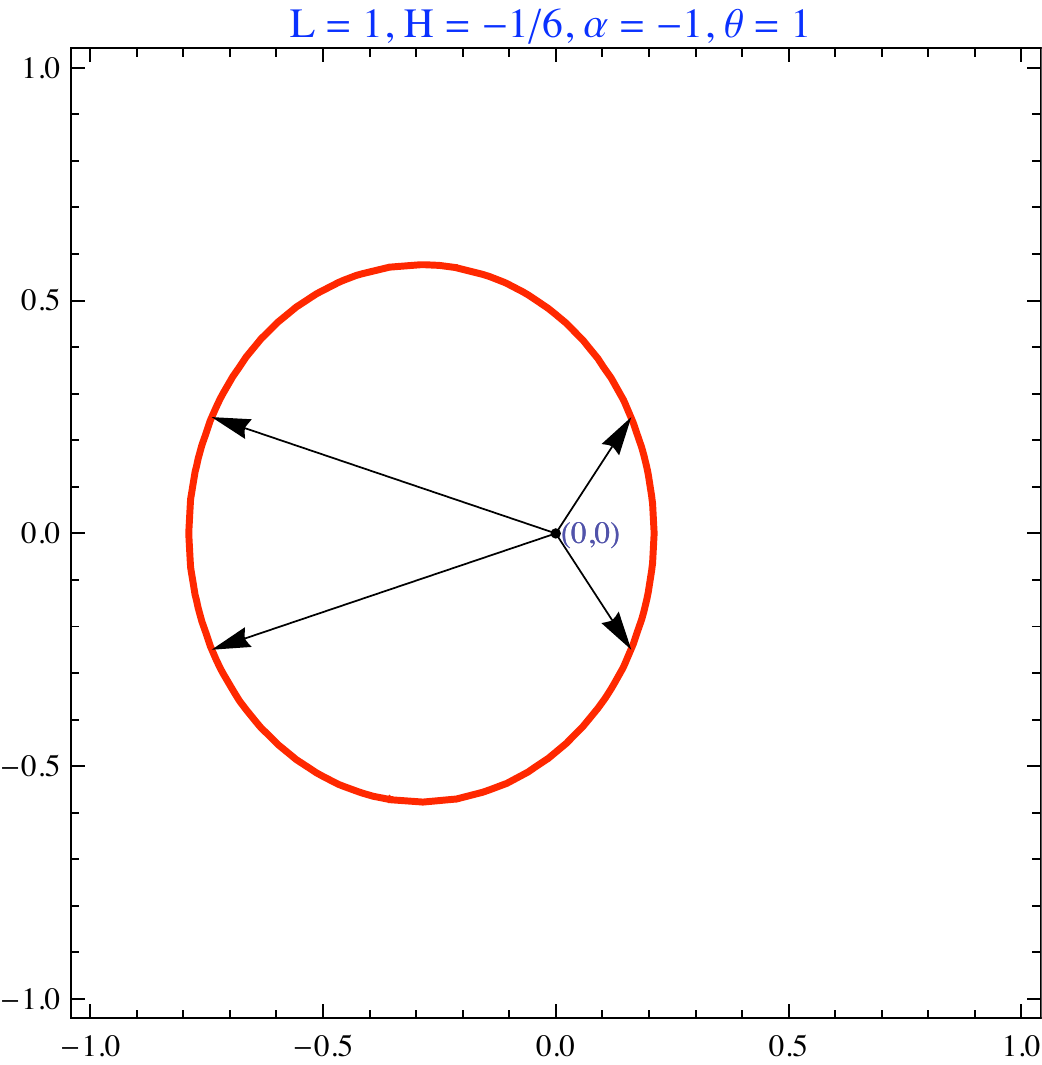}
\caption[]{$\H<0\,$ and the Newtonian potential is attractive $\,\a<0\,$, so the trajectories describe a whole ellipse.}
\end{center}
\end{figure}

When $\H>0$, which is the only possibility in the repulsive case $\,\a>0\,$, the origin is outside the ellipse so that only the right arc [denoted with the heavy line in the left side figure of (\ref{FigEll})] between the tangents drawn from the origin is obtained. However, positive hamiltonian $\,\H>0\,$, is also allowed for attractive Newtonian potential $\,\a<0\,$ but in that event the origin is again outside the ellipse so that the $\,\vx\,$-trajectories are confined on the left arc of the ellipse [denoted with the heavy line in the right side figure of (\ref{FigEll})]~:
\begin{figure}[!h]
\begin{center}
\includegraphics[scale=.56]{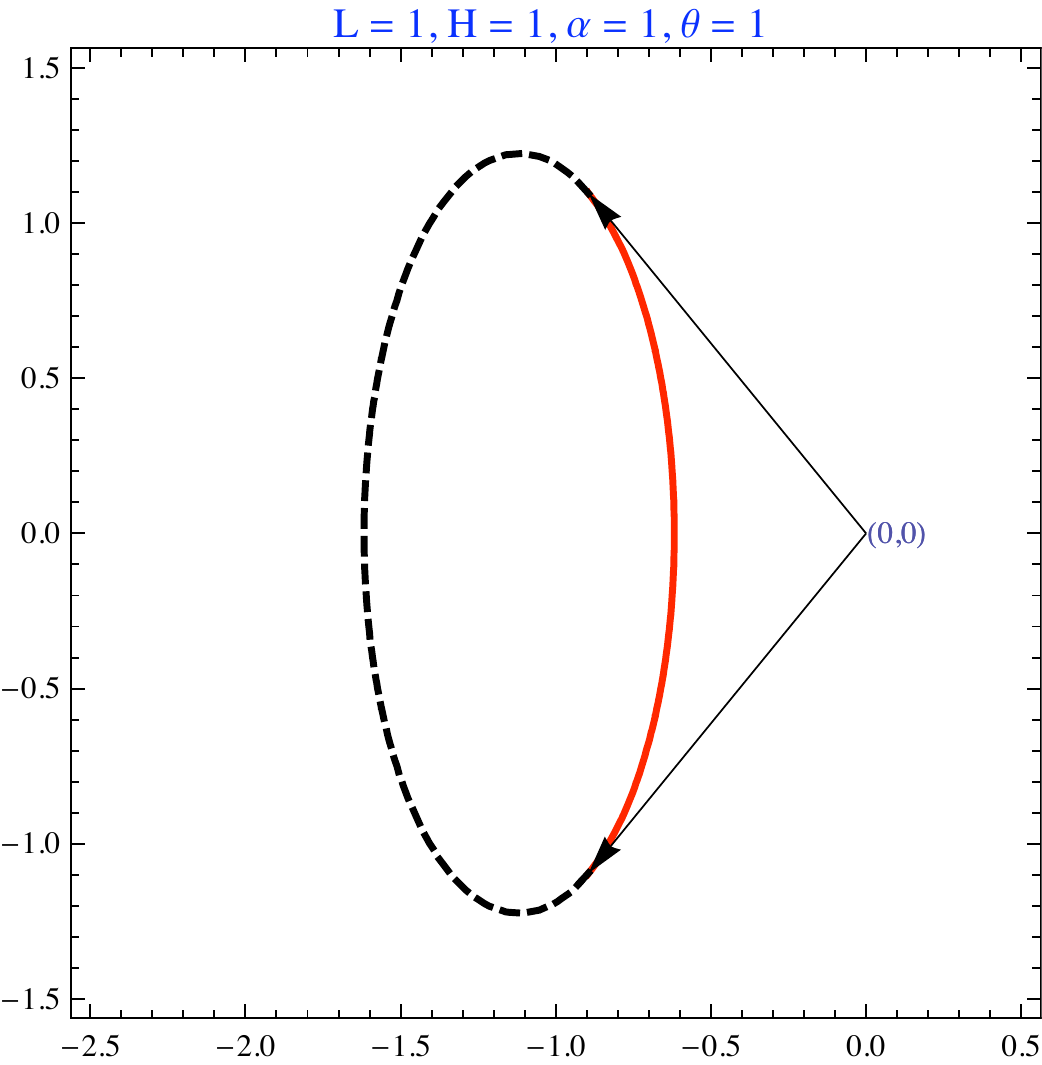}\qquad\qquad\qquad
\includegraphics[scale=.56]{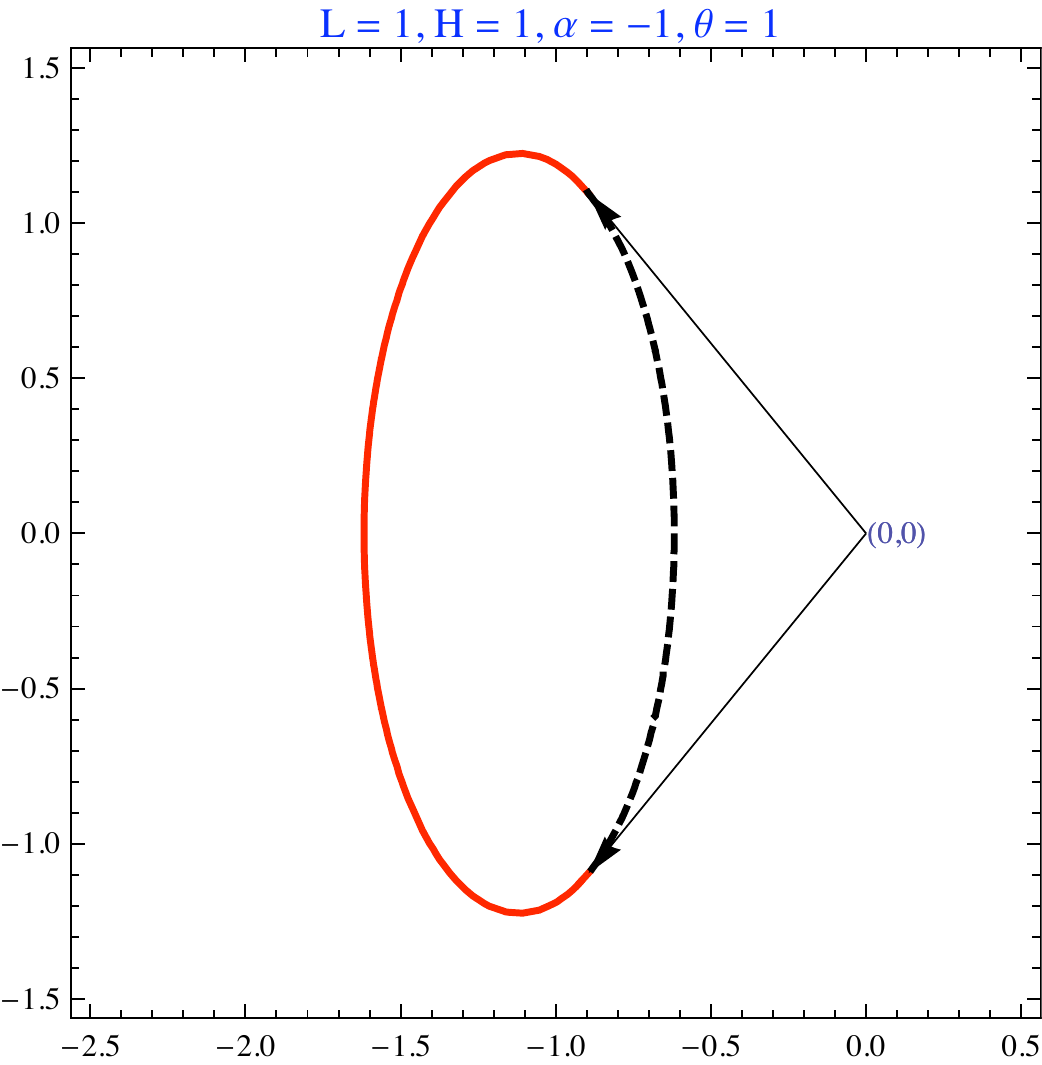}
\caption[]{The left side figure represents a right arc of an ellipse spanned by the $\,\vx\,$-trajectories for $\H>0$ and $\a>0$. While the right side figure represents a left arc of an ellipse spanned by the $\,\vx\,$-trajectories for $\H>0$ and $\a<0$.}\label{FigEll}
\end{center}
\end{figure}

For $\H=0$, the origin lies  on the ellipse, and 
``motion'' reduces to this single point~:
\begin{figure}[!h]
\begin{center}
\includegraphics[scale=.56]{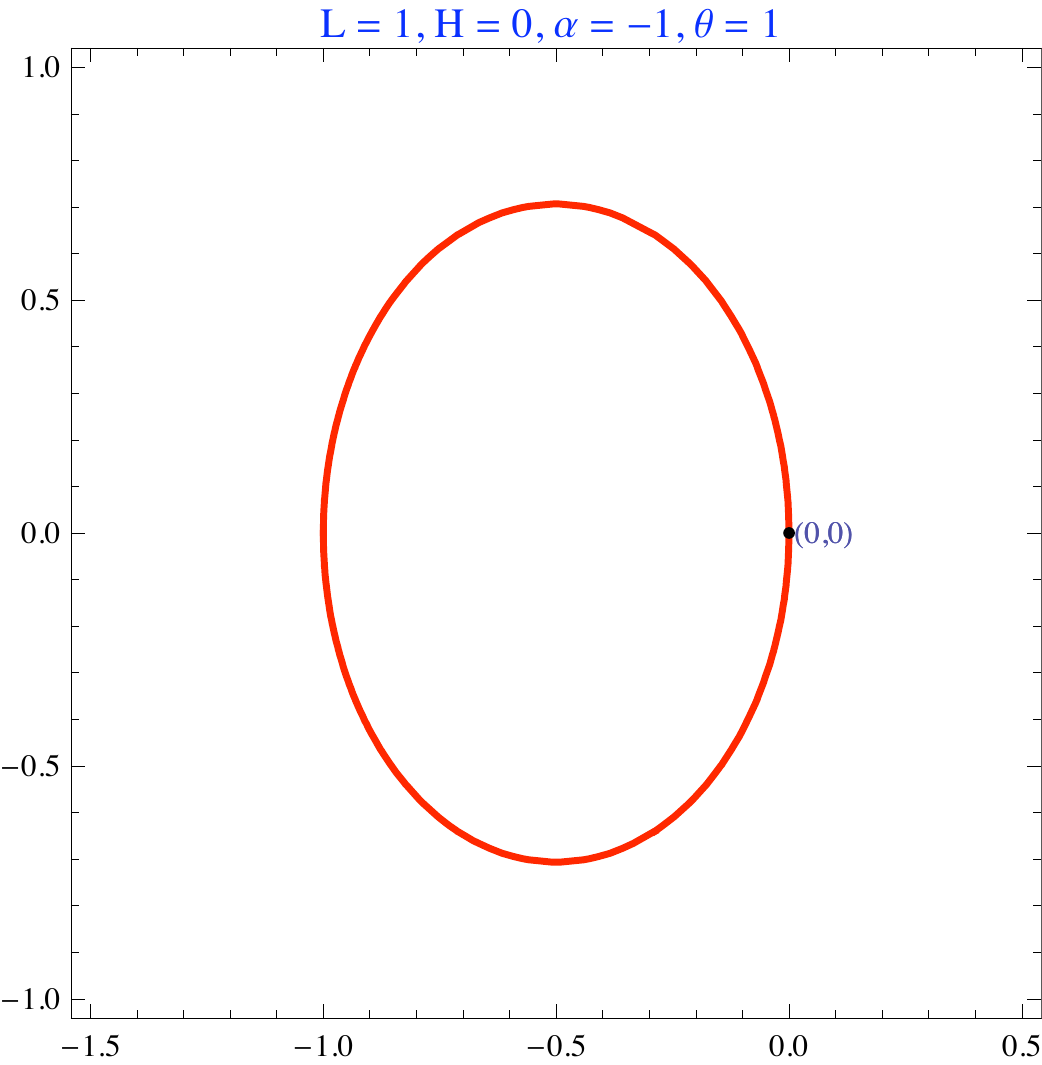}
\caption[]{$\,\vx\,$-trajectories degenerate to one single point for $\H=0$.}
\end{center}
\end{figure}

When the non-commutativity is turned off, $\theta\to0$, the known circular hodographs of the dual Kepler problem are recovered. As $\alpha\to0$, the trajectory becomes unbounded, and follows the $y$-axis.

So far, we only discussed classical mechanics. Quantization is now straightforward using the known group theoretical properties of the 
MICZ problem in dual space. The non-commutativity, alias monopole charge, $\theta$ has to be an integer or  half
 integer. This is indeed the first indication about the quantization of the non-commutative parameter. The wave functions should be chosen in the momentum representation, $\psi(\vp)$.
The angular momentum, $\vJ$, and the rescaled Runge-Lenz vector, $\vK/\sqrt{2|\H|}$, close into $\ort(3,1)/\ort(4)$
depending on the sign of the energy.  In the last case, the representation theory provides us with the discrete energy spectrum, see (\ref{SpectrumDirac}), [in units $\hbar=1$]
\beq
E_n=-\frac{\alpha^2}{2n^2},
\qquad
n=n_r+\frac{1}{2}+(l+\frac{1}{2})\sqrt{1+\frac{4\theta^2}{(2l+1)^2}}\, ,
\eeq
where $n=0,1,\dots,\, l=0,1,\dots$, with degeneracy $$\,n^2-\theta^2=\big(n-\theta\big)\big(n+\theta\big)\,.$$
Note that the degeneracy always takes integer or half-integer value, as it should, since $\,n\,$ and $\,\theta\,$ are simultaneously integer or half-integer.
The same result can plainly be derived directly by solving the Schr\"odinger equation in $\vp$-space \cite{MIC,Zwanziger}. Also related to the MICZ system, calculation of energy levels of hydrogen atom using NC QED theory is discussed in \cite{Chaichian0}.

Moreover, the symmetry extends to the
conformal $\ort(4,2)$  symmetry, due to the fact that the massless Poincar\'e orbits with helicity $\theta$ are in fact orbits of the conformal group, cf. \cite{CFH3}.

Let us observe that in most approaches one studies the properties (like trajectories, symmetries, etc.) of some given physical system. Here we followed the reverse direction: after positing the fundamental commutation relations, we were looking for potentials with remarkable properties. This leads us to the momentum-dependent potentials (\ref{Vee}), realizing a McIntosh-Cisneros-Zwanziger system \cite{MIC,Zwanziger} in dual space. Unlike as in a constant electric field \cite{AHEHPA}, the motions lie in an (oblique) plane. The particle is confined to bounded trajectories, namely to (arcs of) ellipses.

 The best way to figure our motions is to think of them as analogs of the circular hodographs of the Kepler problem to 
which they indeed reduce when the non-commutativity is turned off.
For $\H<0$, for example, the dual motions are bound, and the velocity
turns around the whole ellipse; for $H>0$ instead,
the motion along a finite arc, starting from one extreme point
and tending to the other one at the end of the arc, corresponds to
the variation of the velocity in the course of a hyperbolic motion of a comet, or in Rutherford scattering, but in dual space.
   
Our system, with monopole-type non-commutativity (\ref{monoptheta}), has some remarkable properties~: 

Momentum-dependent potentials are widely used in nuclear physics, namely in the study of heavy ion collisions, where they correspond to non-local interactions \cite{Nuphy,Das,Das1}. Remarkably, in non-commutative field theory, a $1/p^2$ contribution to the propagator emerges from UV-IR mixing.

The absence of a mass term should not be thought of as the system being massless; it  is rather reminiscent of ``Chern-Simons dynamics''  \cite{DJTCS}.

One can be puzzled how the system would look like in configuration space. Trying to eliminate the momentum from the phase-space 
equations (\ref{expleqmot}) in the usual way, which amounts to deriving $\dot{\vx}$ with respect to time and using the equations for $\dot{\vp}$,  fails, however, owing to the presence of underived $\vp$ in the resulting equation. This reflects  the non-local character of the system.

One can, instead,  eliminate $\vx$ using the same procedure, but in dual space. This yields in fact the equations of the motion of MICZ in dual momentum space,
\begin{eqnarray}
\ddot{p}=\frac{J^2}{p^3}+\frac{\alpha}{p^2}\,,    
\qquad
\ddot{\vec{p}}=\frac{\alpha}{p^3}\,\vec{p}-\frac {\theta}{p^3}\,\vec{J}\,.
\label{ppeq}
\end{eqnarray}

Are these equations related to a theory with higher-order derivatives of the type \cite{LSZ,LSZ1}~?
The answer is yes and no. 
The clue is that \emph{time is not a ``good'' parameter}
for Kepler-type problems, owing to the impossibility of expressing it from the Kepler equation \cite{Cordani}. This is also the reason for which we describe the \emph{shape} of the trajectories, but we do not integrate the equations of the motion. A ``better'' parameter can be found along the lines indicated by Souriau \cite{SouriauKepler,Bates} and then, deriving with respect to the new parameter, transforms (\ref{ppeq}) into  a fourth-order linear matrix differential equation, which can be solved.

It is, however, not clear at all if these equations derive from some higher-order Lagrangian, and if they happen to do, what would be the physical meaning of the latter.

The fourth-order equations do certainly \emph{not} come from one of the type stated in Ref. \cite{LSZ,LSZ1}~: the latter lives in fact in two space dimensions and has constant scalar non-commutativity $\theta$, while our system is 3-dimensional and has a momentum-dependent vector $\vTheta(\vp)$, given in (\ref{monoptheta}).

It is tempting to ask if the relation to the ``closest physical theory'' with a momentum-dependent potential, namely nuclear physics, can be further developed and if similar (super)symmetries can be found also in nuclear physics. Once again, the answer seems to be negative, though~: while dynamical symmetries \emph{do}
play a role in nuclear physics \cite{Iachello}, those used so far do not seem to be of a momentum-dependent Keplerian type.

\section{Concluding remarks}\label{chap:Conclusion}

In this thesis, we developed a systematic method to search for hidden and (super)symmetries of several physics system. In some cases, like in the SUSY of the monopole, our recipe needs to be extended to fermionic degrees of freedom. In the case of the momentum-space monopole, we needed to adapt our recipe to the non-commutative structure by interchanging the role of positions and momenta. In these models, as expected, the hidden symmetry of the Kepler-type is always related to the addition into the system of a fine-tuned inverse-square potential. This requirement appears clearly, in the van Holten algorithm, to be a consistency condition on the existence of a conserved Runge-Lenz-type vector.

Having introduced the Abelian Dirac magnetic monopole-field; we studied, in particular, the classical geodesic motion of a particle in Kaluza-Klein-type monopole spaces and its generalization: the Gibbons-Hawking space. We derived the conditions under which the Killing tensors imply the existence of conserved quantities on the dimensionally reduced curved manifold. We observed that the Killing tensor generating the Runge-Lenz-type vector, preserved by the geodesic motion, can be lifted to an extended manifold, namely, (\ref{LKT1}) and (\ref{LlightKillingTensor2}) \cite{DGH}. As an illustration, we have treated, in detail,  the generalized Taub-NUT metric, for which we derived the most general additional scalar potential so that the combined system admits a Runge-Lenz vector \cite{GW}. Another example considered is the multi-center metric where we have found a conserved Runge-Lenz-type scalar (\ref{RLScalar}), in the special case of motions confined onto a particular $2$-sphere. Moreover, from the Theorem \ref{NmetricRL} we deduced, for $N>2$, that no Runge-Lenz vector does exist in the case of $N$-center metrics. It is worth mentioning that apart from the generic importance of constructing constants of motion, namely in the confinement of particle to conic sections; the existence, in particular, of quadratic conserved quantity like Runge-Lenz vector yields the separability of the Hamilton-Jacobi equation for the generalized Taub-NUT metric and for the two-center metric.

In the case of isospin-carrying particle in a non-Abelian Wu-Yang monopole field, we found the most general scalar potential such that the combined system admits a conserved Runge-Lenz vector. Indeed, it generalizes the fine-tuned inverse-square plus Coulomb potential \cite{MIC, Zwanziger}, for a charged particle in the field of a Dirac monopole. Following Feh\'er, the result is interpreted as describing motion in the asymptotic field of a self-dual Prasad-Sommerfield monopole \cite{Feher:1984ik,Feher:1984xc,Feher:1986}.

We also treated the case of the effective ``truly'' non-Abelian monopole-like field generates by nuclear motion in a diatomic molecule. This system is due to Moody, Shapere and Wilczek where \textit{despite the non-conservation of the electric charge (\ref{noecharge})}, we surprisingly constructed, in addition to the ``unusual'' angular momentum (\ref{diatangmomJ}), a new conserved charge (\ref{Gamma}). 
 
We remarked that Runge-Lenz-type vector plays a role also in SUSY. Indeed, we investigated the bosonic symmetries as well as the supersymmetries of a spinning particle coupled to a magnetic monopole field. The gyromagnetic ratio  determines the type of (super)symmetry the system can admit~: for the Pauli-like hamiltonian (\ref{Hamiltonian}) $\,\N=1\,$ SUSY   only arises for gyromagnetic ratio $\gyro=2$ and with no external potential, $V=0$, confirming  Spector's observation \cite{Spector}. We also derived additional supercharges, which are not square roots of the Hamiltonian of the system, though. A Runge-Lenz-type dynamical symmetry  requires instead an anomalous gyromagnetic ratio,
$$\gyro=0\quad\hbox{or}\quad\gyro=4\,,$$
with the additional bonus of an extra ``spin" symmetry.
These particular values of gyro-ratio come from the effective coupling of the form $F_{ij}\mp\epsilon_{ijk}D_k\Phi$, which
add or cancel   for self-dual fields, $F_{ij}=\epsilon_{ijk}D_k\Phi$  \cite{FHsusy}. We found that the super- and the bosonic symmetry can be combined, but the price to pay is, however, to enlarge the fermionic space. This provides us with an $\,\N=2\,$ SUSY.

We also applied the van Holten algorithm to a planar fermion in any  planar magnetic field, i.e. one perpendicular to the plane. We shown, for ordinary gyromagnetic, that in addition to the usual supercharge (\ref{a.b5}) generating the supersymmetry,  the system also admits another square root of the Pauli Hamiltonian \cite{HmonRev} happening due to the existence of a dual Killing tensor.

A three-dimensional non-commutative oscillator with no mass term but with a certain momentum-dependent potential is obtained when studying the hidden symmetry of a monopole-type non-commutativity \cite{ZHN}. This oscillator system exhibits a conserved Runge-Lenz-type vector derived from the dual description in momentum space. The latter corresponds, but in dual space, to a Dirac monopole with a fine-tuned inverse-square plus Newtonian potential, introduced by McIntosh, Cisneros, and by Zwanziger some time ago. The resulting additional Kepler-type symmetry leads to the confinement of the particle's trajectories to bounded trajectories, namely to (arcs of) ellipses. When the non-commutativity is turned off, i.e. in the commutative limit, the motions reduce to the circular hodographs of the Kepler problem. It is worth mentioning that the momentum-dependent potentials which are rather unusual in high-energy physics, however, are widely used in nuclear physics, namely in the study of heavy ion collisions; they correspond to non-local interactions \cite{Nuphy,Das,Das1}. Moreover, in non-commutative field theory, it is remarkable that a $1/p^2$ contribution to the propagator emerges from the UV-IR mixing. See in \cite{Gubser}. The absence of a mass term in the Hamiltonian describing this non-commutative oscillator should not be thought of as the system being massless; it is rather  reminiscent of ``Chern-Simons dynamics''  \cite{DJTCS}.






\end{document}